\documentclass[11pt]{article} 

\usepackage[utf8]{inputenc} 
\usepackage{authblk}
\usepackage{caption}
\usepackage{subcaption}

\usepackage{geometry} 
\usepackage{graphicx} 

\usepackage{booktabs} 
\usepackage{array} 
\usepackage{paralist} 
\usepackage{verbatim} 
\usepackage{comment}
\usepackage{subfig} 
\usepackage{fancyhdr} 

\usepackage{sectsty}
\allsectionsfont{\sffamily\mdseries\upshape} 

\usepackage[nottoc,notlof,notlot]{tocbibind} 
\usepackage[titles,subfigure]{tocloft} 

\title{Beyond Contagion: Reality Mining Reveals Complex Patterns of Social Influence}
\author[1]{Aamena Alshamsi}
\author[2]{Fabio Pianesi}
\author[2,3]{Bruno Lepri}
\author[3]{Alex Pentland}
\author[1,3]{Iyad Rahwan}

\affil[1]{Masdar Institute, UAE}
\affil[2]{Foundation Bruno Kessler, Italy}
\affil[3]{Massachusetts Institute of Technology, USA}

\begin{document}
\maketitle

\section*{Abstract}
Contagion, a concept from epidemiology, has long been used to characterize social influence on people's behavior and affective (emotional) states. While it has revealed many useful insights, it is not clear whether the contagion metaphor is sufficient to fully characterize the complex dynamics of psychological states in a social context. Using wearable sensors that capture daily face-to-face interaction, combined with three daily experience sampling surveys, we collected the most comprehensive data set of personality and emotion dynamics of an entire community of work. From this high-resolution data about actual (rather than self-reported) face-to-face interaction, a complex picture emerges where contagion (that can be seen as adaptation of behavioral responses to the behavior of other people) cannot fully capture the dynamics of transitory states. We found that social influence has two opposing effects on states: \emph{adaptation} effects that go beyond mere contagion, and \emph{complementarity} effects whereby individuals' behaviors tend to complement the behaviors of others.  Surprisingly, these effects can exhibit completely different directions depending on the  stable personality or emotional dispositions (stable traits) of target individuals. Our findings provide a foundation for richer models of social dynamics, and have implications on organizational engineering and workplace well-being.


\section*{Introduction}

Social influence is a fundamental force in society that drives the formation and propagation of opinions \cite{asch1955opinions}, attitudes \cite{wood2000attitude}, behaviors \cite{socialphysics}, social norms \cite{cialdini2004social} and of psychological states \cite{Fleeson2007}. Its power can be exploited to increase political participation \cite{61million}, promote physical activity and personal well-being \cite{mani2013inducing}, and reduce energy consumption \cite{schultz2007constructive}. 

The metaphor of \emph{contagion} provides a powerful framework for modeling social influence \cite{burt1987social,coleman1957diffusion,Christakis2013}. Psychological and behavioral phenomena can be seen to \emph{spread}, like a disease, from one person to another as a result of face-to-face \cite{Christakis2007} or electronic communication \cite{coviello2014detecting,kramer2014experimental}. Recent work showed that contagion can characterize (at least partially) the spread of obesity \cite{Christakis2007}, eating habits \cite{Moturu2010}, cooperative behavior \cite{Fowler2010,jordan2013contagion,suri2011cooperation}, generosity \cite{macy-generosity}, smoking \cite{Christakis2008}, happiness \cite{Fowler2008}, smiling \cite{Pugh2001}, depression \cite{Madan2010,Howes1985}, and emotion more generally \cite{Hatfield1993,Barasade2002}. It is also possible to estimate parameters of epidemic models, such as the Susceptible-Infected-Susceptible (SIS) model, directly from behavioral data \cite{Hill2010,hill2010infectious}.

Recently, the notion of behavioral contagion in social networks has become a subject of heated debate, particularly surrounding the difficulty of differentiating between contagion and homophily from observational data \cite{shalizi2011homophily,Christakis2013, Steeg2013, aral2009distinguishing}. The present article is not a contribution to this debate, but rather raises a number of orthogonal issues that we believe are equally important to our understanding of social dynamics.

Firstly, existing literature has mostly captured the dynamics of long-lasting state, with a temporal resolution of months \cite{coleman1957diffusion} or years \cite{Christakis2013}, with no consideration to the fleeting states that we all go through daily. One possible reason for this could be that traditional sampling methodologies have so far made it difficult to capture those states, as well as the similarly fast-changing situational factors, at a high enough temporal resolution. 


Another limitation of existing work stems from the fact that social and biological contagion are fundamentally different \cite{dodds2005generalized}. In particular, existing contagion models focus on social-situational influences \cite{Fowler2008,Hill2010} and neglect the role of individual differences in psychological state dynamics \cite{friedkin2010attitude, smilkov2014beyond}. Individuals differ in their tendencies to experience given psychological states, since the \emph{dynamics of states} are a result of the interplay between situational factors (e.g., social interaction) and individuals' \emph{stable traits} (also known as \emph{dispositions}) \cite{Zelenski2000}. Thus, one should expect that individual between-subject differences should play an equally important role in social dynamics as situational factors \cite{Fleeson2004, FleesonNoftle2009}.

For example, consider the Big-Five personality model, which measures openness, conscientiousness, extraversion, agreeableness, and neuroticism \cite{costa2008revised}. A person might be high on the extraversion \emph{trait}, but his extraversion \emph{state} fluctuates over time. Indeed, through experience sampling, Fleeson \cite{Fleeson2001} showed that: (a) people pass through all levels of personality states in their daily lives; (b) the central tendency of personality states is stable and reflects the corresponding trait level; (c) within-person variation in personality states can be attributed to the interplay between situational aspects and the stable dispositions captured by traits.

Another source of limitation in the classical contagion model is that it typically deals with transition among discrete conditions -- infected and not infected. Yet, sociological and psychological theory has long recognized that social influence often involves changes across continuous \cite{friedkin2011social} or ordinal state structure \cite{Fleeson2004}, as people transition among levels of the same state (e.g. higher or lower extraversion).

We collected high-resolution data of daily face-to-face interaction among members of an organization, along with detailed experience sampling (3 times per day) of their Big-Five personality states, and their affective (emotional) states using measures of Positive and Negative Affect \cite{Watson1988}. Results reveal a complex picture whereby: (a) contagion plays a marginal role; (b) more nuanced effects like \emph{attraction}, \emph{inertia}, \emph{repulsion} and \emph{push} that are reminiscent of the mimicry/adaptation vs. complementarity distinction \cite{chartrand1999chameleon, tiedens2003power}; (c) these processes are moderated by individual dispositions (traits).

Our findings suggest that the intuitive preventive actions, of avoiding infection by staying away from people in undesirable states (e.g. depression) and seeking people in desirable states (e.g. high positive affect) may not always be effective. A more nuanced approach, which takes into account these subtle social influences, would be necessary. These results are highly relevant to organizational engineering \cite{Pentland2013} and efforts to make work environments, teamwork, and schools more effective \cite{Barrick2001,Judge2002,Bell2007,Barsade2003}.

\section*{Methods}

High-resolution sensors have made collecting and analyzing enormous amount of social interaction data possible \cite{Olguin2009, pentland2010honest, Cattuto2010, Salathe2010, Stehle2011}, alleviating the exclusive reliance on unreliable, subjective self-reports based on people's memory \cite{Todd2007}. Moreover, sensors can  log data at very fine time-scales without interfering with people's routines or consuming their time, making it easier to investigate short-duration phenomena. 

\subsection*{Data Collection}

\begin{figure*}
	\begin{center}
		\centerline{\includegraphics[width=1\textwidth]{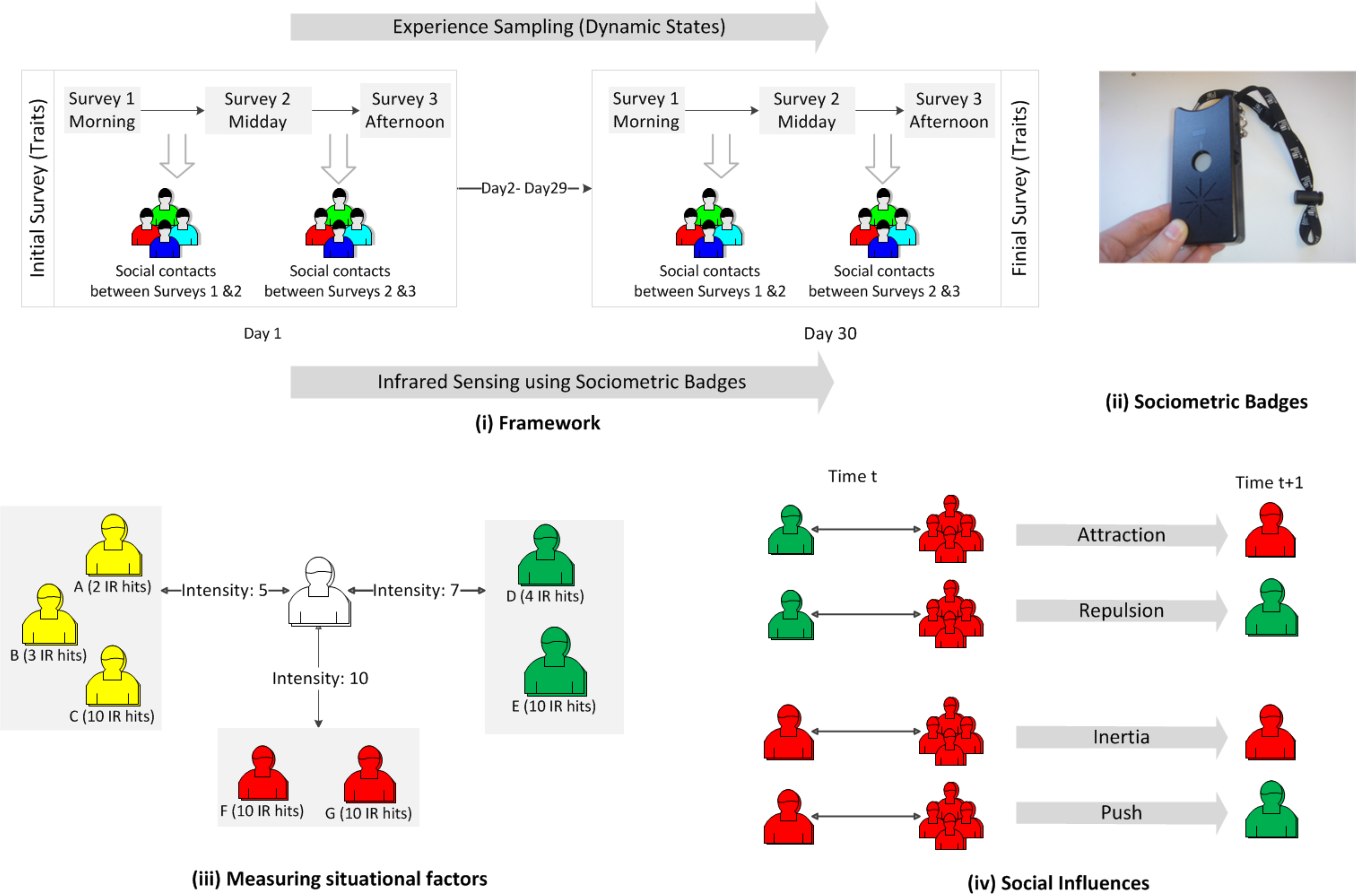}}
		\caption {{\bf (i) Framework of the Study:} First, participants filled a survey capturing personality and affect (stable) traits. Then they filled 3 daily surveys for 30 work days to measure personality and affect (dynamic) states. {\bf (ii) Sociometric Badge:} Each participant's social network between any two consecutive surveys was constructed from infrared sensor data from sociometric badges worn around the neck \cite{Olguin2009}. {\bf (iii) Measuring Situational Factors:} An example of how the situational factors (intensity of contacts) is calculated. The ego's infrared sensor (in the middle) detected 7 alters between two consecutive surveys. Two of the alters were in the high level (green) with 14 infrared hits, leading to intensity of contact $14/2=7$. Similarly, the intensity of contact with three alters in the neutral level (yellow) is $15/3=5$ and that for alters in the low level (red) is $20/2=10$.
			{\bf (iv) Social Influences:} Four possible social influences exemplified. Attraction: an ego in the high level interacts with others in the low level, then moves to the low level to adapt to his peers. Repulsion: a participant in the high level interacts with others in the low level, and consequently remains in the high level in complement to his peers. Inertia: a participant in the low level interacts with others in the same level, who prevent him from moving to a different level, maintaining his adaptation to their level. Push: a participant in the low level interacts with others in the same level, as a result pushing him away to a different, complementary level.}
		\label{framework1}
	\end{center}
\end{figure*}

\emph{Sociometric Badges}, designed and built by author Pentland, are capable of tracking various activities and behaviors of individuals \cite{Olguin2009}. These sensors --Fig. \ref{framework1}(ii)-- track face-to-face interactions by means of infrared (IR) sensors that recognize similar sensors facing them, implying that the two participants wearing them had a conversation or eye contact.

We used these badges to track face-to-face interactions of 52 individuals and conducted three daily experience sampling surveys \cite{csikszentmihalyi2003happiness} to collect information about their personality and emotional states --Fig. \ref{framework1}(i). In addition, affect and personality (stable) traits were measured at the beginning of the study. This methodology was applied to five personality states and their corresponding traits (extraversion, agreeableness, conscientiousness, emotional stability and creativity) and two affect states and their corresponding traits --high positive affect (HPA) and low negative affect (LNA).

Specifically, we followed the well-established procedure proposed by Fleeson \cite{Fleeson2001}. At the beginning and at the end of the experiment the participants filled extended surveys about: (1) dispositional (stable) personality traits (2) dispositional affective traits. These scores are considered as the dispositional factors of participants in our study. During the 30 work days, participants were asked to fill three \emph{experience sampling} surveys about transient psychological states (personality and affect) that they have experienced in the last 30 minutes. The three daily reports were the same format as Big Five scales traditionally used for traits, with the exception that rather than describing themselves in general, participants described their personality-related behaviors during the previous 30 minutes (e.g. during the last 30 minutes, how well does talkative describe you?). It is very unlikely that people would have experienced significantly varying affect or personality states during such a short period of time. The surveys were triggered to be sent via email every working day at (11:00 AM, 2:00 PM and 5:00 PM). The participants were given 2.5 hours to fill the surveys. We refer to the first survey as the morning survey, the second survey as the midday survey and the third survey as the afternoon survey. 

\begin{table}[!ht]
	\resizebox{0.95\textwidth}{!}{\begin{minipage}{\textwidth}
			\begin{tabular}{l l l}
				\hline
				Group  & Survey  & Measurement \\
				\hline
				Personality & \textbf{States} & Extraversion \\
				& Ten-Item Personality Inventory (TIPI) & Agreeableness \\
				&       & Conscientiousness \\
				& \textbf{Traits} & Emotional Stability \\
				& Big Five Marker Scale (BFMS) & Creativity \\
				\hline
				Affect & \textbf{States} & High Positive Affect \\
				& Positive and Negative Affect Schedule (PANAS) & High Negative Affect \\
				& \textbf{Traits} & Low Positive Affect \\
				& Multidimentional Personality Questionnaire (MPQ) & Low Negative Affect \\
				\hline
			\end{tabular}
		\end{minipage}}
		\caption{Surveys for personality and affect states and traits } 
		\label{tab:surveys}
	\end{table}

Table \ref{tab:surveys} summarizes the types of surveys used to capture different groups of states and traits. The Big Five Marker Scale (BFMS) is widely used to assess personality scores for extraversion, agreeableness, conscientiousness, emotional stability and creativity \cite{Perguini2002}. Therefore, BFMS was used in the Sociometric Badge Corpus at the beginning and at the end of the experiment to capture personality traits of participants \cite{Bruno2012}. Similarly, Multidimensional Personality Questionnaire (MPQ) was utilized to measure dispositional affective scores of participants \cite{Tellegen2008}. 

On the other hand, experience sampling surveys elicit transient states of personality and affect (emotions) including questions about BIG5 personality scale and fifteen items concerning affective states. Questions in these surveys report participants' states which were experienced in the last 30 minutes. For transient states of personality, the ten-item personality inventory TIPI was used in the experience sampling \cite{Gosling03avery}. For each personality dimension e.g. extraversion, we recoded the reverse-scored items and then we computed the average of the two items (the standard item and the recoded reverse-scored item) that make up each dimension. 

The short version of Positive and Negative Affect Schedule (PANAS) was used to evaluate the affective states of participants \cite{watson1988development}. High positive affect (HPA) was assessed using 3 items: \emph{enthusiastic}, \emph{interested} and \emph{active}. High negative affect (HNA) was assessed using 3 items: \emph{sad}, \emph{bored} and \emph{sluggish}. Low positive affect was assessed using 2 items: \emph{calm} and \emph{relaxed} while low negative affect (LNA) was assessed using 2 items \emph{lonely} and \emph{isolated}. 

The study lasted 30 working days and was performed during work hours at the premises of a research organization in northern Italy (for more details, see \cite{Bruno2012}). The data comprised 1,426 surveys by the 52 participant (see SI for more details). The total number of Infrared hits detected by participants' sensor while socializing with other participants is 114,642. When discussing the social interaction that a given participant was involved in, we refer to him/her as an \emph{ego} and to the peers he/she interacted with as \emph{alters}. Based on source, destination and time of those Infrared hits, we were able to construct the temporal face-to-face networks for each participant. When an ego's Infrared sensor detects an alter, a directional transient edge is created between ego and alter and the weight of the edge is determined by the number of Infrared hits detected by ego's sensor that are triggered by the recognition of alter's sensor. The temporal networks consisted of 1,643 transient edges.  

Personality and affect states' scores are measured through continuous scales which are quantized into three ordinal levels -- low, medium (neutral) and high -- according to the three quantiles of their distributions in such a way that level L (Low) consisted of cases between the 0-th and the 33rd quantile; level N (Neutral) consisted of cases between the 33rd and the the 66th quantiles, and level H (High) consisted of cases above the latter (see SI for more details). In \cite{Hill2010}, high and low state levels are considered as infectious while the medium (neutral) level is considered susceptible. In our case, state dynamics consist of changes (or lack thereof) of level between two consecutive surveys filled by a participant in a given day. Although our data is temporal, we focused on immediate single transitions within each state rather than looking at longer-term temporal trends. The addressed states fluctuate more than once on daily basis, so we just focused on the transitions that take place between two subsequent time periods and therefore we studied which social-situational factors are associated with those transitions.

Following the Italian regulations all participants were asked to sign an informed consent form and the study was conducted in accordance to them. The form and the general study was also approved by the Ethical Committee of Ca' Foscari University of Venice.

Contact \emph{intensity} is distinguished according to the level that the interacting partners were at the beginning of the relevant time interval (i.e., $t$) as illustrated in Fig. \ref{framework1}(iii). The intensity of contacts with alters in a particular level (e.g. high) of the state is the ratio between the total number of infrared hits with those people and the total number of unique alters in the level. As a consequence, we have three different measures of situational factors: $L$, $N$ and $H$ corresponding to the intensity of contact with people in the low, medium (neutral) and high levels, respectively. Concerning individual dispositions, the level transitions for a given state were associated with the normalized score of the corresponding trait measured at the beginning of the study.  In order to investigate the moderating role each trait plays in the association between transitions between levels of the state and the social-situational factors, we focused only on high and low trait scores ($\pm$ 1 standard deviation).    

\subsection*{Statistical Models}

For each possible transition between levels of a particular state (e.g. from Low to High in extraversion), our model consists of one dependent variable, the transition probability. Independent variables capture the corresponding trait score ($T$), the three situational measures $L$, $N$ and $H$ concerning contact intensity described above, the interaction effects $T*L$, $T*N$ and $T*H$ between trait and situational variables to account for the moderating effect of the former on the latter. The association between the dependent and the independent variables, including interactions, is modeled through logistic regression as shown in Equation \ref{eq:model}. Following Banerjee et al \cite{banerjee2013diffusion}, we used logistic regression instead of OLS regression (used by Hill et al \cite{Hill2010}) because the value of the dependent variable is binary (0 if there is no transition and 1 otherwise). Let $X \to Y$ denote a transition by the ego from level $X$ to level $Y$ of some state $S$ ($X=Y$ denotes stability). Let $p(X \to Y)$ be the probability of this transition between two consecutive surveys. For each dynamic state \emph{S}, we fit the following model:

\begin{equation}
ln (\frac{p(X \to Y)}{1-p(X \to Y}) = \alpha + \beta_{L}L + \beta_{N}N+  \beta_{H}H+ \beta_{T}T + \\
\beta_{T*L}T*L + \beta_{T*N}T*N + \beta_{T*H}T*H + \beta_{C}C 
\label{eq:model}
\end{equation}

where $\alpha$  is a constant (intercept); $\beta_{L}$, $\beta_{N}$, $\beta_{H}$ and  $\beta_{T}$ are  coefficients of the \emph{main} effects; $\beta_{T*L}$, $\beta_{T*N}$ and $\beta_{T*H}$ are coefficients of the interaction effects between the trait $T$ and the situational variables, and $\beta_{C}$ is the coefficient of all the control variables. Our inclusion of individual-level trait effect reduces the likelihood that correlation is driven by choice of social connections, since it accounts for observable homophily \cite{coviello2014detecting, banerjee2013diffusion}. However, latent homophile effects cannot be completely ruled out \cite{shalizi2011homophily}.

The model also contains control variables, denoted $C$, whose role is modelled by parameter $\beta_C$ (see SI for more details). Importantly, it has been shown that the time of day can have a significant, universal (i.e. culture-independent) effect on mood --e.g. people tend to be more positive in the morning \cite{golder2011diurnal}. This \emph{circadian rhythm} can be a major confounding variable in our analysis. By controlling for it, we eliminate a major confounding factor, since transitions between different levels of a particular state $S$ (e.g. HPA) may be correlated due to spontaneous changes due to the time of day.

\subsection*{Model and Parameter Estimation}
Our dataset consists of repeated observations for each participant, so we expected to have correlations within observations of participants. Hence, we used generalized linear models to analyze our longitudinal data using unstructured covariance matrices whereby variances and covariances are estimated directly from the data. Generalized Estimation Equations (GEE) are used to estimate the parameters of our models. For each transition in each state, we used backward elimination that starts with a full model that contains all candidate variables. Then, we tested the effect of deletion of insignificant variables using QICC (Corrected Quasilikelihood under Independence Models Criterion) \cite{pan2001akaike} iteratively until there is no further enhancement in the results.  We evaluated the goodness of fit based on QICC which is an indicator of goodness of fit of models that use generalized estimating equations. Therefore, it can be utilized to choose between two models favoring the one with the smaller QICC. After we end up with the best sub-model for each state transition, we compare its QICC to the QICC of the null model thats contains only the intercept. If the best sub-model is better than the null model, then we retain it. Otherwise, we consider the null model.

Although it is not possible, with current statistical techniques, to check whether a decrease in QICC is statistically significant, the significance of the independent variables remain the same during the backward elimination in the majority of the cases. Moreover, backward elimination is widely used with information criterion for model selection \cite{royston2009prognosis}. Hence, the usage of QICC for model selection, as suggested by Pan\cite{pan2001akaike}, is capable of supporting the significance of our results.


\section*{Results}

We investigated 63 types of transitions by applying Equation \ref{eq:model} (3 transitions for each state level $\times$ three levels per state $\times$ 7 states). We analyzed 2,378 transitions and 3,390 stability cases in total (see Table \ref{tbl:generalStatistics} for other statistics).

	\begin{table*}[!ht]
		\centering
		\begin{tabular}{ccccc}
			\hline
			\textbf{Transition} & \textbf{Max} & \textbf{Min} & \textbf{Median} & \textbf{Mean} \\
			\hline
			L $\to$ L  & 167 & 79  & 100 & 111 \\
			L $\to$ N & 98  & 72  & 81  & 84 \\
			L $\to$ H & 35  & 10  & 18  & 19 \\
			N $\to$ L & 112  & 59  & 95  & 90 \\
			N $\to$ N & 268 & 173 & 240 & 226 \\
			N $\to$ H & 99 & 46  & 57   & 61 \\
			H $\to$ L & 44  & 13  & 21  & 24 \\
			H $\to$ N & 81  & 42 & 67  & 62 \\
			H $\to$ H & 185 & 110 & 153 & 146 
		\end{tabular}
		\caption{{The maximum, minimum, median and mean number of transitions between levels of states for each transition.}}
		\label{tbl:generalStatistics}
	\end{table*}

Here, for simplicity, we discuss the results of conscientiousness state only and we move the results of other states to the supporting information. Table \ref{tbl:cons} shows the detailed results of the conscientiousness state. The search for confirmation of the contagion model involves investigating the transition from the neutral level to another level, and transitions from the high or the low level back to the neutral level. According to Hill et al's SISa model \cite{Hill2010}, a high or a low level is infectious only if (a) transitions to that level are only affected by the presence of people in that level, and (b) recovery is independent of social contact. We could not find any case that conformed to these conditions in conscientiousness state or the remaining six states as shown in SI.

	\begin{table*} 		[!ht]
		\resizebox{0.9\textwidth}{!}{\begin{minipage}{\textwidth}
				
				\begin{tabular}{lll|lll|lll}
					\multicolumn{3}{c}{\textbf{L to L}} & \multicolumn{3}{c}{\textbf{L to N}} & \multicolumn{3}{c}{\textbf{L to H}} \\
					\hline
					Variable  & Coefficient & P-value & Variable  & Coefficient & P-value & Variable  & Coefficient & P-value \\
					\hline
					L     & \multicolumn{1}{l}{0.003} & \multicolumn{1}{l|}{0.0017} & Intercept & \multicolumn{1}{l}{-0.441} & \multicolumn{1}{l|}{0.0000} & Intercept & \multicolumn{1}{l}{-1.727} & \multicolumn{1}{l}{0.0000} \\
					H     & \multicolumn{1}{l}{-0.002} & \multicolumn{1}{l|}{0.0129} & N     & \multicolumn{1}{l}{-0.003} & \multicolumn{1}{l|}{0.0000} & (period=1) & \multicolumn{1}{l}{0.715} & \multicolumn{1}{l}{0.0130} \\
					N     & \multicolumn{1}{l}{0.003} & \multicolumn{1}{l|}{0.0000} &       &       & \multicolumn{1}{l|}{} & L     & \multicolumn{1}{l}{-0.034} & \multicolumn{1}{l}{0.0000} \\
					T     & \multicolumn{1}{l}{-0.269} & \multicolumn{1}{l|}{0.0286} &       &       & \multicolumn{1}{l|}{} & N     & \multicolumn{1}{l}{-0.009} & \multicolumn{1}{l}{0.0041} \\
					L*T   & \multicolumn{1}{l}{0.003} & \multicolumn{1}{l|}{0.0127} &       &       & \multicolumn{1}{l|}{} & T     & \multicolumn{1}{l}{0.751} & \multicolumn{1}{l}{0.1131} \\
					H*T   & \multicolumn{1}{l}{-0.002} & \multicolumn{1}{l|}{0.0060} &       &       & \multicolumn{1}{l|}{} & L*T   & \multicolumn{1}{l}{0.013} & \multicolumn{1}{l}{0.0000} \\
					&       & \multicolumn{1}{r}{} &       &       & \multicolumn{1}{r}{} & N*T   & \multicolumn{1}{l|}{-0.008} & \multicolumn{1}{l}{0.0012} \\
					\multicolumn{3}{c}{\textbf{N to L}} & \multicolumn{3}{c}{\textbf{N to N}} & \multicolumn{3}{c}{\textbf{N to H}} \\
					\hline
					Variable  & Coefficient & P-value & Variable  & Coefficient & P-value & Variable  & Coefficient & P-value \\
					\hline
					Intercept & \multicolumn{1}{l}{-0.408} & \multicolumn{1}{l|}{0.0005} & H     & \multicolumn{1}{l}{-0.001} & \multicolumn{1}{l|}{0.0447} & Intercept & \multicolumn{1}{l}{-1.564} & \multicolumn{1}{l}{0.00} \\
					L     & \multicolumn{1}{l}{-0.0025} & \multicolumn{1}{l|}{0.0231} & T     & \multicolumn{1}{l}{-0.406} & \multicolumn{1}{l|}{0.0005} & (period=1)     & \multicolumn{1}{l}{0.726} & \multicolumn{1}{l}{0.00} \\
					L*T   & \multicolumn{1}{l}{-0.006} & \multicolumn{1}{l|}{0.0000} & L*T   & \multicolumn{1}{l}{0.0025} & \multicolumn{1}{l|}{0.0007} & T     & \multicolumn{1}{l}{0.405} & \multicolumn{1}{l}{0.013} \\
					H*T   & \multicolumn{1}{l}{-0.002} & \multicolumn{1}{l|}{0.0357} & H*T   & \multicolumn{1}{l}{0.003} & \multicolumn{1}{l|}{0.0010} & &  &  \\
					&       &       &       & \multicolumn{1}{l}{} & \multicolumn{1}{l|}{} & &  &  \\
					&       & \multicolumn{1}{r}{} &       &       & \multicolumn{1}{l}{} &       & \multicolumn{1}{l}{} & \multicolumn{1}{l}{} \\
					\multicolumn{3}{c}{\textbf{H to L}} & \multicolumn{3}{c}{\textbf{H to N}} & \multicolumn{3}{c}{\textbf{H to H}} \\
					\hline
					Variable  & Coefficient & P-value & Variable  & Coefficient & P-value & Variable  & Coefficient & P-value \\
					\hline
					Intercept & \multicolumn{1}{l}{-1.579} & \multicolumn{1}{l|}{0.0000} & Intercept & \multicolumn{1}{l}{-0.483} & \multicolumn{1}{l|}{0.0002} & N     & \multicolumn{1}{l}{0.003} & \multicolumn{1}{l}{0.0088} \\
					L     & \multicolumn{1}{l}{0.0024} & \multicolumn{1}{l|}{0.0294} & N     & \multicolumn{1}{l}{-0.0039} & \multicolumn{1}{l|}{0.0004} & T     & \multicolumn{1}{l}{0.5106} & \multicolumn{1}{l}{0.0000} \\
					H     & \multicolumn{1}{l}{-0.008} & \multicolumn{1}{l|}{0.0255} & T     & \multicolumn{1}{l}{-0.270} & \multicolumn{1}{l|}{0.0050} & N*T   & \multicolumn{1}{l}{-0.003} & \multicolumn{1}{l}{0.0144} \\
					L*T   & \multicolumn{1}{l}{-0.003} & \multicolumn{1}{l|}{0.0006} &       & \multicolumn{1}{l}{} & \multicolumn{1}{l|}{} &       & \multicolumn{1}{l}{} & \multicolumn{1}{l}{} \\
				\end{tabular}%
			\end{minipage}}
			\caption{{\bf Results of Conscientiousness State.} The mere effects of social-situational factors (intensity with alters in each level: L, N and H) and corresponding traits of egos (T) are reported in the table, if they are statistically significant. The interaction results between the two effect are reported also ($L*T$, $N*T$ and $H*T$), if they are statistically significant. The coefficients of the control variables are reported also: the main effect of the time of the day (period) and the interaction between the time of the day and the trait (period*T).  Some reported coefficients are relatively small, therefore we used a threshold of 0.001 to consider them relevant. We focus more on the direction of the effect (increase or decrease in the probability) rather than the actual value of the effect.}
			\label{tbl:cons}
		\end{table*}

\begin{figure*}
	\begin{center}
		{\includegraphics[width=1.05\textwidth]{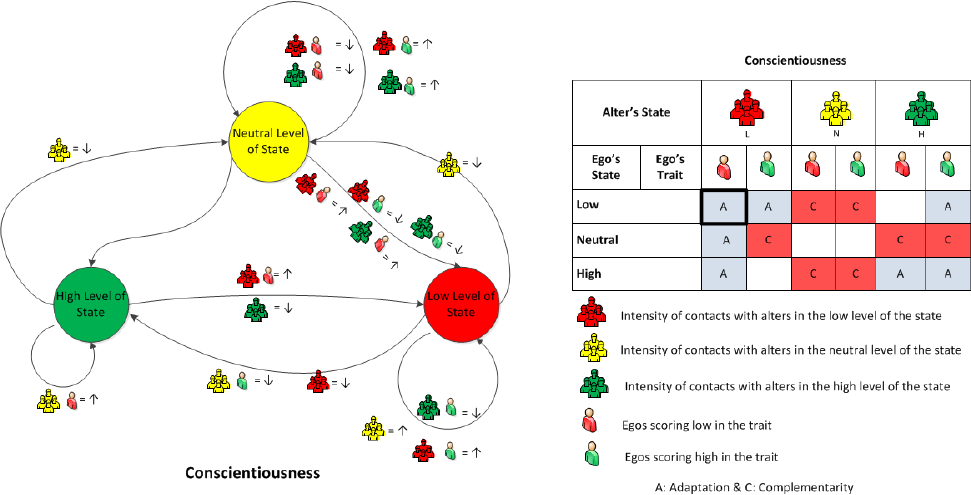}}
		\caption {{\bf (Left) Transition graph for conscientiousness:} Nodes represent conscientiousness level of the ego. Arrows represent transitions from one level to another. Transitions are labeled with conditions that affect the corresponding probabilities. Icons represent the conscientiousness levels of alters and ego's trait level. Symbol $\uparrow$ (respectively $\downarrow$) indicates an increase (respectively decrease) in transition probability associated with the given combination of alters' state level and ego's trait level. For example, if the ego high in the conscientiousness state, then the probability of the ego transitioning to the high level decreases with ego's contact with alters in the neutral level of the state. Another example is the transition from neutral to low level of conscientiousness, which is moderated by the ego's trait score. If the ego is high in trait, then the probability of transition from a neutral state to a low state decreases with his contact with alters in high and low levels. But if the ego is low in the trait, then the probability increases instead.  {\bf (Right) Social influences:} The table summarizes the level transition graph by means of adaptation (A) and complementarity (C). Rows represent ego's state levels; columns are labeled with alters' state levels and sub-labeled with ego's trait level (Low or High). Cells report the effects observed when egos in the corresponding state level and trait level interact with alters in the corresponding state level. For example, the square with thick border indicates that when the ego is low in conscientiousness state and also low in conscientiousness trait, contact with alters who are also low in conscientiousness state results in an adaptation effect. Empty cells lack statistically significant effects in a given combination.}
		\label{cons}
	\end{center}
\end{figure*}

We did, however, find evidence for \emph{conditional contagion}. These are cases where the satisfaction of Hill et al's definition was contingent on trait level. We found that transitions $(N \to L)$  do satisfy these requirements for people who have low scores in the conscientiousness trait. The same pattern was not observed for egos who have high scores in the conscientiousness trait; in this case, the probabilities of ($N \to L$) decrease when the contact intensity with alters in state level $L$ increases. This is illustrated in the state diagram in Fig. \ref{cons} (left). No other cases conformed to the contagion model. 

One reason for the failure of the contagion model is the fact that the role of alters in the neutral state level is not as passive as required by Hill et al \cite{Hill2010} to define the recovery from infection. However, the presence of alters in the neutral level can be associated with changing probabilities of moving towards that level. This is what happens with alters in the neutral level whose presence was associated with a decrease in the probability of $(H \to N)$ and $(L \to N)$ transitions.

We observed also other roles of alters. For example, the presence of alters in the low level is associated with an increase in the probability of $(H \to L)$ for people who have low scores in the conscientiousness trait; while contact with alters in a low level of the state is associated with a decrease in the ego's probability of transition $(N \to N)$. 

Putting together all these observations, we introduce \emph{attraction}, as a generalization of contagion. Intuitively, attraction obtains whenever increased interaction with alters in a state level different from the ego's corresponds to either an increased probability for the ego to move towards that level or a decreased probability to move away from that level.  Two things should be noted about this definition: (a) it does not require the ``behavior'' of the ego to fully conform to that of his/her alter, but simply to become more similar to it; (b) nothing is assumed concerning the recovery mechanism.

Attraction, by itself, is not enough, though. Often, we observed \emph{repulsion} -- cases in which increased interaction with people in a certain state level corresponds to decreasing probabilities of transition towards that level. This is what happens, for examples, with transitions $(L \to N)$ and $(H \to N)$. In the trait-conditional mode, it applies to the $(N \to L)$  transition of conscientiousness state for people who have high scores in the corresponding trait.

\begin{figure*}
	\begin{center}
		\centerline{\includegraphics[width=1.1\textwidth]{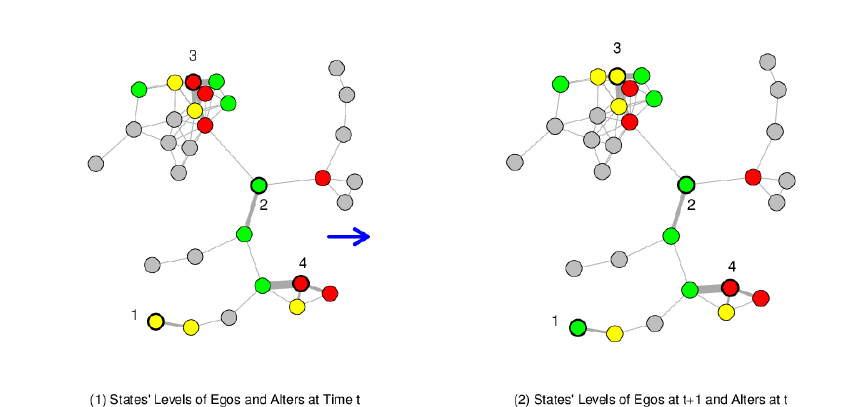}}
		\caption{{\bf Two consecutive snapshots of 1st to 2nd survey in day 1, with the transient social network.} Nodes are participants who filled both surveys. Edge thickness is proportional to contact intensity (IR hits) between surveys. Colors represent state levels (red:low, yellow:medium, green:high) and are shown only for egos 1--4 and their alters. The four types of social influence discussed in the text can be seen: ego 1 moved from the neutral level to the high level in the presence of interaction with an alter in the neutral level (push). Ego 2 remained in the high level in correspondence to intense contact with an alter in the high level (inertia). Ego 3 was in the low level and moved to neutral after intense contact with alters in the high level (attraction). Remarkably, the three egos have low scores in their corresponding traits.  Ego 4 remained in the low level after contact with an alter in the high level (repulsion). The represented states are creativity for node 1, extraversion for nodes 2 and 4 and agreeableness for node 3.}
		\label{temporalNetwork}
	\end{center}
\end{figure*}

Attraction and repulsion cover the associations between the intensity of contacts with alters in a level \emph{different} from the ego's initial one. But there is no reason not to expect that the intensity of contacts with people in the \emph{same} level as the ego's can also affect his/her transition probabilities. For instance, Ego 1 of Fig. \ref{temporalNetwork}, who was in the neutral level and interacted with an alter in the same level, was \emph{pushed} to the high level at the later sampling. Ego 2, who was in the high level and interacted with people in that same level, did not move, a situation we call \emph{inertia}.

At the level of the whole sample, inertia is exemplified by increasing probability for people who are in the low level of the state but have high scores in the conscientiousness trait to remain at that level when they interact with alters in the low level of state ($L \to L$). Similarly, the intensity of contacts with alters in the high level of the state is associated with a decrease in the probability of people in the same level to go to the low level ($H \to L$). There are push effects in the conscientiousness state. However, there are some cases in other states (for push effects, see SI).

In \emph{attraction}, alters are in levels that are different from egos and therefore they encourage people to move to their levels; whereas, in \emph{inertia}, alters are in the same levels of egos and encourage them to stay at the same levels. Both attraction and inertia represent a tendency of egos to \emph{conform} to their alters. In \emph{repulsion}, alters are in levels different from egos and therefore discourage people to move to their levels; whereas, in \emph{push}, alters are in the same levels of egos and push people away from their levels. Therefore, repulsion and push represent the tendency to \emph{diverge} from alters. We can, therefore, subsume our four effects under  the \emph{mimicry/adaptation} vs. \emph{complementarity} distinction  \cite{chartrand1999chameleon, tiedens2003power, pentland2010honest}.

So far, we addressed one transition at a time. For a more complete picture, observe the state diagram of conscientiousness in Fig. \ref{cons} (left) that summarizes the effect of the interplay between social-situational factors and individual dispositional factors on each transition. Due to space limitation, detailed results of other states can be found in SI. We present our results in a more compact format in Fig. \ref{cons} (right). Alters in the low level of the state attract egos to stay in the low level or switch to lower levels if those egos are already not conscientious by nature. Nevertheless, the alters cannot drag conscientious egos by nature towards their level unless the egos are already in the low level. Alters in the neutral level repulse egos to stay in their transient levels. Alters in the high level help egos to stay in the high level or move to their level except egos who are in the neutral level and have low trait scores.  

\section*{Discussion}

\subsection*{Summary and Implications}

From a methodological perspective, our study shows how the combination of automated sensing of social interaction with high-frequency experience sampling \cite{killingsworth2010wandering} can build a detailed picture of the dynamics of personality and affect states in a sizable work community. This provides a significantly finer grained perspective compared to methodologies that exclusively rely on surveys and self-reported social interaction \cite{Fowler2008,Hill2010}. This methodology can be applied well beyond the present study, e.g. to study the spread of healthy behavior or productive work practices in an organization.

Our methodology quantified complex patterns of social influence that go beyond the contagion metaphor \cite{Fowler2008,Hill2010}. The associations we identified between social-situational aspects and transition probabilities -- attraction, repulsion, inertia and push --  account for a consistent majority (70\%, 43 out of 63 ) of the transition types (3 transitions for each state level $\times$ three levels per state $\times$ 7 states) within the seven personality and affect states addressed in this work. They and their grouping under the headings of adaptation/mimicry and complementarity constitute an alternative and, we think, more appropriate taxonomy of social influence, one that is better suited to the ordinal nature of our psychological states. The model that emerges is one in which personality and affect states are not caught from someone else; they are not the result of mere contact, but from the ways egos (possibly unconsciously) respond to other people's behaviors by either adapting to it -- as discussed by social psychologists under the rubric of the ``perception behavior link'' or ``chamaleon effect'' \cite{chartrand1999chameleon} -- or diverging from it. From this perspective, the ubiquitous moderating effect of individual differences (traits) corresponds to differential dispositions to respond to external solicitations.

Our observations suggest that interventions should not simply increase the prevalence of desired behavior. For example, while extrovertedly acting peers bring an introvert out of his shell (contagion), they push already extroverted persons towards introversion. So, simply adding extroverts to a group may not lead to an increase in overall extraversion. 

In this work, we examined the role of corresponding traits in moderating the relationship between social-situational factors and variability within personality and affective states. Nevertheless, it was previously found that some traits such as extroversion and emotional stability traits are associated with fleeting affective states \cite{tellegen1985structures, watson1984negative}.  In the future, it would be interesting to conduct similar investigations using all personality and affective traits.

\subsection*{The Issue of Causality}

Recently, the notion of contagion has been subject to considerable debate. In a seminal paper, Shalizi and Thomas identified situations in which latent homophily (similarity among interaction partners) together with causal effects from the homophilous trait cannot be distinguished, observationally, from contagion \cite{shalizi2011homophily}. They argue that this raises barriers to many inferences social scientists would like to make about the underlying causal mechanisms. They further argue that these barriers can only be overcome by making exogenous assumptions about the causal architecture of the process in question.

While the contagion versus homophily debate is still ongoing \cite{Christakis2013, Steeg2013}, it is important to clarify that our present paper is neutral on this particular issue. Rather, our results call for attention to a fundamental orthogonal issue: even if we take contagion as a given, it would still fail to account for the majority of potential social influence effects (i.e. effects that \emph{may} be attributed to contact among individuals). 

That is, regardless of whether true (causal) contagion exists, we show that all kinds of other associations (not just positive) can be observed when psychological dynamics are measured at higher resolution, and that these associations are moderated by individual stable traits. But the nature of our data still restricts us to correlational conclusions. For instance, just as contagion and homophily may be confounded in observational data \cite{shalizi2011homophily}, it may very well be the case that our observed `repulsion' or `push,' may be confounded with corresponding phenomena like `heterophily' (the tendency to interact preferentially with people who exhibit opposing behavior or traits). Whether these associations reflect new causal mechanisms of influence remains an open question, and claims of causality must be based on well-justified assumptions \cite{Christakis2013}. The goal of this paper is precisely to motivate further investigation of these issues.

Said differently, we hope that our findings will encourage the broadening of the present research agenda on social dynamics \cite{shalizi2011homophily, Steeg2013, aral2009distinguishing, aral2012identifying, Hill2010, hill2010infectious, muchnik2013social, coviello2014detecting}, from the specific idea of `contagion' to the broader notion of `social influence' which manifests itself through other psychological mechanisms (like mimicry/adaptation and complementarity). If true social influence exists, contagion is only a small part of it, and more complex interpersonal psychological dynamics are likely at play. Recognizing this is, we believe, a necessary prerequisite to unravelling the true nature of social dynamics.


\section*{Competing Financial Interests}

The author declare no competing interests.

\section*{Author Contributions}

IR, BL, FP conceived, designed, and coordinated the study; BL and FP collected the data; AA carried out data processing, statistical analysis, and visualization of results. AP, BL, FP and IR interpreted the results; All authors wrote the manuscript and gave final approval for publication.




\newpage 

\section*{Supporting Information}

\section*{Methods}
\subsection*{Data Collection}
Through an experiment, the SocioMetric Badges Corpus tracked the activities of 52  participants \cite{Bruno2012}. The fifty two participants are employees in a research institution in Italy who volunteered to participate in the experiment for six weeks (working days are considered only). They belong to five units whereby all the employees of these units participated in the experiment along with the heads of these units. Their ages range from 23 to 53 with an average of 36. Forty seven participants are men (90.3\%) and five are women (9\%). Forty four participants are Italian (84\%)  and eight participants are from other countries (15.3\%). Forty-six out of the fifty-three participants were researchers in computer science belonging to four research groups; the remaining six participants were part of the full-time IT support staff. 

\subsection*{Procedure}

At the beginning and at the end of the experiment, the participants filled extended surveys about: (1) dispositional (stable) personality traits (2) dispositional affective traits. These scores are considered as the dispositional factors of participants in our study. During the 6 weeks, participants were asked to fill three \emph{experience sampling} surveys about transient psychological states (personality and affect) that they have experienced in the last 30 minutes. It is very unlikely that people would have experienced significantly varying affect or personality states during such a short period of time. The surveys were triggered to be sent via email every working day at (11:00 AM, 2:00 PM and 5:00 PM). The participants were given 2.5 hours to fill the surveys. We refer to the first survey as the morning survey, the second survey as midday survey and the third survey as the afternoon survey.

Table \ref{tab:surveys} summarizes the types of surveys used to capture different groups of states and traits. The Big Five Marker Scale (BFMS) is widely used to assess personality scores for extraversion, agreeableness, conscientiousness, emotional stability and creativity \cite{Perguini2002}. Therefore, BFMS was used in the Sociometric Badge Corpus at the beginning and at the end of the experiment to capture personality traits of participants \cite{Bruno2012}. Similarly, Multidimensional Personality Questionnaire (MPQ) was utilized to measure dispositional affective scores of participants \cite{Tellegen2008}. 

On the other hand, experience sampling surveys elicit transient states of personality and affect (emotions) including questions about BIG5 personality scale and fifteen items concerning affective states. Questions in these surveys report participants' states which were experienced in the last 30 minutes. For transient states of personality, the ten-item personality inventory TIPI was used in the experience sampling \cite{Gosling03avery}.  For each personality dimension e.g. extraversion, we recoded the reverse-scored items and then we computed the average of the two items (the standard item and the recoded reverse-scored item) that make up each dimension. 

The short version of Positive and Negative Affect Schedule (PANAS) was used to evaluate the affective states of participants \cite{watson1988development}. High positive affect (HPA) was assessed using 3 items: \emph{enthusiastic}, \emph{interested} and \emph{active}. High negative affect (HNA) was assessed using 3 items: \emph{sad}, \emph{bored} and \emph{sluggish}. Low positive affect was assessed using 2 items: \emph{calm} and \emph{relaxed} while low negative affect (LNA) was assessed using 2 items \emph{lonely} and \emph{isolated}.

After we calculate the scores of personality and affective states of each participant in each filled survey, we centered all the scores of personality and affect dynamic states using the median of each state. We generated the quantiles for each state, discretizing the scores of personality and affect dynamic states into three ordered levels (Low, Neutral and High).The three levels were identified on the basis of the 33rd and 66th quantiles of the state scores distribution in such a way that level $L$ (low) consisted of cases between the 0-th and the 33rd quantile; level $N$ (neutral) consisted of cases between the 33rd and the the 66th quantiles, and level $H$ (high) consisted of cases above the latter.

Note that the experience sampling method has a long history and is highly reliable in measuring dynamics of psychological states within individuals \cite{conner2009experience}. For those interested in the caveats around the use of experience sampling, we also point to extensive discussions elsewhere \cite{csikszentmihalyi1987validity}.

The participants wore SocioMetric Badges every working day within the institution. These sensors are equipped with accelerometers, audio, Bluetooth and Infrared to respectively capture: body movements, prosodic speech features, co-location with other individuals and face-to-face interactions \cite{Olguin2009}. We harnessed Infrared (IR) transmissions to detect face-to-face interactions between people. In order for a badge to be detected through IR, two of them must have a direct line of sight and the receiving badge's IR must be within the transmitting badge's IR signal cone of height $h <= 1$ meter and a radius of $ r <= htan\theta$, where $\theta$ = $15^\circ$ degrees; the infrared transmission rate (TRir) was set to 1Hz. 

\subsection*{Preprocessing the surveys}

The data comprises 1,426 surveys by the 52 participants. Ideally, the number of filled surveys should be 4,680 (52 participants $\times$ 3 daily surveys $\times$ 30 working days). However, participants reported absence from work 536 times \cite{Bruno2012}. This reduces the number of expected responses to 4,144. We addressed transitions in states, the levels between daily first surveys and second surveys and transitions in states, the levels between the second and the third surveys. Therefore, the number of daily expected transitions is 2. This further reduces the number of records (transitions) by 1,560 (52 participants $\times$ 30 days) to reach 2,582 expected records. Also, it has been observed that the majority of participants used to leave the organization before 5PM on Fridays afternoon. This further reduces the number of expected responses. The response rate of filling surveys is 83.9\% according to Lepri et al \cite{Bruno2012} which means that participants skip some surveys despite their availability at work. Unfilling surveys has an impact on the number of transitions. For example, if a participant missed the second daily survey, then two transitions will be missing: transition from first to second survey and transition from second survey. Also, we did not consider the transitions of egos in which the alters have not filled the corresponding surveys. Therefore, we ended up having only 1,426 surveys.

\paragraph{Transient states of personality and affect}
First we centered all the scores of personality and affect dynamic states using the median of each state. Then, we generated the quantiles for each state, discretized the scores of personality and affect dynamic states into three ordered levels (Low, Neutral and High).
 The three levels were identified on the basis of the 33rd and 66th quantiles of the state scores distribution in such a way that level L (low) consisted of cases between the 0-th and the 33rd quantile; level N (neutral) consisted of cases between the 33rd and the the 66th quantiles, and level H (high) consisted of cases above the latter.  We discarded both high negative affect and low positive affect because their distributions are very skewed making it difficult to identify appropriate intervals for each level of these states. Hence, high negative affect and low positive affect were not discussed in the results. 
 
 \begin{figure*}[ht]
 	\begin{center}
 		\centerline{\includegraphics{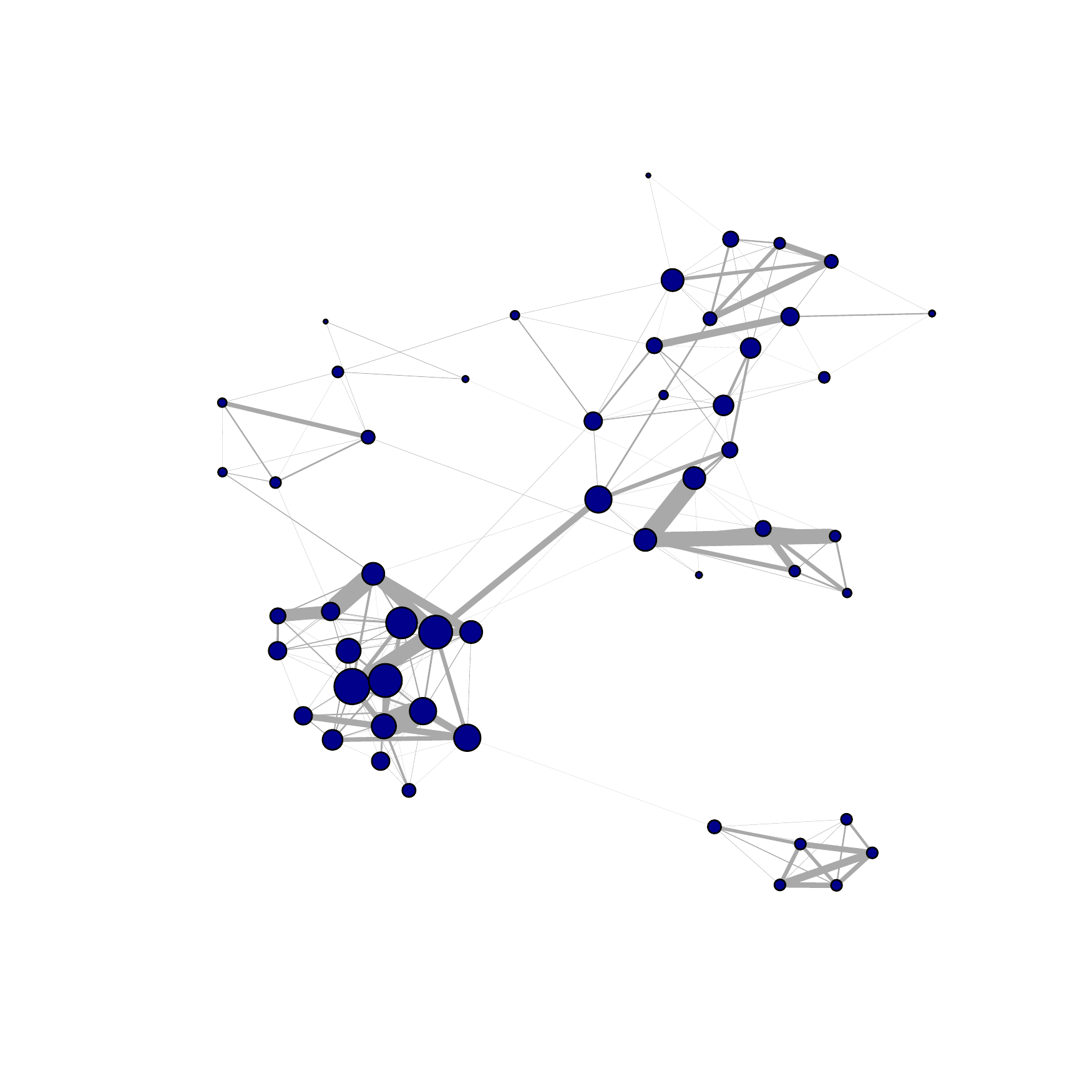}}
 		\caption{
 			{\bf The composite social network of participants.}  The aggregate social network is plotted showing the social ties among participants where the nodes are the participants and the edges are the IR hits detected between each pair of participants. The size of the nodes indicates their degree while the thickness of the edges is proportional number of total number of IR hits between a pair of participants. \textit {Remarks: We considered only IR hits that are in total more than 10 to create the social ties between participants.}
 		}
 		\label{fig:socialnetwork}
 	\end{center}
 \end{figure*}  

Accordingly, the level of each participant in each dynamic state recorded at a given survey is identified to be one of the three above levels. We leveraged those levels to extract the transitions in levels of dynamic states between two consecutive survey in the same day. Also, we used those levels to extract the social-situational factors by categorizing the social ties (other participants) whom the Infrared sensor of a particular participant detected between two consecutive surveys to be one of the three predefined levels. Remarkably, we used the lagged levels of the social ties. For example, the transition in question is taking place between the morning survey (time $t$) and midday survey (time $t+1$) for a particular participant. Therefore, we consider the scores of social ties recorded in the morning survey (time $t$). The statistics of transitions in each level is provided in Table \ref{tbl:trans}.

\begin{table}[h]
\resizebox{0.755\textwidth}{!}{\begin{minipage}{\textwidth}

\begin{tabular}{lrrrrrrr}
\hline
\multicolumn{1}{c}{\textbf{Transition}} & \multicolumn{1}{c}{\textbf{Extraversion}} & \multicolumn{1}{c}{\textbf{Agreeableness}} & \multicolumn{1}{c}{\textbf{Conscientiousness}} & \multicolumn{1}{c}{\textbf{Emotional Stability}} & \multicolumn{1}{c}{\textbf{Creativity}} & \multicolumn{1}{c}{\textbf{HPA}} & \multicolumn{1}{c}{\textbf{LNA}} \\
\hline
\multicolumn{1}{l}{L $\to$ L} & \multicolumn{1}{c}{79} & \multicolumn{1}{c}{100} & \multicolumn{1}{c}{167} & \multicolumn{1}{c}{147} & \multicolumn{1}{c}{107} & \multicolumn{1}{c}{91} & \multicolumn{1}{c}{89} \\
\multicolumn{1}{l}{L $\to$ N} & \multicolumn{1}{c}{77} & \multicolumn{1}{c}{98} & \multicolumn{1}{c}{83} & \multicolumn{1}{c}{72} & \multicolumn{1}{c}{97} & \multicolumn{1}{c}{81} & \multicolumn{1}{c}{79} \\
\multicolumn{1}{l}{L $\to$ H} & \multicolumn{1}{c}{35} & \multicolumn{1}{c}{25} & \multicolumn{1}{c}{21} & \multicolumn{1}{c}{10} & \multicolumn{1}{c}{14} & \multicolumn{1}{c}{18} & \multicolumn{1}{c}{13} \\
\multicolumn{1}{l}{N $\to$ L} & \multicolumn{1}{c}{59} & \multicolumn{1}{c}{90} & \multicolumn{1}{c}{112} & \multicolumn{1}{c}{79} & \multicolumn{1}{c}{102} & \multicolumn{1}{c}{95} & \multicolumn{1}{c}{95} \\
\multicolumn{1}{l}{N $\to$ N} & \multicolumn{1}{c}{240} & \multicolumn{1}{c}{254} & \multicolumn{1}{c}{173} & \multicolumn{1}{c}{187} & \multicolumn{1}{c}{246} & \multicolumn{1}{c}{217} & \multicolumn{1}{c}{268} \\
\multicolumn{1}{l}{N $\to$ H} & \multicolumn{1}{c}{99} & \multicolumn{1}{c}{51} & \multicolumn{1}{c}{47} & \multicolumn{1}{c}{57} & \multicolumn{1}{c}{60} & \multicolumn{1}{c}{65} & \multicolumn{1}{c}{46} \\
\multicolumn{1}{l}{H $\to$ L} & \multicolumn{1}{c}{44} & \multicolumn{1}{c}{25} & \multicolumn{1}{c}{18} & \multicolumn{1}{c}{20} & \multicolumn{1}{c}{21} & \multicolumn{1}{c}{13} & \multicolumn{1}{c}{25} \\
\multicolumn{1}{l}{H $\to$ N} & \multicolumn{1}{c}{76} & \multicolumn{1}{c}{71} & \multicolumn{1}{c}{50} & \multicolumn{1}{c}{67} & \multicolumn{1}{c}{45} & \multicolumn{1}{c}{81} & \multicolumn{1}{c}{42} \\
\multicolumn{1}{l}{H $\to$ H} & \multicolumn{1}{c}{115} & \multicolumn{1}{c}{110} & \multicolumn{1}{c}{153} & \multicolumn{1}{c}{185} & \multicolumn{1}{c}{132} & \multicolumn{1}{c}{163} & \multicolumn{1}{c}{167} \\
      &       &       &       &       &       &       &  \\
\hline
\multicolumn{1}{c}{\textbf{Transition to}} & \multicolumn{1}{c}{\textbf{Extraversion}} & \multicolumn{1}{c}{\textbf{Agreeableness}} & \multicolumn{1}{c}{\textbf{Conscientiousness}} & \multicolumn{1}{c}{\textbf{Emotional Stability}} & \multicolumn{1}{c}{\textbf{Creativity}} & \multicolumn{1}{c}{\textbf{HPA}} & \multicolumn{1}{c}{\textbf{LNA}} \\
\hline
L     & 22.087 & 26.092 & 36.044 & 29.854 & 27.913 & 24.150 & 25.364 \\
N     & 52.752 & 58.425 & 46.575 & 48.154 & 54.114 & 51.705 & 52.925 \\
H     & 37.275 & 29.712 & 38.502 & 41.653 & 33.226 & 37.730 & 34.451 \\
\end{tabular}%

\end{minipage}}
\caption{Statistics of transitions between levels of each state: The first sub-table presents the number of cases per transition in each state. The second sub-table presents the percentage of transitions to each target level of each state.}
\label{tbl:trans}
\end{table}

\paragraph{Dispositional Traits of Personality and Affect}
We considered the trait scores that were reported by participants at the beginning of the experiment. Then, we normalized the trait scores of participants using the mean and the standard deviation. To discuss the statistical interaction between traits and social-situational factors associated with a given transition, we focused on only on participants with high scores in the trait (+1 standard deviation) and participants with low scores in the trait (-1 standard deviation). By using this method, we are able to know how levels of traits moderates the association between the social-situational factors and transitions in states. For example, we are interested to know how introverts respond to an increase in the intensity of contacts with others in a certain extraversion level in comparison to extroverts' response to the same increase. 

One might argue that there might be a systematic difference in how emotive marketing and sales people are compared to engineers. However, our control for personality traits would capture these variations. In fact, our control for traits goes significantly beyond broad classifications based on profession, since it captures underlying personality types using a systematic approach founded in well-established dimensions of personality.


\begin{table}[h]
\resizebox{0.95\textwidth}{!}{\begin{minipage}{\textwidth}
\begin{tabular}{l l l}
\hline
Group  & Survey  & Measurement \\
\hline
Personality & \textbf{States} & Extraversion \\
   & Ten-Item Personality Inventory (TIPI) & Agreeableness \\
   &       & Conscientiousness \\
   & \textbf{Traits} & Emotional Stability \\
      & Big Five Marker Scale (BFMS) & Creativity \\
\hline
Affect & \textbf{States} & High Positive Affect \\
      & Positive and Negative Affect Schedule (PANAS) & High Negative Affect \\
     & \textbf{Traits} & Low Positive Affect \\
      & Multidimentional Personality Questionnaire (MPQ) & Low Negative Affect \\
\hline
\end{tabular}
\end{minipage}}
\caption{Surveys for personality and affect states and traits.} 
\label{tab:surveys}
\end{table}

\subsection*{Dynamic Social networks}
We created dynamic temporal networks of face-to-face interaction for each participant . Between each two subsequent surveys a participant filled, we created the participant's temporal social network based on the social ties that the Infrared sensor had detected. We considered only social networks that were formed between morning and midday daily surveys or between midday and afternoon daily surveys. Based on this, we considered the transition in personality and emotional states of egos between these surveys (time $t$ and time $t+1$) and compared them to lagged states' levels of alters and traits of egos at time $t+1$. To generate our social networks, we exploited the Sociometric Badges Corpus, first introduced by Lepri et al. \cite{Bruno2012}.

We harnessed the experience sampling data and Infrared readings to create the dynamic social networks for participants. The time boundary of each social network is delineated by the time of subsequent surveys a participant filled. Between the two surveys, all face-to-face interactions detected via Infrared sensors are considered to be the social-situational factors for the participant. Particularly, what matters is the level at which each alter was at time $t$ (morning survey if the transition was initiated at the morning survey and midday surveys if the transition was initiated at midday survey).

\subsection*{Model and Parameter Estimation}
Our dataset consists of repeated observations for each participant, so we expected to have correlations within observations of participants. Hence, we used generalized linear models to analyze our longitudinal data using unstructured covariance matrices whereby variances and covariances are estimated directly from the data. Generalized Estimation Equations (GEE) are used to estimate the parameters of our models. For each transition in each state, we used backward elimination that starts with a full model that contains all candidate variables. Then, we tested the effect of deletion of insignificant variables using QICC (Corrected Quasilikelihood under Independence Models Criterion) iteratively until there is no further enhancement in the results.  We evaluated the goodness of fit based on QICC which is an indicator of goodness of fit of models that use generalized estimating equations. Therefore, it can be utilized to choose between two models favoring the one with the smaller QICC. After we end up with the best sub-model for each state transition, we compare its QICC to the QICC of the null model thats contains only the intercept. If the best sub-model is better than the null model, then we retain it. Otherwise, we consider the null model. Table \ref{tbl:QICC} compares between the QICC of our best sub-model and the null model for each transition in each state. If the QICC of the null model is better, then we report only the QICC of the null model.

In other words, we used QICC to penalize having complex models that might cause overfitting. Therefore, for each transition in each state, we start with having the complete model $Model1$ that includes all of the candidate independent variables and calculate the QICC of the model. Then, each time we encounter a statistically insignificant independent variable, we drop the variable, run the model again without this variable and calculate the QICC of the reduced model $Model2$. If the QICC of $Model2$ is higher than the QICC of $Model1$, we consider the results of Model1 and report them in the paper. If there is any statistically insignificant variable, we drop the variable from the model, run the model again without this variable $Model3$ and calculate the QICC of $Model3$. Then, we compare QICC of $Model2$ and QICC of $Model3$. If the QICC of $Model2$ is lower, then we report the results of $Model2$. Otherwise, we repeat the same process until there is no further reduction in QICC.

Although we do not have any statistical test to check whether the decrease in QICC is statistically significant, the significance of the independent variables remain the same during the backward elimination in the majority of the cases. Hence, we believe that using QICC for model selection is capable of supporting the significance of our results.

\begin{table}[h]
  \resizebox{0.77\textwidth}{!}{\begin{minipage}{\textwidth}
\begin{tabular}{rrrrrrrrrrr}
\multicolumn{3}{c}{\textbf{Extraversion}} &       & \multicolumn{3}{c}{\textbf{Agreeableness}} &       & \multicolumn{3}{c}{\textbf{Conscientiousness}} \\
\cline{1-3}\cline{5-7}\cline{9-11}\multicolumn{1}{l}{Transition} & \multicolumn{1}{c}{Our Model} & \multicolumn{1}{c}{Null Model} & \multicolumn{1}{c}{} & Transition & \multicolumn{1}{c}{Our Model} & \multicolumn{1}{c}{Null Model} &       & \multicolumn{1}{l}{Transition} & \multicolumn{1}{c}{Our Model} & \multicolumn{1}{c}{Null Model} \\
\cline{1-3}\cline{5-7}\cline{9-11}L to L & 258.823 & 261.053 &       & L to L & 302.3 & 309.7 &       & L to L & 355.184 & 367.347 \\
L to N &       & 260.987 &       & L to N & 303 & 308.9 &       & L to N & 336.434 & 340.500 \\
L to H & 193.796 & 194.776 &       & L to H & 167.87 & 171.7 &       & L to H & 156.765 & 167.498 \\
N to L & 346.415 & 348.420 &       & N to L &       & 435.9 &       & N to L & 420.844 & 430.617 \\
N to N & 536.751 & 540.690 &       & N to N & 519.7 & 520.5 &       & N to N & 455.696 & 463.190 \\
N to H &       & 460.236 &       & N to H & 314.4 & 316.1 &       & N to H & 273.47 & 286.849 \\
H to L & 231.949 & 239.164 &       & H to L &       & 171.5 &       & H to L & 140.832 & 142.188 \\
H to N &       & 303.378 &       & H to N & 262.045 & 270.8 &       & H to N & 236.687 & 247.471 \\
H to H & 321.325 & 329.069 &       & H to H & 271.5 & 290.359 &       & H to H & 263.305 & 281.446 \\
      &       &       &       &       &       &       &       &       &       &  \\
\multicolumn{3}{c}{\textbf{Emotional Stability}} &       & \multicolumn{3}{c}{\textbf{Creativity}} &       & \multicolumn{3}{c}{\textbf{High Positive Affect (HPA)}} \\
\cline{1-3}\cline{5-7}\cline{9-11}Transition & \multicolumn{1}{c}{Our Model} & \multicolumn{1}{c}{Null Model} &       & Transition & \multicolumn{1}{c}{Our Model} & \multicolumn{1}{c}{Null Model} &       & Transition & \multicolumn{1}{c}{Our Model} & \multicolumn{1}{c}{Null Model} \\
\cline{1-3}\cline{5-7}\cline{9-11}L to L &       & 305.981 &       & L to L & 299.310 & 304.301 &       & L to L & 400.737 & 406.689 \\
L to N &       & 293.489 &       & L to N &       & 302.174 &       & L to N &       & 310.175 \\
L to H & 91.054 & 101.587 &       & L to H & 111.789 & 126.067 &       & L to H & 294.914 & 306.805 \\
N to L & 359.896 & 367.961 &       & N to L &       & 463.480 &       & N to L & 196.9 & 206.045 \\
N to N & 434.818 & 442.637 &       & N to N &       & 551.166 &       & N to N & 175.454 & 176.941 \\
N to H & 305.636 & 310.707 &       & N to H & 348.1 & 356.400 &       & N to H &       & 203.480 \\
H to L & 161.530 & 163.776 &       & H to L &       & 153.455 &       & H to L & 355.656 & 374.595 \\
H to N &       & 315.597 &       & H to N & 205.969 & 226.689 &       & H to N & 307.305 & 316.248 \\
H to H &       & 354.350 &       & H to H & 251.582 & 270.706 &       & H to H & 443.738 & 481.045 \\
      &       &       &       &       &       &       &       &       &       &  \\
\multicolumn{3}{c}{\textbf{Low Negative Affect (LNA)}} &       &       &       &       &       &       &     \\
\cline{1-3}Transition & \multicolumn{1}{c}{Our Model} & \multicolumn{1}{c}{Null Model} &       &       &       &       &       &       &       \\
\cline{1-3}L to L & 251.309 & 252.927 &       &       &       &       &       &       &       \\
L to N &       & 250.422 &       &       &       &       &       &       &       &  \\
L to H & 107.607 & 111.148 &       &       &       &       &       &       &       &  \\
N to L & 449.815 & 452.999 &       &       &       &       &       &       &       &  \\
N to N & 529.541 & 531.894 &       &       &       &       &       &       &       &  \\
N to H & 302.882 & 308.433 &       &       &       &       &       &       &       &  \\
H to L & 169.170 & 184.558 &       &       &       &       &       &       &       &  \\
H to N & 232.818 & 242.474 &       &       &       &       &       &       &       &  \\
H to H & 263.323 & 302.914 &       &       &       &       &       &       &       &  \\
\end{tabular}%
\end{minipage}}
\caption{Comparison of Goodness of Fit: We compare between the QICC of our best sub-model and the QICC of the null model of each transition in each state. Only the QICC of the null model is reported if it is less (better) than the QICC of the sub-model.}
\label{tbl:QICC}
\end{table}

For each possible transition between levels of a particular state, our model consists of: one dependent variable, the transition probability;  nine independent variables that capture the corresponding trait score ($T$) and the three situational measures concerning contact intensity described above: $L$, $N$ and $H$. The model also contains the interaction effects between trait and situational variables, in order to account for the moderating effect of the former on the latter: $T*L$, $T*N$ and $T*H$. The level transitions might take place spontaneously due to the time of the day as highlighted by Golder and Macy \cite{golder2011diurnal}. Therefore, we added a time period variable that can act as a control variable to capture possible diurnal rhythms $P$. Particularly, we are interested to study the effect of time of day at time $t+1$ for the transition that takes place between time $t$ and time $t+1$. Therefore, the time period could be either midday or afternoon. Also, we expect to have an interaction between the time of the day and the corresponding trait. Therefore, we added one more control variable $P*T$.  The association between the dependent and the independent variables, including interactions, is modeled through logistic regression as shown in Equation \ref{eq:model}. We used logistic regression instead of OLS regression (used by Hill et al \cite{Hill2010}) because the value of the dependent variable is binary (0 if there is no transition and 1 otherwise). Let $X \to Y$ denotes a transition by the ego from level $X$ to level $Y$ of some state $S$ (we permit $X=Y$ denoting stability). Let $p(X \to Y)$ be the probability of this transition between two consecutive surveys. For a given dynamic personality or affect state \emph{S}:
\begin{equation}
ln (\frac{p(X \to Y)}{1-p(X \to Y}) = \alpha + \beta_{L}L + \beta_{N}N+  \beta_{H}H+ \beta_{T}T + \\
\beta_{T*L}T*L + \beta_{T*N}T*N + \beta_{T*H}T*H + \beta_{P}P +  \beta_{P*T}P*T \\
\label{eq:model}
\end{equation}

where $\alpha$  is a constant (intercept); $T$, $L$, $N$ and $H$ are as explained above. $\beta_{L}$, $\beta_{N}$, $\beta_{H}$ and  $\beta_{T}$ are the coefficients of the \emph{main} effects $L$, $N$, $H$ and $T$, respectively. $\beta_{T*L}$, $\beta_{T*N}$ and $\beta_{T*H}$ are the coefficients of the interaction effects between the trait $T$ and the situational variables $L$, $N$ and $H$, respectively. $\beta_{P}$ is the coefficient of time of the day  and $\beta_{P*T}$ is the coefficient of the interaction between the corresponding trait and the time of the day. 


 \section*{Results}
First, we discuss some descriptive statistics about network dynamics and potential homophily in the networks of participants. Second, we describe the four social influence processes that we identified in our model. Third, we describe our results in the context of these processes.

 \subsection*{Descriptive Statistics}
\begin{figure*}[h]
	\begin{center}
		\centerline{\includegraphics[width=0.7\textwidth]{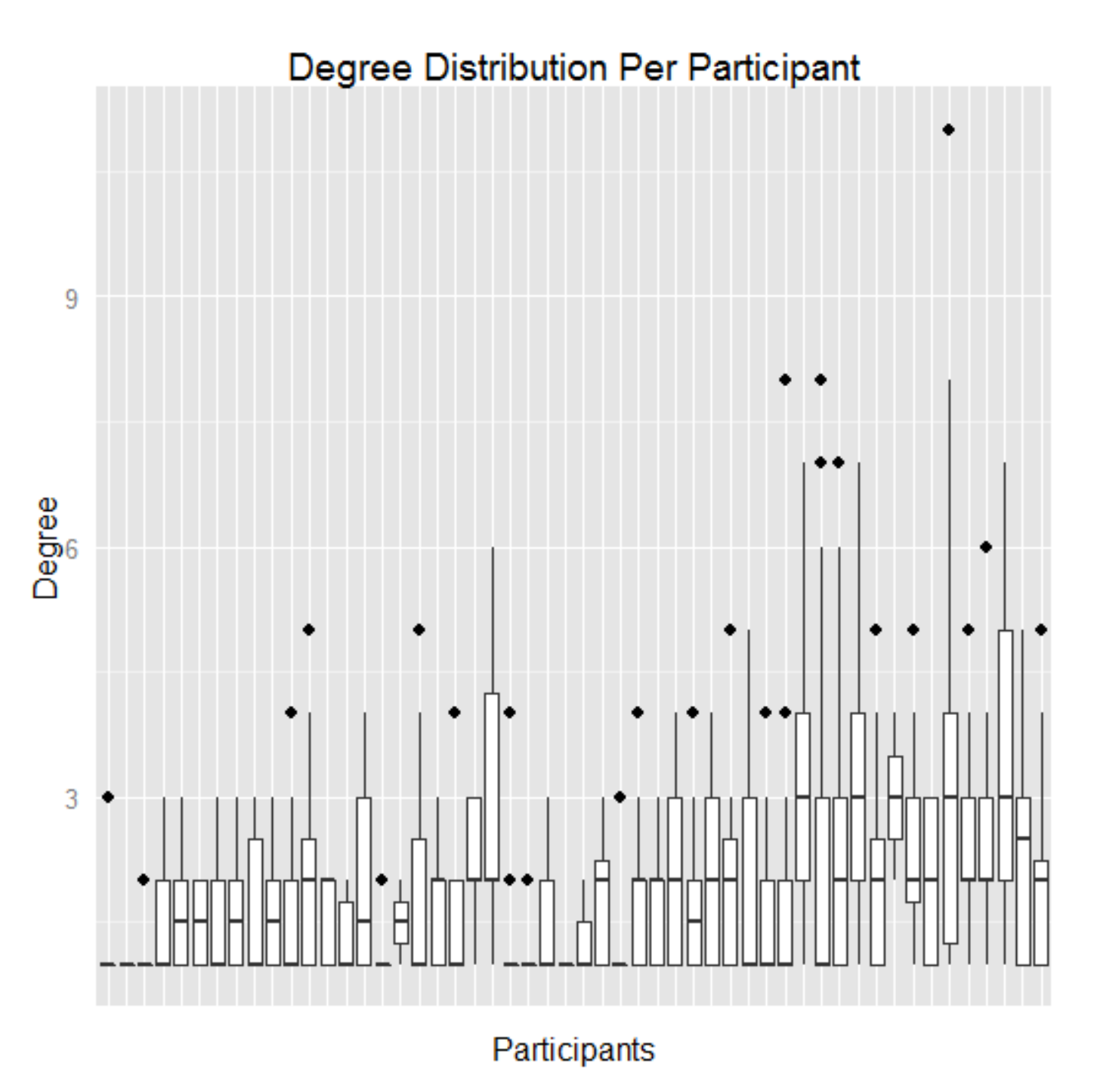}}
		\caption{
			{\bf The Degree Distribution for each Participant}  Each box plot shows the variation in the number of alters in dynamic networks of participants. We can observe that there is a variation in the number of alters in the dynamic networks of participants.}
		\label{fig:degree}
	\end{center}
\end{figure*}

\begin{figure*}[h]
	\begin{center}
		\centerline{\includegraphics[width=0.7\textwidth]{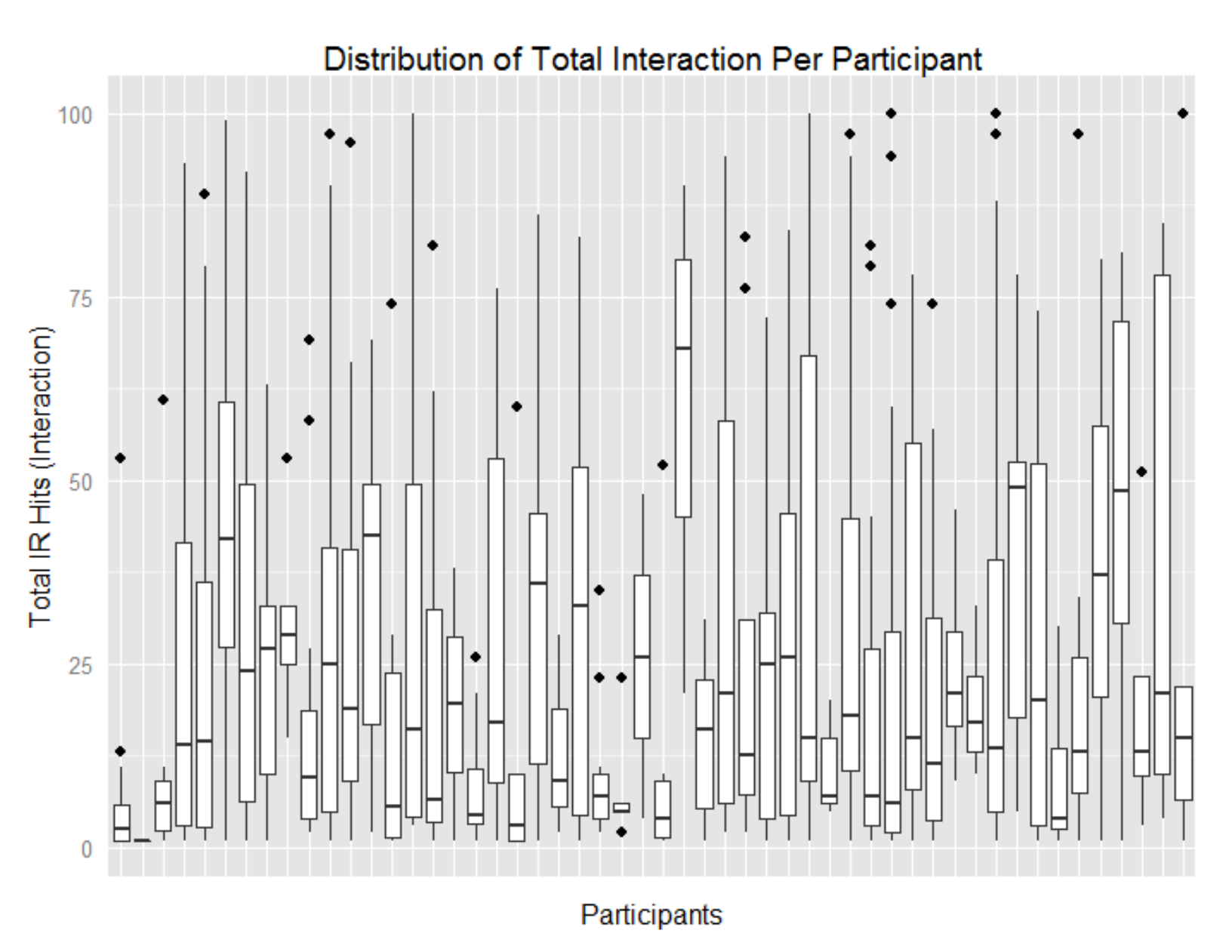}}
		\caption{
		{\bf The Interaction Distribution for each Participant}  Each box plot shows the variation in the total number of interactions (IR hits) in dynamic networks of participants. We can observe that there is a variation in the number of interactions in the dynamic networks of participants.}
		\label{fig:interaction}
	\end{center}
\end{figure*}

We provide some descriptive statistics about social network dynamics.  Fig. \ref{fig:degree} and Fig. \ref{fig:interaction} show the degree distribution and interaction distribution. In most cases, the dynamic networks of participants vary in terms of the number of alters and the total number of interactions. 

Also, we quantify the similarity between egos and alters in terms of their state levels. We calculated the ratio between the number of alters in the same level of egos and the total number of alters using the following equation $similarity_s = \frac{n_s}{n}$ where $similarity_s$ is the quantified similarity between egos at level $s$ in a given state, $n_s$ is the number of alters at level $s$ for a given state and $n$ is the total number of alters. We calculated this ratio for each dynamic network of participants and considered the levels of egos and alters at the same time. Then, we took the average of all calculated ratios. In all cases, we found that the percentage of similar alters does not exceed 50\% which means that interaction between people does not necessarily take place based on homophily (Check Table \ref{tbl:homophily}). We repeated the same process to quantify the similarity between people with respect to the number of interaction. We calculated the ratio between the number of interactions with alters in the same level of egos and the total number of interactions using the following equation $similarity_s = \frac{i_s}{i}$ where $similarity_s$ is the quantified similarity between egos at level $s$ in a given state, $i_s$ is the total interaction with alters at level $s$ for a given state and $i$ is the total number of interactions with alters. Again, we calculated this ratio for each dynamic network of participants and considered the levels of egos and alters at the same time. Then, we took the average of all calculated ratios. Again, we found that the percentage of similar alters does not exceed 50\% (Check Table \ref{tbl:homophily}).

\begin{table}[htbp]
	\centering
	\begin{tabular}{rrrrr}
		\toprule
		\multicolumn{1}{c}{\textbf{State}} & \multicolumn{2}{c}{\textbf{Number of Alters}} & \multicolumn{2}{c}{\textbf{Total Interaction}} \\
		\midrule
		\multicolumn{1}{l}{\textbf{Extroversion}} & \multicolumn{1}{c}{\textbf{Mean}} &
		\multicolumn{1}{c}{\textbf{Standard Deviation}} & \multicolumn{1}{c}{\textbf{Mean}} & \multicolumn{1}{c}{\textbf{Standard Deviation}} \\
		\multicolumn{1}{l}{Low} & 0.28  & 0.38  & 0.266 & 0.4 \\
		\multicolumn{1}{l}{Neutral} & 0.49  & 0.41  & 0.199 & 0.366 \\
		\multicolumn{1}{l}{High} & 0.28  & 0.38  & 0.49  & 0.45 \\
		\multicolumn{1}{l}{\textbf{Conscientiousness }} & \multicolumn{1}{c}{\textbf{Mean}} & \multicolumn{1}{c}{\textbf{Standard Deviation}} & \multicolumn{1}{c}{\textbf{Mean}} & \multicolumn{1}{c}{\textbf{Standard Deviation}} \\
		\multicolumn{1}{l}{Low} & 0.34  & 0.4   & 0.335 & 0.43 \\
		\multicolumn{1}{l}{Neutral} & 0.44  & 0.39  & 0.42  & 0.45 \\
		\multicolumn{1}{l}{High} & 0.3   & 0.388 & 0.29  & 0.417 \\
		\multicolumn{1}{l}{\textbf{Agreeableness}} & \multicolumn{1}{c}{\textbf{Mean}} & \multicolumn{1}{c}{\textbf{Standard Deviation}} & \multicolumn{1}{c}{\textbf{Mean}} & \multicolumn{1}{c}{\textbf{Standard Deviation}} \\
		\multicolumn{1}{l}{Low} & 0.29  & 0.35  & 0.31  & 0.41 \\
		\multicolumn{1}{l}{Neutral} & 0.45  & 0.41  & 0.45  & 0.45 \\
		\multicolumn{1}{l}{High} & 0.27  & 0.37  & 0.29  & 0.422 \\
		\multicolumn{1}{l}{\textbf{Emotional Stability}} & \multicolumn{1}{c}{\textbf{Mean}} & \multicolumn{1}{c}{\textbf{Standard Deviation}} & \multicolumn{1}{c}{\textbf{Mean}} & \multicolumn{1}{c}{\textbf{Standard Deviation}} \\
		\multicolumn{1}{l}{Low} & 0.35  & 0.39  & 0.36  & 0.44 \\
		\multicolumn{1}{l}{Neutral} & 0.43  & 0.39  & 0.43  & 0.44 \\
		\multicolumn{1}{l}{High} & 0.39  & 0.4   & 0.38  & 0.44 \\
		\multicolumn{1}{l}{\textbf{Creativity}} & \multicolumn{1}{c}{\textbf{Mean}} & \multicolumn{1}{c}{\textbf{Standard Deviation}} & \multicolumn{1}{c}{\textbf{Mean}} & \multicolumn{1}{c}{\textbf{Standard Deviation}} \\
		\multicolumn{1}{l}{Low} & 0.3   & 0.38  & 0.31  & 0.42 \\
		\multicolumn{1}{l}{Neutral} & 0.54  & 0.4   & 0.55  & 0.45 \\
		\multicolumn{1}{l}{High} & 0.346 & 0.4   & 0.344 & 0.445 \\
		\multicolumn{1}{l}{\textbf{High PA (HPA)}} & \multicolumn{1}{c}{\textbf{Mean}} & \multicolumn{1}{c}{\textbf{Standard Deviation}} & \multicolumn{1}{c}{\textbf{Mean}} & \multicolumn{1}{c}{\textbf{Standard Deviation}} \\
		\multicolumn{1}{l}{Low} & 0.38  & 0.4   & 0.387 & 0.44 \\
		\multicolumn{1}{l}{Neutral} & 0.21  & 0.36  & 0.22  & 0.39 \\
		\multicolumn{1}{l}{High} & 0.5   & 0.42  & 0.5   & 0.45 \\
		\multicolumn{1}{l}{\textbf{Low NA (LNA)}} & \multicolumn{1}{c}{\textbf{Mean}} & \multicolumn{1}{c}{\textbf{Standard Deviation}} & \multicolumn{1}{c}{\textbf{Mean}} & \multicolumn{1}{c}{\textbf{Standard Deviation}} \\
		\multicolumn{1}{l}{Low} & 0.27  & 0.37  & 0.25  & 0.4 \\
		\multicolumn{1}{l}{Neutral} & 0.48  & 0.4   & 0.51  & 0.45 \\
		\multicolumn{1}{l}{High} & 0.4   & 0.39  & 0.43  & 0.44 \\
		\bottomrule
	\end{tabular}%
	\caption{{\bf Quantifying similarity between people according to their states.} We quantified the similarity between egos and alters in terms of their state levels. We calculated the ratio between the number of alters in the same level of egos and the total number of alters. Then, we took the mean and the standard deviation of those quantities. Also, we calculated the ratio between the total number of interactions with alters in the same level of egos and the total number of interactions. Then, we took the mean and the standard deviation of those quantities. In both cases, the similarity does not exceed 50\%.}
	\label{tbl:homophily}%
\end{table}%

\subsection*{Social Influence Processes}
In our results, we identified four social influence processes : (1) attraction (2) repulsion (3) inertia (4) push.  

\subsubsection*{Attraction and Repulsion}
\begin{figure*}[h]
\begin{center}

\centerline{\includegraphics[width=0.55\textwidth]{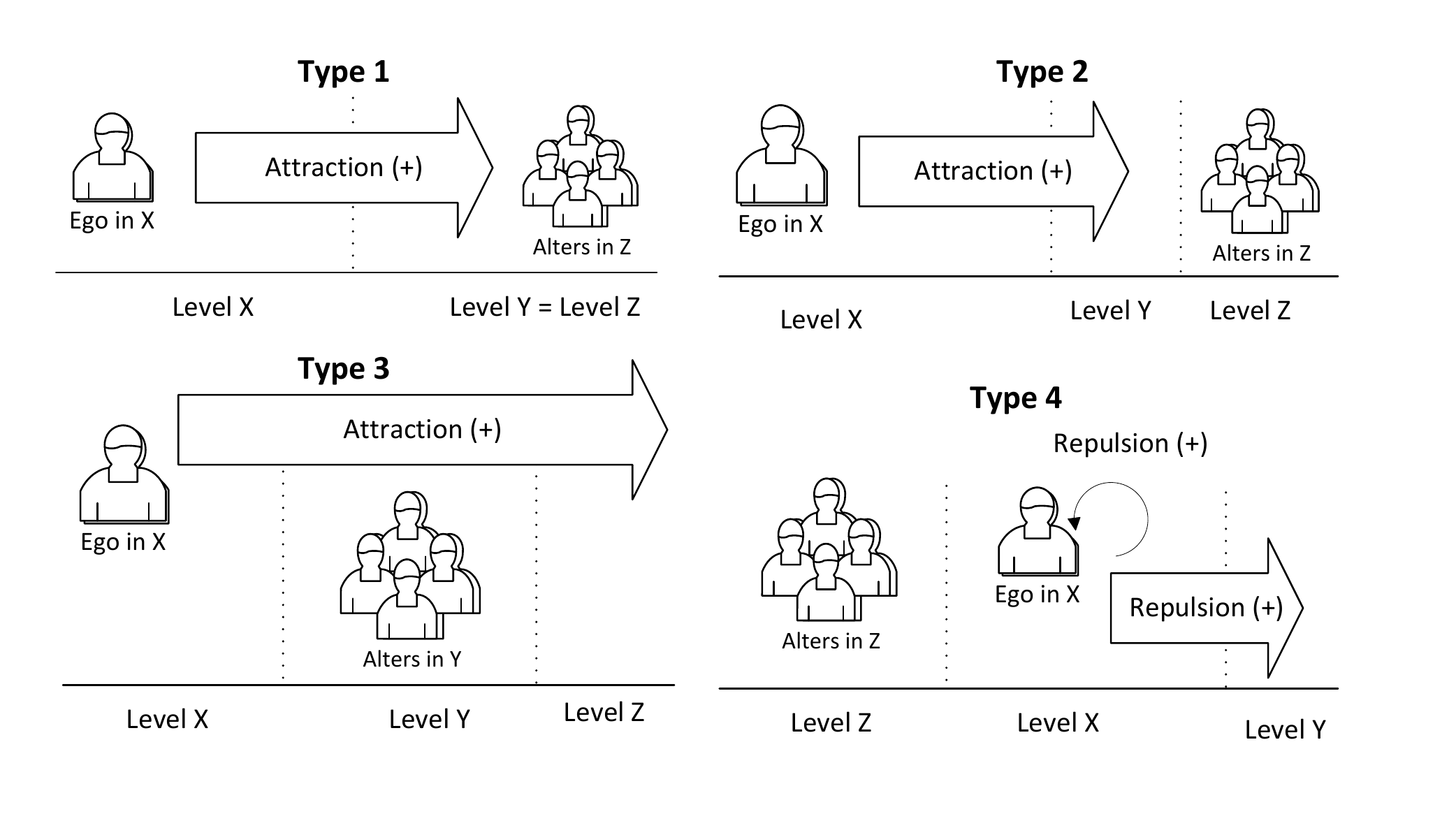}}
\caption{Attraction and Repulsion. Case 1-3 exemplify different ways which attraction can manifest. In Case 1, the presence of alters in level Z is associated with ego's switching to that level. In Case 2, the ego moves to level Y which is though lower than that of the attracting alters but higher than his/her original level.  In Case 3, the ego moves to a level that is higher of both his/her original one and that of the attracting alters. Case 4 exemplifies repulsion: the presence of alters in a lower level is associated with the ego remaining at his/her current state or moving to a higher level. Attraction to lower levels works in the same way. }
\label{fig:attraction}
\end{center}
\end{figure*}

Formally, let $P(X \to Y | K)$ be the probability of transition $X \to Y$ conditional on intensity of contacts $K$ with alters in level $Z$ of a state $S$:
\begin{description}
   \item[Attraction] by $K$ on egos in $X$ level iff, for $X$ different from $Y$, $P(X \to Y| K)$ is increasing with $K$ and either $Z \textless X$ and $Y<X$, or $Z>X$ and $Y>X$. 
 
   \item[Repulsion] we have repulsion by $K$ on people in $X$ iff, for $X$ different from $Y$, $P(X \to Y | K)$ is decreasing with $K$ and either $Z<X$ and $Y<X$, or $Z>X$ and $Y>X$. Equivalently, we have repulsion by $K$ on people in $X$ iff, for $X$ different from $Y$, $P(X \to Y | K)$ is increasing with $K$ and either $Z<X$ and $Y \geq X$ or $Z>X$ and $Y \leq X$
 \end{description}

\begin{figure*}[h]
\begin{center}
\centerline{\includegraphics[width=.4\textwidth]{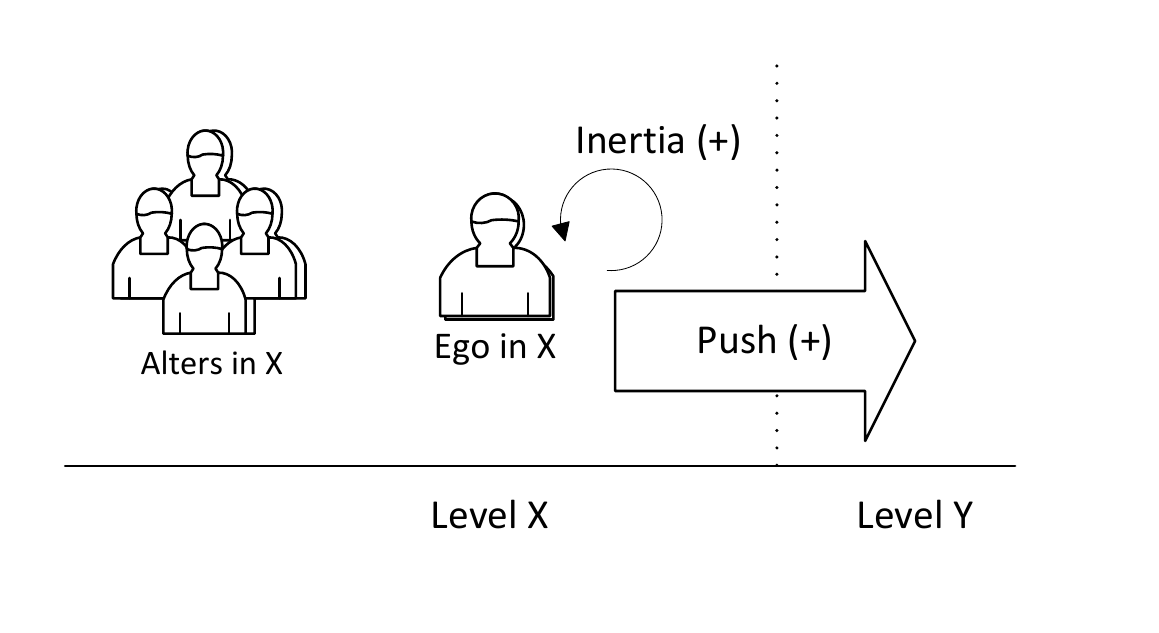}}
\caption{Inertia and Push: Alters are inertial for egos to move away from the alters' levels by either lowering the transition probability to another level or increasing the probability of staying at the alter's level. Vice versa for the push affect}
\label{fig:inertia}
\end{center}
\end{figure*}

Fig. \ref{fig:attraction} exemplifies some ways in which attraction and repulsion can manifest. We already gave examples of Case 1 in the main paper. Concerning Case 2, it was observed with agreeableness state and transition $(L \to N)$: the intensity of contacts with alters in the high level is associated with egos scoring high in the trait to upgrade the level from low to neutral. 

Case 3 is exemplified by transition $(H \to L)$ of conscientiousness state. Intensity of contacts with alters in the neutral level is associated with shifting egos who have high trait score to the low level. 

Case 4 is exemplified by transition $(N \to N)$ of conscientiousness. Intensity of contacts with alters in the low level is associated with increased probabilities of that stability of egos with high scores in the trait. All detailed examples are provided in the supporting information.

\subsubsection*{Inertia and Push}
Formally, let P($X \to Y | K$) be the probability of transition $X \to Y$ conditional to contacts with people $K$ in level $X$ of a state $S$:
\begin{description}
\item[Inertia]: $K$ is inertial for people in level $X$ iff $P(X \to X|K)$ increases with $K$. Equivalently, $X$ is inertial iff, for $X$ different from $Y$, $P(X \to Y |K)$ decreases with $K$.
 \item[Push]: $K$ pushes away egos in level $X$ iff, for $X$ different from $Y$, $P(X \to Y |X)$ increase with $K$. Equivalently, $K$ pushes away alters in level $X$ iff $P(X \to X |K)$ decreases with $K$.
 \end{description}

Fig. \ref{fig:inertia} depicts the inertia and push influences triggered by intensity contacts with alters in egos' initial levels. 

\subsection*{Control Variables}
We believe that there might be diurnal rhythms in the between-subject correlations observed in our data \cite{golder2011diurnal}. Hence, transitions between levels of states can be attributed to the day of the week or the time of the day and not only attributed to social-situational factors and dispositional traits. To test for the existence of diurnal rhythm, we ran an ANOVA to check if the average level of each personality and affect measure changes systematically across days of the week or times of the day. Then, we controlled for the statistically significant diurnal rhythms by running regressions which controls for those rhythms.

Analysis of variance (ANOVA) is used to analyze the differences between group means. In our case, the groups contain either the days of the week or the times of the day (morning, midday and afternoon).  In the case of week days, we found that there are no statistically significant differences of affect and personality states between weekdays except for creativity (p-value 0.00006). We took one more step to pairwise compare means of affect and personality states using Tukey's test \cite{yandell1997practical}. Tukey HSD is a statistical test that is used with an ANOVA to do pairwise comparisons between the means of different treatments. In our case, we have seven weekdays and thus we have 10 pairwise comparisons. We found that people tend to be more creative at the end of the week than the beginning of the week: (1) Monday-Friday (diff: -0.23, p-value: 0.003) (2) Thursday-Friday (diff: 0.27, p-value: 0.0001). Also, there is a difference in the mean of creativity between Wednesday and Thursday (diff: -0.19, p-value: 0.018). If we were addressing transitions in creativity states between days, then controlling for the day of the week would be essential. However, we address transitions between times of the day. Therefore, the days of the week are not suitable to be used as control variables in our model.

When we ran the ANOVA test to investigate whether the means of personality and affect states changes systematically across times of the day, we found that those differences are statistically significant for most of the states.  Therefore, we ran Tukey's test to check for pairwise comparisons across different times of the day. With respect to extraversion, people are less extrovert on average at the end of the day in comparison to their extraversion state's mean score at the beginning of the day (diff: -0.136, p-value: 0.029) or at midday (diff: -0.2, p-value: 0.0003). With respect to agreeableness, people got less agreeable on average in comparison to their mean score at the beginning of the day (-0.113, p-value: 0.023). With respect to conscientiousness, people got less conscientious on average at midday in comparison to their mean score at the beginning of the day (diff: -0.2, p-value: 0.0000006) and they are less conscientious on average at the end of the day in comparison to their score at the beginning of the day (diff: -0.118, p-value: 0.016). However, people are more conscientious on average at the end of the day in comparison to their mean score at midday. With respect to emotional stability, people are less emotionally stable on average at the end of the day in comparison to midday (diff: -0.117, p-value: 0.033) and beginning of the day (diff: -0.14, p-value: 0.0078). Creativity on average doesn't change according to the time period. With respect to high positive affect state, people's mean score is lower at midday in comparison to the beginning of the day (-0.177, p-value = 0). Also, their mean score in HPA is lower at the end of the day in comparison to the beginning of the day. However, people's mean score at the end of the day is higher at the end of the day than the mean score at midday (diff: 0.088, p-value). With respect to low negative affect state, people's mean score is lower at the end of the day in comparison to their mean score at midday (diff: -0.099, p-value: 0.018) and the mean score at the beginning of the day (diff: -0.095, p-value: 0.0245). In our model, we address the transitions in states' level from mornings to middays and from middays to afternoons. Thus, we are interested in the mean differences in states in those two types of transitions. 

\begin{figure*}[h]
\begin{center}
\centerline{\includegraphics[width=.95\textwidth]{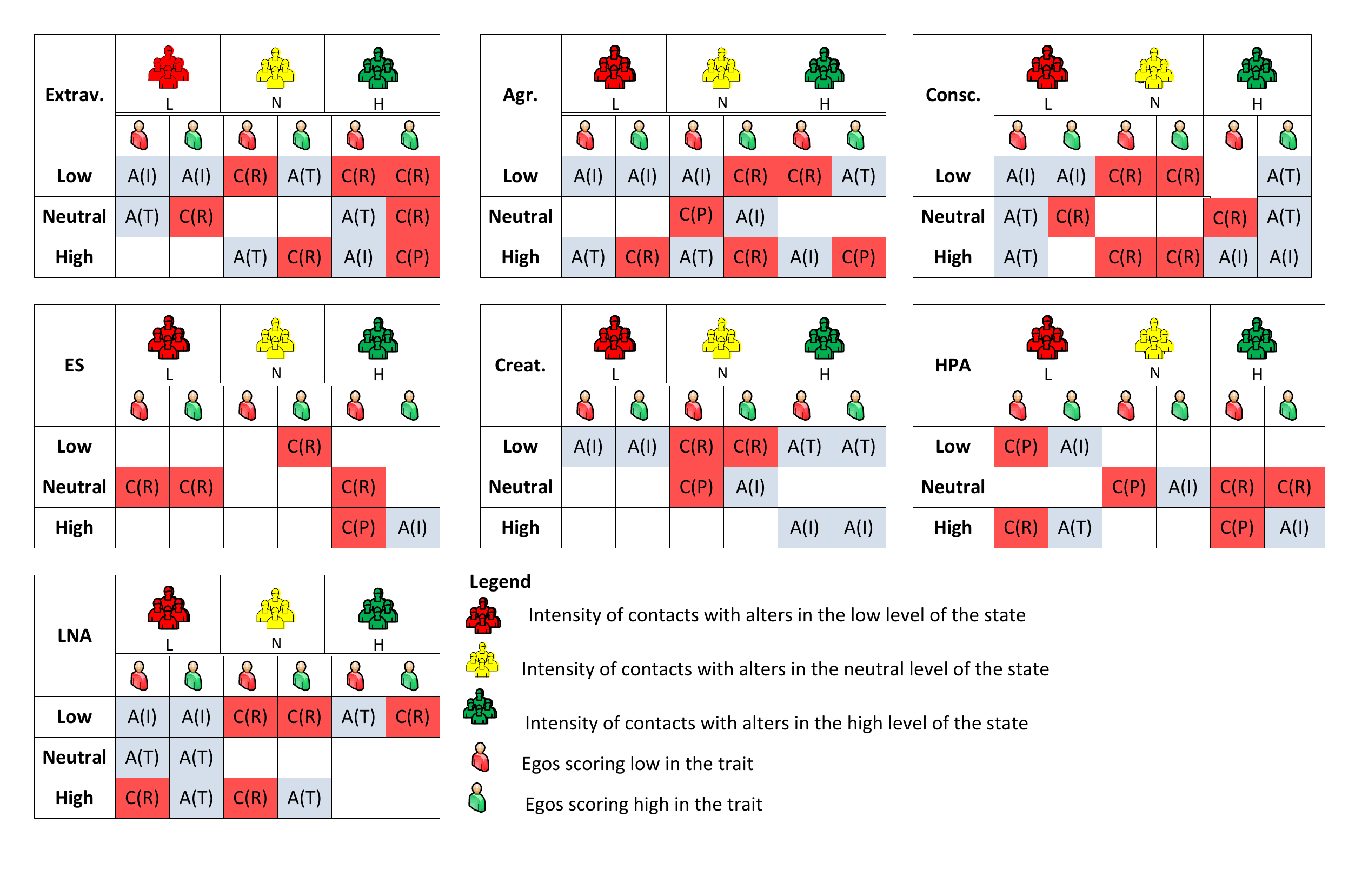}}
\caption{Social Influences: The table summarizes each state's results  by means of the two adverse effects introduced in the text: adaptation (A) and complementarity (C). Also, the detailed effects are listed between the brackets: (1) attraction (T) (2) repulsion (R) (3) inertia (I) and  (4) push (P).  Rows represent ego's state levels; columns are labeled with alters' levels and sub-labeled with ego's trait level (Low or High). Cells report the effects observed when egos in the corresponding state level and trait level interact with alters in the corresponding state level.}
\label{statePatterns}
\end{center}
\end{figure*}

\begin{figure*}
\begin{center}
{\includegraphics[width=.7\textwidth]{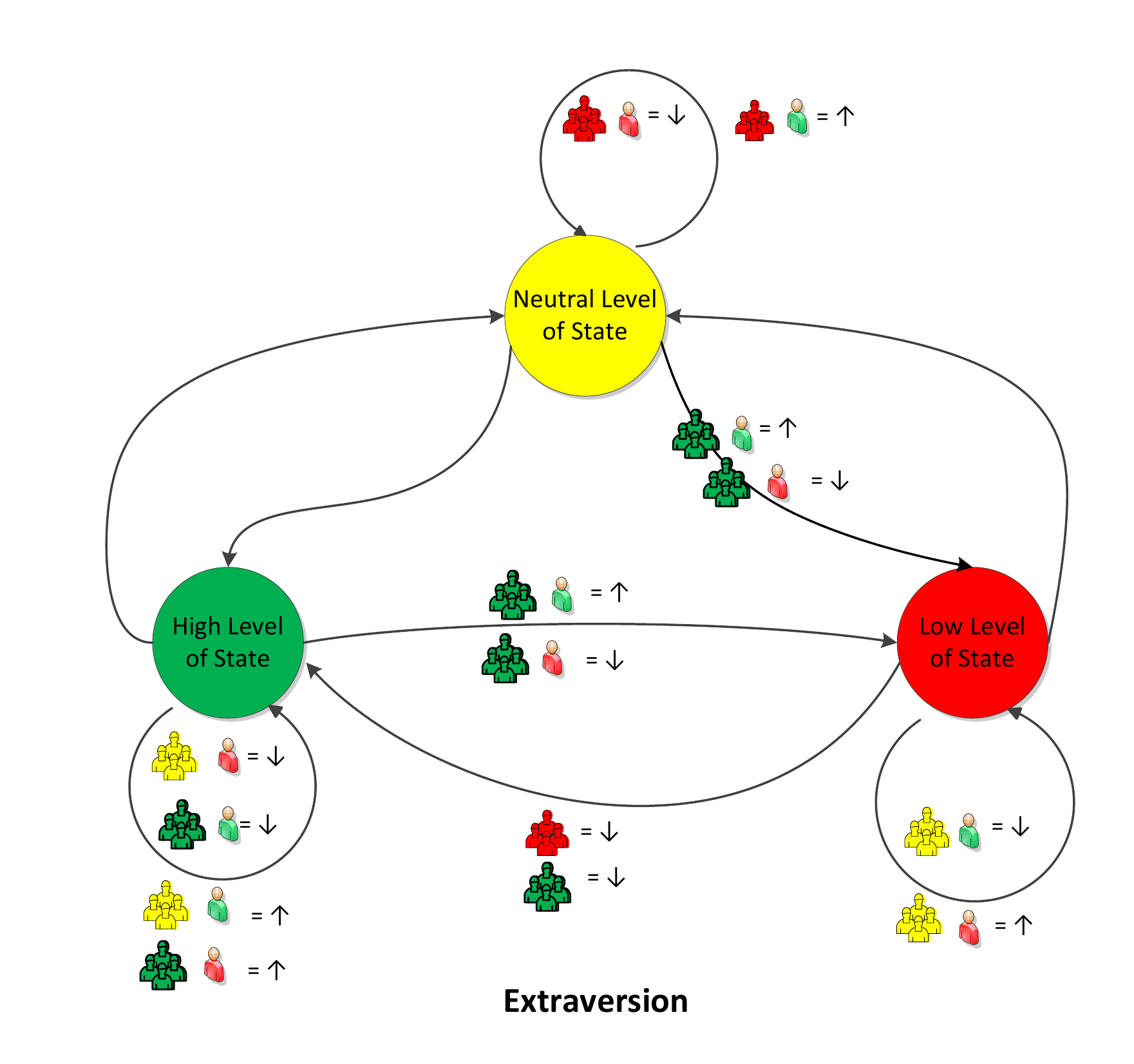}}
\caption{Level transition graph for extraversion state: Nodes represent extraversion level of the ego. An arrow between two circles represents the transition from one level to another. These transitions are labeled with conditions that affect the corresponding probabilities. Icons represent the extraversion levels of alters and ego's trait level. Symbol $\uparrow$ (respectively $\downarrow$) indicates an increase (respectively decrease) in transition probability associated with the given combination of alters state level and ego trait level. For example, if the ego is in the low level of the extraversion state, then the probability of him/her transitioning to the high level decreases with his/her interactions with alters in the high level.}
\label{ext_statePatterns}
\end{center}
\end{figure*}

\begin{figure*}
\begin{center}
{\includegraphics[width=.7\textwidth]{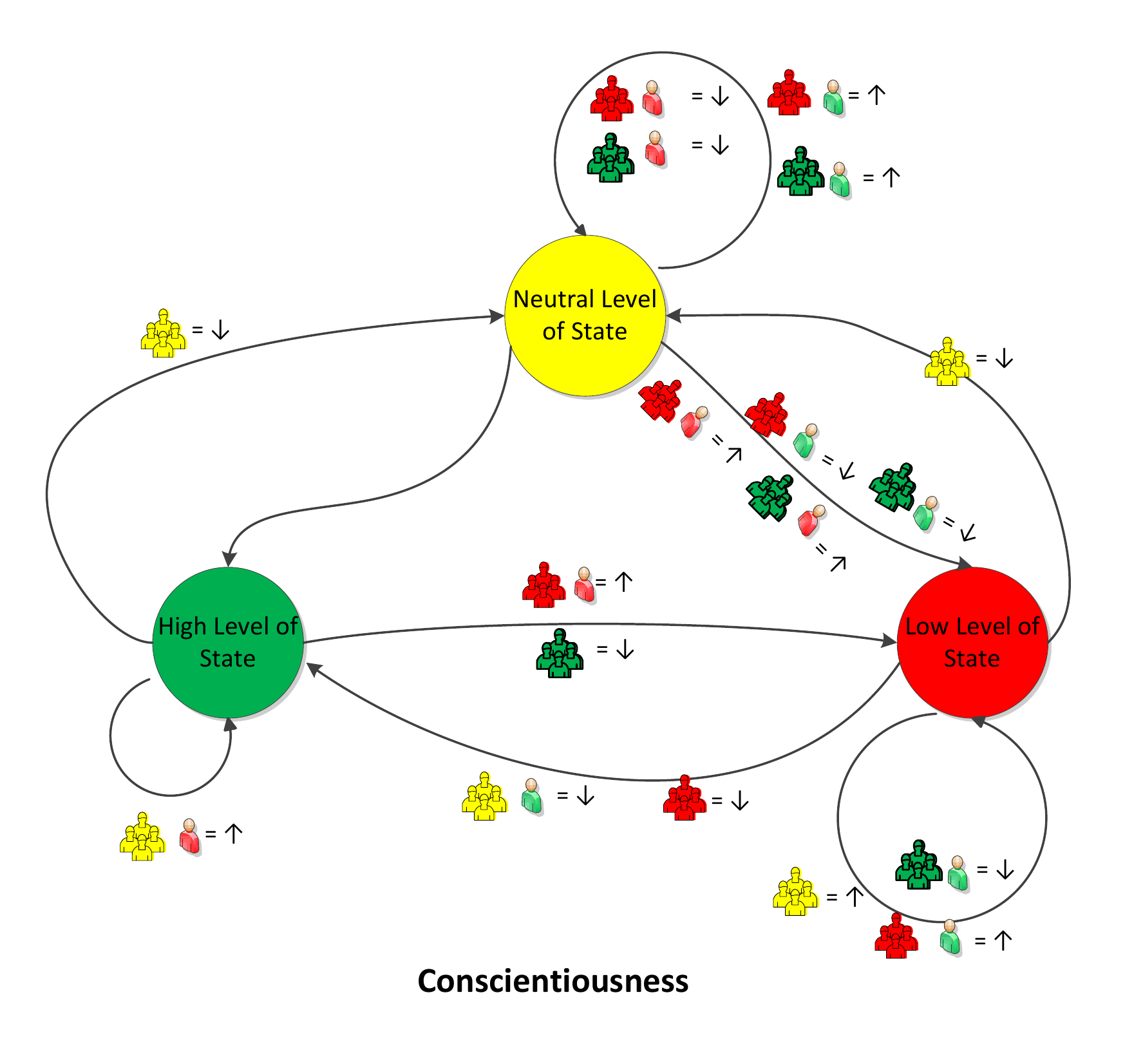}}
\caption{Level transition graph for conscientiousness state: Nodes represent conscientiousness level of the ego. An arrow between two circles represents the transition from one level to another. These transitions are labeled with conditions that affect the corresponding probabilities. Icons represent the conscientiousness levels of alters and ego's trait level. Symbol $\uparrow$ (respectively $\downarrow$) indicates an increase (respectively decrease) in transition probability associated with the given combination of alters state level and ego trait level. For example, if the ego is in the low level of the conscientiousness state, then the probability of him/her interactions with alters in the high level.}
\label{consc_statePatterns}
\end{center}
\end{figure*}

\clearpage

\begin{figure*}
\begin{center}
{\includegraphics[width=.7\textwidth]{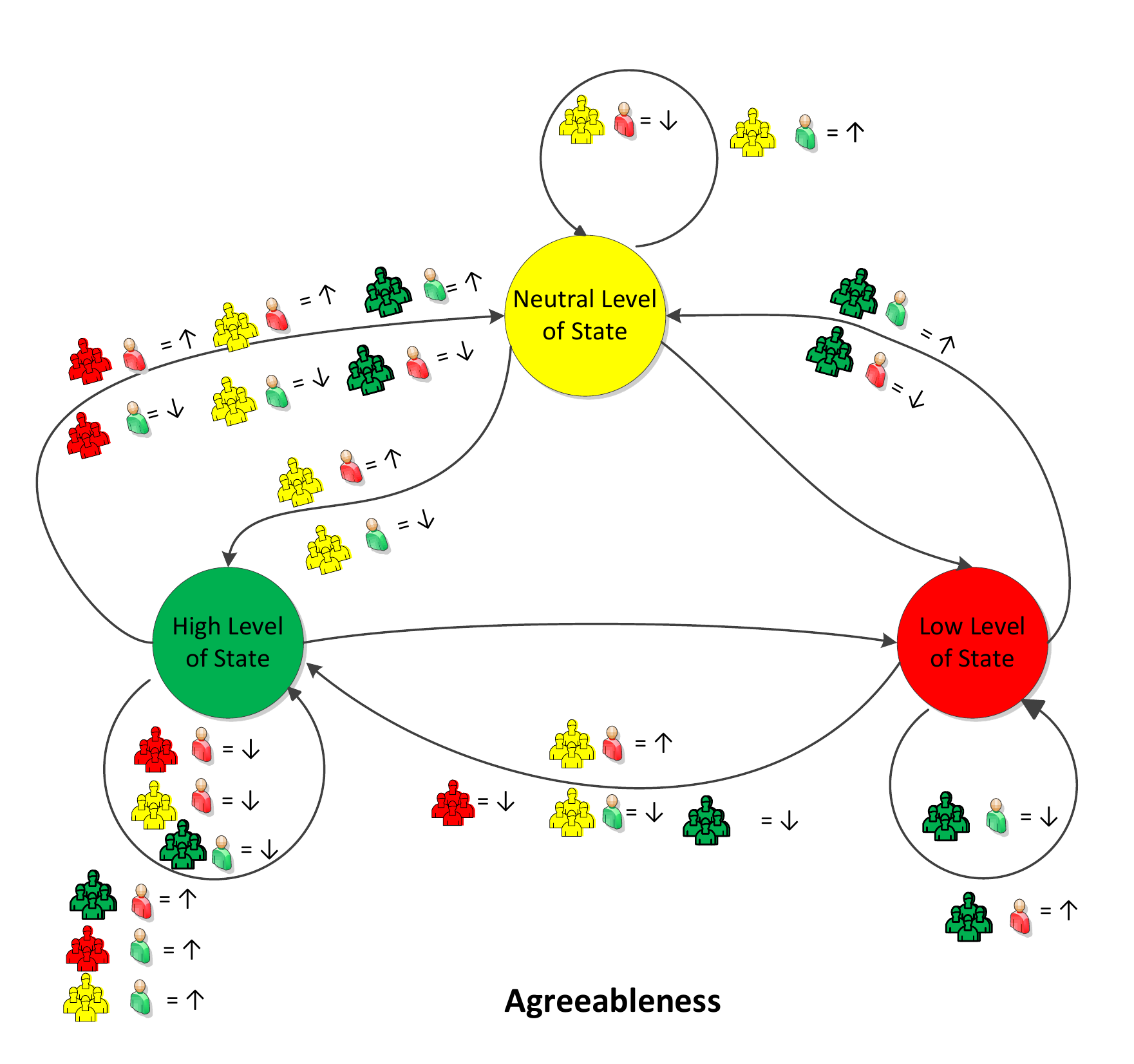}}
\caption{Level transition graph for agreeableness state: Nodes represent agreeableness level of the ego. An arrow between two circles represents the transition from one level to another. These transitions are labeled with conditions that affect the corresponding probabilities. Icons represent the agreeableness levels of alters and ego's trait level. Symbol $\uparrow$ (respectively $\downarrow$) indicates an increase (respectively decrease) in transition probability associated with the given combination of alters state level and ego trait level. For example, if the ego is in the low level of the agreeableness state, then the probability of him/her transitioning to the high level decreases with his/her interactions with alters in the high level. }
\label{agr_statePatterns}
\end{center}
\end{figure*}

\begin{figure*}
\begin{center}
{\includegraphics[width=.7\textwidth]{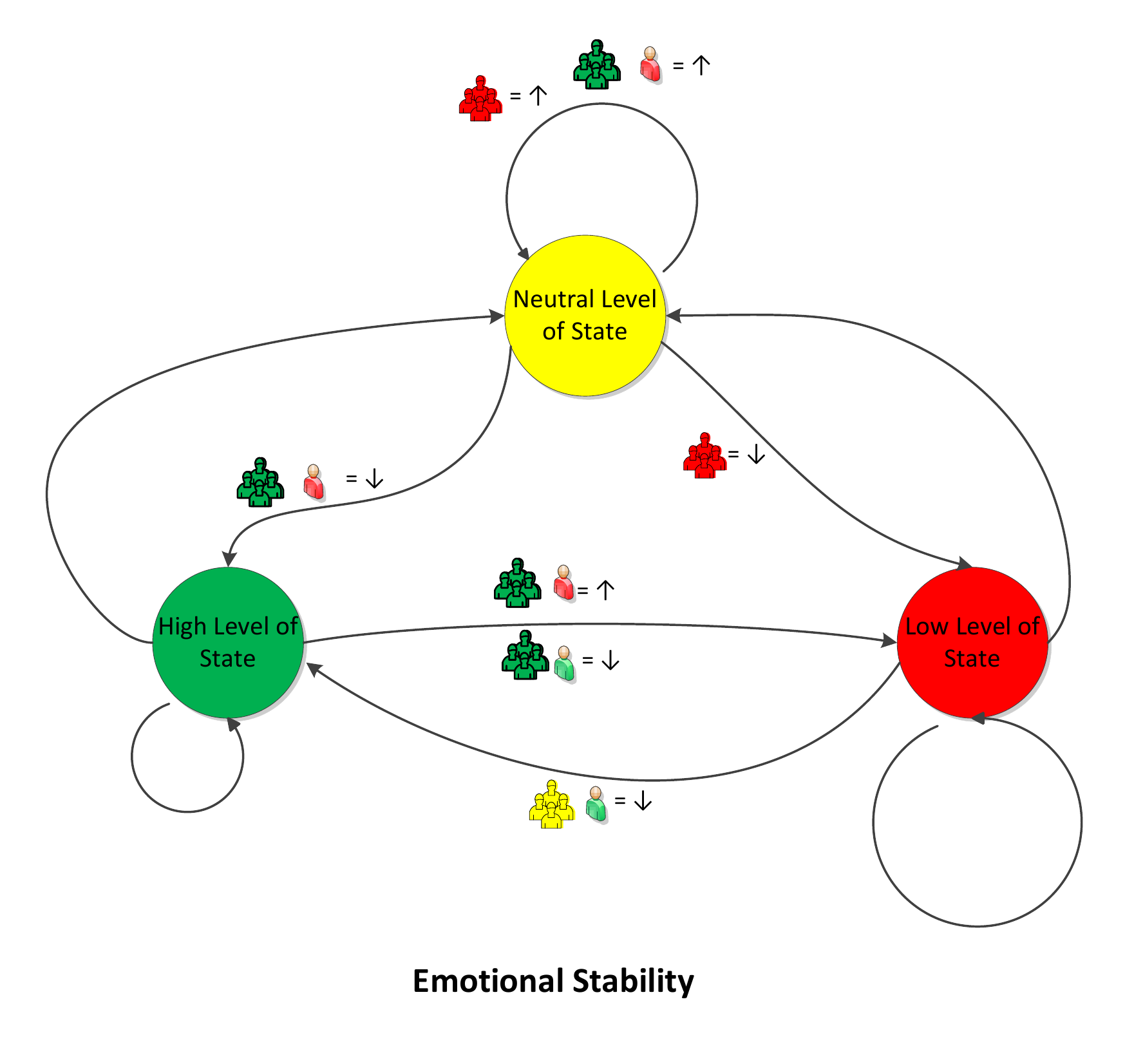}}
\caption{Level transition graph for emotional stability state: Nodes represent  emotional stability level of the ego. An arrow between two circles represents the transition from one level to another. These transitions are labeled with conditions that affect the corresponding probabilities. Icons represent the emotional stability level of alters and ego's trait level. Symbol $\uparrow$ (respectively $\downarrow$) indicates an increase (respectively decrease) in transition probability associated with the given combination of alters state level and ego trait level. For example, if the ego is in the low level of the  emotional stability states, then the probability of him/her transitioning to the high level decreases with his/her interactions with alters in the high level. }
\label{es_statePatterns}
\end{center}
\end{figure*}

\begin{figure*}
\begin{center}
{\includegraphics[width=.7\textwidth]{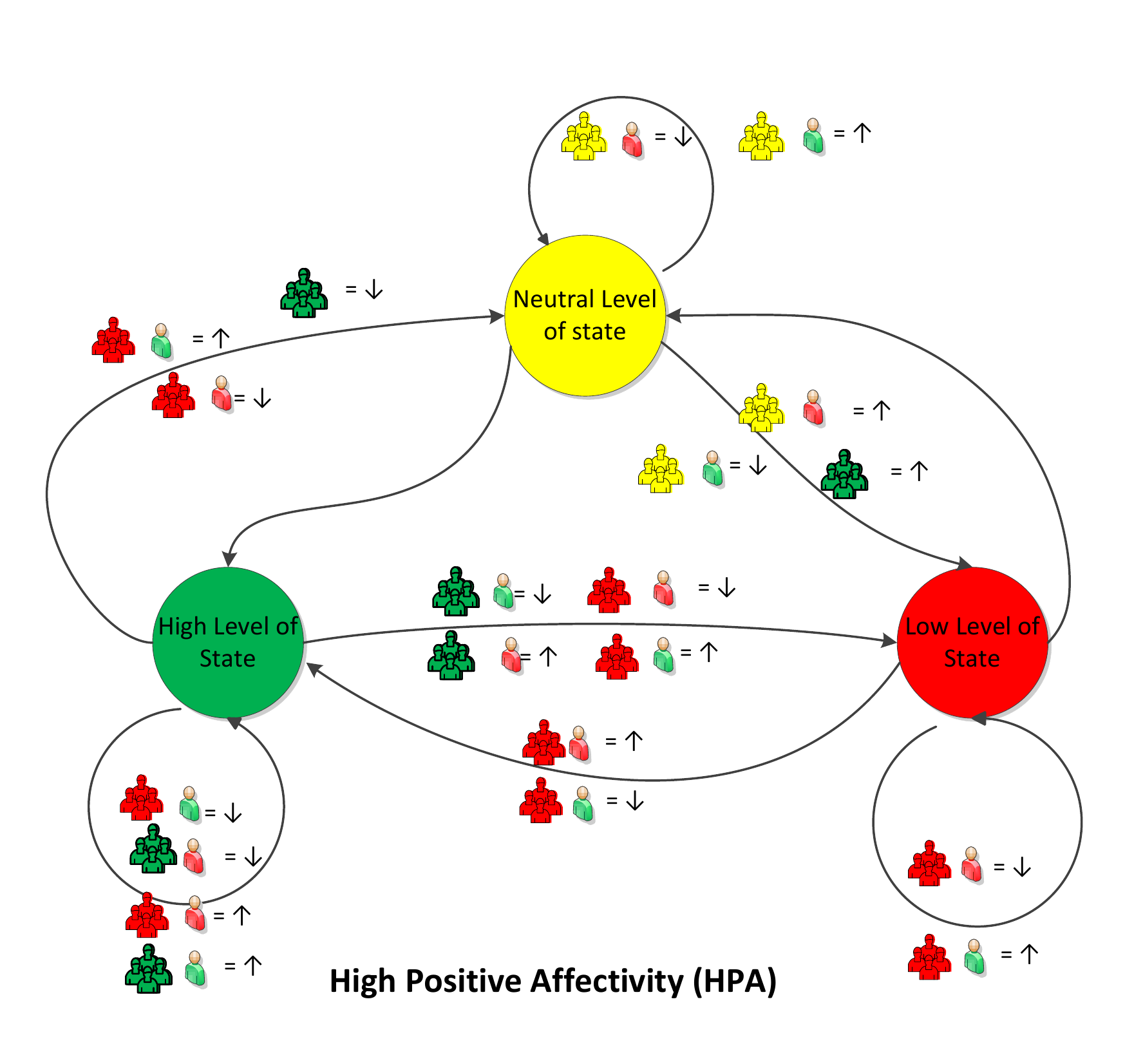}}
\caption{Level transition graph for high positive affect state: Nodes represent  high positive affect level of the ego. An arrow between two circles represents the transition from one level to another. These transitions are labeled with conditions that affect the corresponding probabilities. Icons represent the  high positive affect levels of alters and ego's trait level. Symbol $\uparrow$ (respectively $\downarrow$) indicates an increase (respectively decrease) in transition probability associated with the given combination of alters state level and ego trait level. For example, if the ego is in the low level of the  high positive affect state, then the probability of him/her transitioning to the high level decreases with his/her interactions with alters in the high level. }
\label{hpa_statePatterns}
\end{center}
\end{figure*}

\begin{figure*}
\begin{center}
{\includegraphics[width=.7\textwidth]{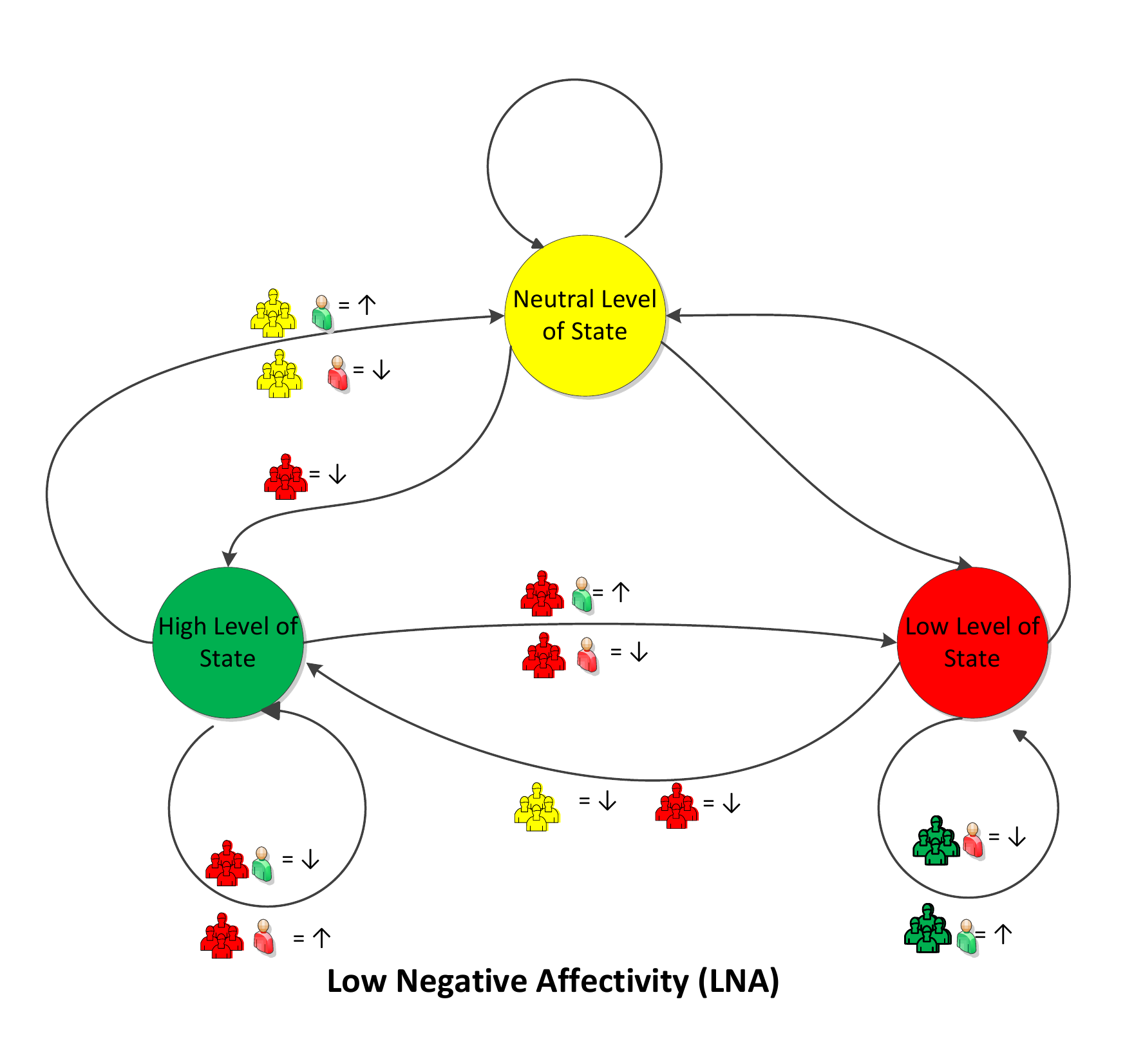}}
\caption{Level transition graph for low negative affect state: Nodes represent low negative affect level of the ego. An arrow between two circles represents the transition from one level to another. These transitions are labeled with conditions that affect the corresponding probabilities. Icons represent the low negative affect level of alters and ego's trait level. Symbol $\uparrow$ (respectively $\downarrow$) indicates an increase (respectively decrease) in transition probability associated with the given combination of alters state level and ego trait level. For example, if the ego is in the low level of the low negative affect states, then the probability of him/her transitioning to the high level decreases with his/her interactions with alters in the high level.}
\label{lna_statePatterns}
\end{center}
\end{figure*}

We added the time of day as a diurnal control variable in our model besides the social-situational factors and dispositional traits to capture between-subject effect. We are interested to know the effect of the time period to which the transition takes place: midday (second surveys) in case of transitions from mornings to middays and the afternoon and afternoons (third surveys) in case of transitions from middays to afternoons. We represent the time of day as a dummy variable: 0 for middays (reference group) and 1 for afternoons. 

First, we present the effects of time of the day and the interaction between time of the day and the corresponding trait. The coefficients of time of the day and the interactions are reported in tables \ref{tbl:extr}, \ref{tbl:agr}, \ref{tbl:cons}, \ref{tbl:es}, \ref{tbl:creat},  \ref{tbl:hpa} and \ref{tbl:lna}.

\begin{table*}[!ht]
 \resizebox{0.9\textwidth}{!}{\begin{minipage}{\textwidth}
\begin{tabular}{lll|rrr|rrr}
\multicolumn{3}{c}{\textbf{L to L}} & \multicolumn{3}{c}{\textbf{L to N}} & \multicolumn{3}{c}{\textbf{L to H}}  \\
\hline
Variable  & Coefficient & P-value & Variable  & Coefficient & P-value & Variable  & Coefficient & P-value \\
\hline
Intercept & \multicolumn{1}{l}{-0.363} & \multicolumn{1}{l|}{0.0016} & Intercept & \multicolumn{1}{l}{-0.218} & \multicolumn{1}{l|}{0.0382} & Intercept & \multicolumn{1}{l}{-0.758} & \multicolumn{1}{l}{0.0000} \\
N*T   & \multicolumn{1}{l}{-0.002} & \multicolumn{1}{l|}{0.0011} &       & \multicolumn{1}{l}{} &       & H     & \multicolumn{1}{l}{-0.005} & \multicolumn{1}{l}{0.0000} \\
      &       &       &       &       &       & L     & \multicolumn{1}{l}{-0.005} & \multicolumn{1}{l}{0.0016} \\
      &       & \multicolumn{1}{r}{} &       &       & \multicolumn{1}{r}{} &       &       &  \\
\multicolumn{3}{c}{\textbf{N to L}} & \multicolumn{3}{c}{\textbf{N to N}} & \multicolumn{3}{c}{\textbf{N to H}} \\
\hline
Variable  & Coefficient & P-value & Variable  & Coefficient & P-value & Variable  & Coefficient & P-value \\
\hline
Intercept & \multicolumn{1}{l}{-1.508} & \multicolumn{1}{l|}{0.0000} & Intercept & \multicolumn{1}{l}{0.177} & \multicolumn{1}{l|}{0.0193} & Intercept & \multicolumn{1}{l}{-0.720} & \multicolumn{1}{l}{0.0000} \\
N     & \multicolumn{1}{l}{0.0009} & \multicolumn{1}{l|}{0.0155} & T     & \multicolumn{1}{l}{-0.284} & \multicolumn{1}{l|}{0.0382} &       & \multicolumn{1}{l}{} & \multicolumn{1}{l}{} \\
H*T   & \multicolumn{1}{l}{0.003} & \multicolumn{1}{l|}{0.0100} & H*T   & \multicolumn{1}{l}{-0.0009} & \multicolumn{1}{l|}{0.0015} &       & \multicolumn{1}{l}{} & \multicolumn{1}{l}{} \\
(Period=1) & \multicolumn{1}{l}{0.352} & \multicolumn{1}{l|}{0.0258} & L*T   & \multicolumn{1}{l}{0.002} & \multicolumn{1}{l|}{0.0034} &       &       &  \\
      &       &       & period*T & \multicolumn{1}{l}{0.239} & \multicolumn{1}{l|}{0.0065} &       &       &  \\
      &       & \multicolumn{1}{r}{} &       &       & \multicolumn{1}{r}{} &       &       &  \\
\multicolumn{3}{c}{\textbf{H to L}} & \multicolumn{3}{c}{\textbf{H to N}} & \multicolumn{3}{c}{\textbf{H to H}} \\
\hline
Variable  & Coefficient & P-value & Variable  & Coefficient & P-value & Variable  & Coefficient & P-value \\
\hline
Intercept & \multicolumn{1}{l}{-1.503} & \multicolumn{1}{l|}{0.0000} & Intercept & \multicolumn{1}{l}{-0.415} & \multicolumn{1}{l|}{0.0002} & H* T  & \multicolumn{1}{l}{-0.007} & \multicolumn{1}{l}{0.0001} \\
(period=1) & \multicolumn{1}{l}{0.792} & \multicolumn{1}{l|}{0.0000} &       &       &       & N*T   & \multicolumn{1}{l}{0.004} & \multicolumn{1}{l}{0.0006} \\
T     & \multicolumn{1}{l}{-0.466} & \multicolumn{1}{l|}{0.0042} &       &       &       &       &       &  \\
H*T   & \multicolumn{1}{l}{0.005} & \multicolumn{1}{l|}{0.0008} &       &       &       &       &       &  \\
period*T & \multicolumn{1}{l}{0.487} & \multicolumn{1}{l|}{0.0112} &       &       &       &       &       &  \\
\end{tabular}%
\end{minipage}}
\caption{Extraversion Results}
\label{tbl:extr}
\end{table*}

\begin{table*}[!ht]
 \resizebox{0.9\textwidth}{!}{\begin{minipage}{\textwidth}
\begin{tabular}{lll|lll|lll}
\multicolumn{3}{c}{\textbf{L to L}} & \multicolumn{3}{c}{\textbf{L to N}} & \multicolumn{3}{c}{\textbf{L to H}} \\
\hline
Variable  & Coefficient & P-value & Variable  & Coefficient & P-value & Variable  & Coefficient & P-value \\
\hline
Intercept & \multicolumn{1}{l}{-0.350} & \multicolumn{1}{l|}{0.003} & H*T     & \multicolumn{1}{l}{0.007} & \multicolumn{1}{l|}{0.004} & Intercept & \multicolumn{1}{l}{-1.033} & \multicolumn{1}{l}{0.0000} \\
H*T     & \multicolumn{1}{l}{-0.007} & \multicolumn{1}{l|}{0.001} &       &       &       & L     & \multicolumn{1}{l}{-0.009} & \multicolumn{1}{l}{0.0000} \\
   &  &  &       &       &       & N     & \multicolumn{1}{l}{-0.007} & \multicolumn{1}{l}{0.0014} \\
      &       &       &       &       &       & H     & \multicolumn{1}{l}{-0.002} & \multicolumn{1}{l}{0.0020} \\
      &       & \multicolumn{1}{r}{} &       &       & \multicolumn{1}{r}{} & N*T   & \multicolumn{1}{l}{-0.011} & \multicolumn{1}{l}{0.0000} \\
      &       & \multicolumn{1}{r}{} &       &       & \multicolumn{1}{r}{} &       & \multicolumn{1}{l}{} & \multicolumn{1}{l}{} \\
      &       & \multicolumn{1}{r}{} &       &       & \multicolumn{1}{r}{} &       &       &  \\
\multicolumn{3}{c}{\textbf{N to L}} & \multicolumn{3}{c}{\textbf{N to N}} & \multicolumn{3}{c}{\textbf{N to H}} \\
\hline
Variable  & Coefficient & P-value & Variable  & Coefficient & P-value & Variable  & Coefficient & P-value \\
\hline
Intercept & \multicolumn{1}{l}{0.854} & \multicolumn{1}{l|}{0.0000} & Intercept & \multicolumn{1}{l}{0.365} & \multicolumn{1}{l|}{0.0000} & Intercept & \multicolumn{1}{l}{-1.472} & \multicolumn{1}{l}{0.000} \\
      & \multicolumn{1}{l}{} & \multicolumn{1}{l|}{} & N*T   & \multicolumn{1}{l}{0.002} & \multicolumn{1}{l|}{0.0187} & N*T   & \multicolumn{1}{l}{-0.003} & \multicolumn{1}{l}{0.015} \\
      & \multicolumn{1}{l}{} & \multicolumn{1}{l|}{} & (period=1) * T & \multicolumn{1}{l}{0.205} & \multicolumn{1}{l|}{0.0149} &       &       &  \\
      & \multicolumn{1}{l}{} & \multicolumn{1}{l|}{} & (period=0) * T & \multicolumn{1}{l}{-0.170} & \multicolumn{1}{l|}{0.0841} &       &       &  \\
      &       & \multicolumn{1}{r}{} &       &       & \multicolumn{1}{r}{} &       &       &  \\
\multicolumn{3}{c}{\textbf{H to L}} & \multicolumn{3}{c}{\textbf{H to N}} & \multicolumn{3}{c}{\textbf{H to H}} \\
\hline
Variable  & Coefficient & P-value & Variable  & Coefficient & P-value & Variable  & Coefficient & P-value \\
\hline
Intercept & \multicolumn{1}{l}{-1.177} & \multicolumn{1}{l|}{0.0000} & \multicolumn{1}{l}{Intercept} & \multicolumn{1}{l}{-0.282} & \multicolumn{1}{l|}{0.053} & \multicolumn{1}{l}{N} & -.004 & .004 \\
      &       &       & \multicolumn{1}{l}{H} & \multicolumn{1}{l}{-0.004} & \multicolumn{1}{l|}{0.055} & \multicolumn{1}{l}{L*T} & .005  & .000 \\
      &       &       & \multicolumn{1}{l}{L*T} & \multicolumn{1}{l}{-0.003} & \multicolumn{1}{l|}{0.000} & \multicolumn{1}{l}{N*T} & .008  & .000 \\
      &       &       & \multicolumn{1}{l}{N*T} & \multicolumn{1}{l}{-0.005} & \multicolumn{1}{l|}{0.002} & \multicolumn{1}{l}{H* T} & -.004 & .001 \\
      &       &       & \multicolumn{1}{l}{H*T} & \multicolumn{1}{l}{0.009} & \multicolumn{1}{l|}{0.001} &       &       &  \\
\end{tabular}%
\end{minipage}}
\caption{Agreeableness Results}
\label{tbl:agr}
\end{table*}

\begin{table*}[!ht]
 \resizebox{0.85\textwidth}{!}{\begin{minipage}{\textwidth}
\begin{tabular}{lll|lll|lll|}
\multicolumn{3}{c}{\textbf{L to L}} & \multicolumn{3}{c}{\textbf{L to N}} & \multicolumn{3}{c}{\textbf{L to H}}  \\
\hline
Variable  & Coefficient & P-value & Variable  & Coefficient & P-value & Variable  & Coefficient & \multicolumn{1}{r}{P-value} \\
\hline
Intercept & \multicolumn{1}{l}{0.273} & \multicolumn{1}{l|}{0.0252} & Intercept & \multicolumn{1}{l}{-0.428} & \multicolumn{1}{l|}{0.0001} & Intercept & \multicolumn{1}{l}{-2.451} & \multicolumn{1}{l}{0.0000} \\
      & \multicolumn{1}{l}{} & \multicolumn{1}{l|}{} &       & \multicolumn{1}{l}{} & \multicolumn{1}{l|}{} & N     & \multicolumn{1}{l}{-0.005} & \multicolumn{1}{l}{0.0208} \\
      & \multicolumn{1}{l}{} & \multicolumn{1}{l|}{} &       & \multicolumn{1}{l}{} & \multicolumn{1}{l|}{} & T     & \multicolumn{1}{l}{1.102} & \multicolumn{1}{l}{0.0079} \\
      & \multicolumn{1}{l}{} & \multicolumn{1}{l|}{} &       & \multicolumn{1}{l}{} & \multicolumn{1}{l|}{} & T*N   & \multicolumn{1}{l}{-0.006} & \multicolumn{1}{l}{0.0097} \\
      & \multicolumn{1}{l}{} & \multicolumn{1}{l|}{} &       & \multicolumn{1}{l}{} & \multicolumn{1}{l|}{} & (period=1) * T & \multicolumn{1}{l}{-3.001} & \multicolumn{1}{l}{0.0000} \\
      &       & \multicolumn{1}{r}{} &       &       & \multicolumn{1}{r}{} &       &       & \multicolumn{1}{r}{} \\
      &       & \multicolumn{1}{r}{} &       &       & \multicolumn{1}{r}{} &       &       & \multicolumn{1}{r}{} \\
\multicolumn{3}{c}{\textbf{N to L}} & \multicolumn{3}{c}{\textbf{N to N}} & \multicolumn{3}{c}{\textbf{N to H}} \\
\hline
Variable  & Coefficient & P-value & Variable  & Coefficient & P-value & Variable  & Coefficient & \multicolumn{1}{r}{P-value} \\
\hline
\multicolumn{1}{l}{Intercept} & \multicolumn{1}{l}{-1.005} & \multicolumn{1}{l|}{0.000} & \multicolumn{1}{l}{L} & \multicolumn{1}{l}{0.001} & \multicolumn{1}{l|}{0.008} & \multicolumn{1}{l}{Intercept} & \multicolumn{1}{l}{-1.001} & \multicolumn{1}{l}{0.000} \\
\multicolumn{1}{l}{(period=1)} & \multicolumn{1}{l}{0.548} & \multicolumn{1}{l|}{0.002} & \multicolumn{1}{l}{H} & \multicolumn{1}{l}{0.005} & \multicolumn{1}{l|}{0.000} & \multicolumn{1}{l}{H} & \multicolumn{1}{l}{-0.012} & \multicolumn{1}{l}{0.002} \\
\multicolumn{1}{l}{L} & \multicolumn{1}{l}{-0.002} & \multicolumn{1}{l|}{0.000} & \multicolumn{1}{l}{T} & \multicolumn{1}{l}{0.379} & \multicolumn{1}{l|}{0.001} & \multicolumn{1}{l}{H*T} & \multicolumn{1}{l}{0.011} & \multicolumn{1}{l}{0.002} \\
\multicolumn{1}{l}{T} & \multicolumn{1}{l}{0.271} & \multicolumn{1}{l|}{0.056} & \multicolumn{1}{l}{H*T} & \multicolumn{1}{l}{-0.007} & \multicolumn{1}{l|}{0.000} &       &       & \multicolumn{1}{r}{} \\
\multicolumn{1}{l}{(period=1) * T} & \multicolumn{1}{l}{-0.907} & \multicolumn{1}{l|}{0.002} &       &       &       &       &       & \multicolumn{1}{r}{} \\
      &       & \multicolumn{1}{r}{} &       &       & \multicolumn{1}{r}{} &       &       & \multicolumn{1}{r}{} \\
\multicolumn{3}{c}{\textbf{H to L}} & \multicolumn{3}{c}{\textbf{H to N}} & \multicolumn{3}{c}{\textbf{H to H}} \\
\hline
Variable  & Coefficient & P-value & Variable  & Coefficient & P-value & Variable  & Coefficient & \multicolumn{1}{r}{P-value} \\
\hline
\multicolumn{1}{l}{Intercept} & \multicolumn{1}{l}{-1.398} & \multicolumn{1}{l|}{0.000} & \multicolumn{1}{l}{Intercept} & \multicolumn{1}{l}{-0.690} & \multicolumn{1}{l|}{0.000} & \multicolumn{1}{l}{Intercept} & \multicolumn{1}{l}{0.325} & \multicolumn{1}{l}{0.044} \\
\multicolumn{1}{l}{(period=1)} & \multicolumn{1}{l}{-0.503} & \multicolumn{1}{l|}{0.006} &       &       &       &       &       & \multicolumn{1}{r}{} \\
\multicolumn{1}{l}{H} & \multicolumn{1}{l}{-0.014} & \multicolumn{1}{l|}{0.034} &       &       &       &       &       & \multicolumn{1}{r}{} \\
\multicolumn{1}{l}{H*T} & \multicolumn{1}{l}{-0.018} & \multicolumn{1}{l|}{0.021} &       &       &       &       &       & \multicolumn{1}{r}{} \\
\multicolumn{1}{l}{(period=1) *T} & \multicolumn{1}{l}{-0.771} & \multicolumn{1}{l|}{0.014} &       &       &       &       &       & \multicolumn{1}{r}{} \\
\multicolumn{1}{l}{(period=0) *T} & \multicolumn{1}{l}{0.035} & \multicolumn{1}{l|}{0.890} &       &       &       &       &       & \multicolumn{1}{r}{} \\
\end{tabular}%
\end{minipage}}
\caption{Emotional Stability Results}
\label{tbl:es}
\end{table*}

\begin{table*}[!ht]
 \resizebox{0.88\textwidth}{!}{\begin{minipage}{\textwidth}
\begin{tabular}{lll|lll|lll}
\multicolumn{3}{c}{\textbf{L to L}} & \multicolumn{3}{c}{\textbf{L to N}} & \multicolumn{3}{c}{\textbf{L to H}} \\
\hline
Variable  & Coefficient & P-value & Variable  & Coefficient & P-value & Variable  & Coefficient & P-value \\
\hline
H & \multicolumn{1}{l}{-0.010} & \multicolumn{1}{l|}{0.0009} & Intercept & \multicolumn{1}{l}{-0.115} & \multicolumn{1}{l|}{0.1284} & Intercept & \multicolumn{1}{l}{-1.125} & \multicolumn{1}{l}{0.0001} \\
      & \multicolumn{1}{l}{} & \multicolumn{1}{l|}{} &       & \multicolumn{1}{l}{} & \multicolumn{1}{l|}{} & L     & \multicolumn{1}{l}{-0.0432} & \multicolumn{1}{l}{0.0013} \\
      & \multicolumn{1}{l}{} & \multicolumn{1}{l|}{} &       & \multicolumn{1}{l}{} & \multicolumn{1}{l|}{} & N     & \multicolumn{1}{l}{-0.009} & \multicolumn{1}{l}{0.0001} \\
      & \multicolumn{1}{l}{} & \multicolumn{1}{l|}{} &       & \multicolumn{1}{l}{} & \multicolumn{1}{l|}{} & T     & \multicolumn{1}{l}{1.910} & \multicolumn{1}{l}{0.0000} \\
      & \multicolumn{1}{l}{} & \multicolumn{1}{l|}{} &       & \multicolumn{1}{l}{} & \multicolumn{1}{l|}{} & L*T   & \multicolumn{1}{l}{-0.022} & \multicolumn{1}{l}{0.0019} \\
      & \multicolumn{1}{l}{} & \multicolumn{1}{l|}{} &       & \multicolumn{1}{l}{} & \multicolumn{1}{l|}{} & N*T   & \multicolumn{1}{l}{-0.0067} & \multicolumn{1}{l}{0.0001} \\
      & \multicolumn{1}{l}{} & \multicolumn{1}{l|}{} &       & \multicolumn{1}{l}{} & \multicolumn{1}{l|}{} & H*T   & \multicolumn{1}{l}{-0.027} & \multicolumn{1}{l}{0.0000} \\
      & \multicolumn{1}{l}{} & \multicolumn{1}{l|}{} &       & \multicolumn{1}{l}{} & \multicolumn{1}{l|}{} & (period=1) * T & \multicolumn{1}{l}{-1.085} & \multicolumn{1}{l}{0.0025} \\
      &       & \multicolumn{1}{r}{} &       &       & \multicolumn{1}{r}{} &       &       &  \\
\multicolumn{3}{c}{\textbf{N to L}} & \multicolumn{3}{c}{\textbf{N to N}} & \multicolumn{3}{c}{\textbf{N to H}} \\
\hline
Variable  & Coefficient & P-value & Variable  & Coefficient & P-value & Variable  & Coefficient & P-value \\
\hline
Intercept & \multicolumn{1}{l}{-0.916} & \multicolumn{1}{l|}{0.0000} & Intercept & \multicolumn{1}{l}{0.318} & \multicolumn{1}{l|}{0.0000} & Intercept & \multicolumn{1}{l}{-1.53} & \multicolumn{1}{l}{0.0000} \\
      & \multicolumn{1}{l}{} & \multicolumn{1}{l|}{} &       & \multicolumn{1}{l}{} & \multicolumn{1}{l|}{} & (period=1) & \multicolumn{1}{l}{0.466} & \multicolumn{1}{l}{0.009} \\
      & \multicolumn{1}{l}{} & \multicolumn{1}{l|}{} &       & \multicolumn{1}{l}{} & \multicolumn{1}{l|}{} & T     & \multicolumn{1}{l}{0.689} & \multicolumn{1}{l}{0.0000} \\
      & \multicolumn{1}{l}{} & \multicolumn{1}{l|}{} &       & \multicolumn{1}{l}{} & \multicolumn{1}{l|}{} & N*T   & \multicolumn{1}{l}{-0.002} & \multicolumn{1}{l}{0.003} \\

      &       & \multicolumn{1}{r}{} &       &       & \multicolumn{1}{r}{} &       &       &  \\
      &       & \multicolumn{1}{r}{} &       &       & \multicolumn{1}{r}{} &       &       &  \\
\multicolumn{3}{c}{\textbf{H to L}} & \multicolumn{3}{c}{\textbf{H to N}} & \multicolumn{3}{c}{\textbf{H to H}} \\
\hline
Variable  & \multicolumn{1}{l}{Coefficient} & \multicolumn{1}{l|}{P-value} & Variable  & \multicolumn{1}{l}{Coefficient} & \multicolumn{1}{l|}{P-value} & Variable  & \multicolumn{1}{l}{Coefficient} & \multicolumn{1}{l}{P-value} \\
\hline
Intercept & \multicolumn{1}{l}{-1.268} & \multicolumn{1}{l|}{0.0000} & H     & \multicolumn{1}{l}{-0.026} & \multicolumn{1}{l|}{0.0000} & T     & \multicolumn{1}{l}{0.533} & \multicolumn{1}{l}{0.0012} \\
      & \multicolumn{1}{l}{} & \multicolumn{1}{l|}{} & T     & \multicolumn{1}{l}{-0.728} & \multicolumn{1}{l|}{0.0000} &       & \multicolumn{1}{l}{} & \multicolumn{1}{l}{} \\
      & \multicolumn{1}{l}{} & \multicolumn{1}{l|}{} & H*T   & \multicolumn{1}{l}{0.013} & \multicolumn{1}{l|}{0.0035} &       & \multicolumn{1}{l}{} & \multicolumn{1}{l}{} \\
      & \multicolumn{1}{l}{} & \multicolumn{1}{l|}{} & (period=1) & \multicolumn{1}{l}{-0.660} & \multicolumn{1}{l|}{0.0015} &       & \multicolumn{1}{l}{} & \multicolumn{1}{l}{} \\

\end{tabular}%
\end{minipage}}
\caption{Creativity Results}
\label{tbl:creat}
\end{table*}

\begin{table*}[!ht]
 \resizebox{0.88\textwidth}{!}{\begin{minipage}{\textwidth}
\begin{tabular}{lll|lll|lll}
\multicolumn{3}{c}{\textbf{L to L}} & \multicolumn{3}{c}{\textbf{L to N}} & \multicolumn{3}{c}{\textbf{L to H}} \\
\hline
Variable  & Coefficient & P-value & Variable  & Coefficient & P-value & Variable  & Coefficient & P-value \\
\hline
T     & \multicolumn{1}{l}{-0.424} & \multicolumn{1}{l|}{0.0002} & Intercept & \multicolumn{1}{l}{-0.928} & \multicolumn{1}{l|}{0.0000} & Intercept & \multicolumn{1}{l}{-1.027} & \multicolumn{1}{l}{0.0000}\\
L*T   & \multicolumn{1}{l}{0.0013} & \multicolumn{1}{l|}{0.0256} &       & \multicolumn{1}{l}{} & \multicolumn{1}{l|}{} & T     & \multicolumn{1}{l}{0.4947} & \multicolumn{1}{l}{0.0015} \\
      & \multicolumn{1}{l}{} & \multicolumn{1}{l|}{} &       & \multicolumn{1}{l}{} & \multicolumn{1}{l|}{} & L*T   & \multicolumn{1}{l}{-0.002} & \multicolumn{1}{l}{0.0000} \\
      & \multicolumn{1}{l}{} & \multicolumn{1}{l|}{} &       & \multicolumn{1}{l}{} & \multicolumn{1}{l|}{} & period & \multicolumn{1}{l}{0.511} & \multicolumn{1}{l}{0.0005} \\
      &       & \multicolumn{1}{r}{} &       &       & \multicolumn{1}{r}{} &       &       &  \\
\multicolumn{3}{c}{\textbf{N to L}} & \multicolumn{3}{c}{\textbf{N to N}} & \multicolumn{3}{c}{\textbf{N to H}} \\
\hline
Variable  & Coefficient & P-value & Variable  & Coefficient & P-value & Variable  & Coefficient & P-value \\
\hline
Intercept & -0.439 & 0.001 & Intercept & -0.73 & 0 & Intercept & -0.34 & 0.006 \\
H    & \multicolumn{1}{l}{0.002} & \multicolumn{1}{l|}{0.037} & N     & \multicolumn{1}{l}{-0.005} & \multicolumn{1}{l|}{0.0486} &       & \multicolumn{1}{l}{} & \multicolumn{1}{l}{} \\
  N & \multicolumn{1}{l}{0.007} & \multicolumn{1}{l|}{0.00} & N*T   & \multicolumn{1}{l}{0.0196} & \multicolumn{1}{l|}{0.0704} &      & \multicolumn{1}{l}{} & \multicolumn{1}{l}{} \\
   N*T   & \multicolumn{1}{l}{-0.015} & \multicolumn{1}{l|}{0.05} & (period=1) *T & \multicolumn{1}{l}{-0.356} & \multicolumn{1}{l|}{0.0198} &       & \multicolumn{1}{l}{} & \multicolumn{1}{l}{} \\
      & \multicolumn{1}{l}{} & \multicolumn{1}{l|}{} & (period=0)*T & \multicolumn{1}{l}{0.015} & \multicolumn{1}{l|}{0.9122} &       & \multicolumn{1}{l}{} & \multicolumn{1}{l}{} \\
      &       & \multicolumn{1}{r}{} &       &       & \multicolumn{1}{r}{} &       &       &  \\
\multicolumn{3}{c}{\textbf{H to L}} & \multicolumn{3}{c}{\textbf{H to N}} & \multicolumn{3}{c}{\textbf{H to H}}\\
\hline
Variable  & Coefficient & P-value & Variable  & Coefficient & P-value & Variable  & Coefficient & P-value \\
\hline
Intercept & \multicolumn{1}{l}{-1.134} & \multicolumn{1}{l|}{0.0000} & Intercept & \multicolumn{1}{l}{-1.180} & \multicolumn{1}{l|}{0.0000} & Intercept & \multicolumn{1}{l}{0.416} & \multicolumn{1}{l}{0.0016} \\
L     & \multicolumn{1}{l}{0.0011} & \multicolumn{1}{l|}{0.0306} & H     & \multicolumn{1}{l}{-0.0042} & \multicolumn{1}{l|}{0.0016} & L     & \multicolumn{1}{l}{-0.0013} & \multicolumn{1}{l}{0.0484} \\
T     & \multicolumn{1}{l}{-0.440} & \multicolumn{1}{l|}{0.0359} & T     & \multicolumn{1}{l}{-0.448} & \multicolumn{1}{l|}{0.0080} & T     & \multicolumn{1}{l}{0.594} & \multicolumn{1}{l}{0.0002} \\
L*T   & \multicolumn{1}{l}{0.002} & \multicolumn{1}{l|}{0.0451} & L*T   & \multicolumn{1}{l}{0.002} & \multicolumn{1}{l|}{0.0249} & L*T   & \multicolumn{1}{l}{-0.003} & \multicolumn{1}{l}{0.0019} \\
H*T   & \multicolumn{1}{l}{-0.006} & \multicolumn{1}{l|}{0.002} &       &       &       & H*T   & 0.003387911 & 0.0014 \\
\end{tabular}%
\end{minipage}}
\caption{High Positive Affect (HPA) Results}
\label{tbl:hpa}
\end{table*}

\clearpage

\begin{table*}[!ht]
 \resizebox{0.88\textwidth}{!}{\begin{minipage}{\textwidth}

\begin{tabular}{lll|lll|lll}
\multicolumn{3}{c}{\textbf{L to L}} & \multicolumn{3}{c}{\textbf{L to N}} & \multicolumn{3}{c}{\textbf{L to H}}\\
\hline
Variable  & \multicolumn{1}{l}{Coefficient} & \multicolumn{1}{l|}{P-value} & Variable  & \multicolumn{1}{l}{Coefficient} & \multicolumn{1}{l|}{P-value} & Variable  & \multicolumn{1}{l}{Coefficient} & \multicolumn{1}{l}{P-value} \\
\hline
H*T   & \multicolumn{1}{l}{0.003} & \multicolumn{1}{l|}{0.0215} & Intercept & \multicolumn{1}{l}{-0.157} & \multicolumn{1}{l|}{0.1354} & Intercept & \multicolumn{1}{l}{-1.151} & \multicolumn{1}{l}{0.0000}\\
      & \multicolumn{1}{l}{} & \multicolumn{1}{l|}{} &       & \multicolumn{1}{l}{} & \multicolumn{1}{l|}{} & L     & \multicolumn{1}{l}{-0.003} & \multicolumn{1}{l}{0.0059} \\
      & \multicolumn{1}{l}{} & \multicolumn{1}{l|}{} &       & \multicolumn{1}{l}{} & \multicolumn{1}{l|}{} & N     & \multicolumn{1}{l}{-0.017} & \multicolumn{1}{l}{0.0000} \\
      & \multicolumn{1}{l}{} & \multicolumn{1}{l|}{} &       & \multicolumn{1}{l}{} & \multicolumn{1}{l|}{} &       & \multicolumn{1}{l}{} & \multicolumn{1}{l}{} \\
\multicolumn{3}{c}{\textbf{N to L}} & \multicolumn{3}{c}{\textbf{N to N}} & \multicolumn{3}{c}{\textbf{N to H}} \\
\hline
Variable  & \multicolumn{1}{l}{Coefficient} & \multicolumn{1}{l|}{P-value} & Variable  & \multicolumn{1}{l}{Coefficient} & \multicolumn{1}{l|}{P-value} & Variable  & \multicolumn{1}{l}{Coefficient} & \multicolumn{1}{l}{P-value} \\
\hline
Intercept & \multicolumn{1}{l}{-1.156} & \multicolumn{1}{l|}{0.0000} & Intercept & \multicolumn{1}{l}{0.440} & \multicolumn{1}{l|}{0.0000} & Intercept & \multicolumn{1}{l}{-1.317} & \multicolumn{1}{l}{0.0000}\\
N*T   & \multicolumn{1}{l}{0.001} & \multicolumn{1}{l|}{0.0481} & N     & \multicolumn{1}{l}{0.0005} & \multicolumn{1}{l|}{0.0602} & L     & \multicolumn{1}{l}{-0.009} & \multicolumn{1}{l}{0.0034} \\
period & \multicolumn{1}{l}{0.615} & \multicolumn{1}{l|}{0.0000} & (period=1) *T & \multicolumn{1}{l}{0.248} & \multicolumn{1}{l|}{0.0265} & N     & \multicolumn{1}{l}{-0.0007} & \multicolumn{1}{l}{0.0305} \\
      & \multicolumn{1}{l}{} & \multicolumn{1}{l|}{} & (period=0) *T & \multicolumn{1}{l}{-0.267} & \multicolumn{1}{l|}{0.0567} &       & \multicolumn{1}{l}{} & \multicolumn{1}{l}{} \\
      &       & \multicolumn{1}{r}{} &       &       & \multicolumn{1}{r}{} &       &       &  \\
      &       & \multicolumn{1}{r}{} &       &       & \multicolumn{1}{r}{} &       &       &  \\
\multicolumn{3}{c}{\textbf{H to L}} & \multicolumn{3}{c}{\textbf{H to N}} & \multicolumn{3}{c}{\textbf{H to H}} \\

Variable  & \multicolumn{1}{l}{Coefficient} & \multicolumn{1}{l|}{P-value} & Variable  & \multicolumn{1}{l}{Coefficient} & \multicolumn{1}{l|}{P-value} & Variable  & \multicolumn{1}{l}{Coefficient} & \multicolumn{1}{l}{P-value} \\
\hline
Intercept & \multicolumn{1}{l}{-1.193} & \multicolumn{1}{l|}{0.0000} & Intercept & \multicolumn{1}{l}{-0.876} & \multicolumn{1}{l|}{0.0000} & (period=1) & \multicolumn{1}{l}{0.011} & \multicolumn{1}{l}{0.9501} \\
L     & \multicolumn{1}{l}{-0.002} & \multicolumn{1}{l|}{0.0490} & T     & \multicolumn{1}{l}{-0.403} & \multicolumn{1}{l|}{0.0018} & (period=0) & \multicolumn{1}{l}{0.321} & \multicolumn{1}{l}{0.0587} \\
T     & \multicolumn{1}{l}{-0.594} & \multicolumn{1}{l|}{0.0072} & N*T   & \multicolumn{1}{l}{0.002} & \multicolumn{1}{l|}{0.0601} & T     & \multicolumn{1}{l}{0.708} & \multicolumn{1}{l}{0.0000} \\
L*T   & \multicolumn{1}{l}{0.004} & \multicolumn{1}{l|}{0.0004} & period & \multicolumn{1}{l}{0.403} & \multicolumn{1}{l|}{0.0011} & L*T   & \multicolumn{1}{l}{-0.003} & \multicolumn{1}{l}{0.0228} \\
\end{tabular}%
\end{minipage}}
\caption{Low Negative Affect (LNA) Results}
\label{tbl:lna}
\end{table*}

\begin{figure*}
\begin{center}
{\includegraphics[width=.7\textwidth]{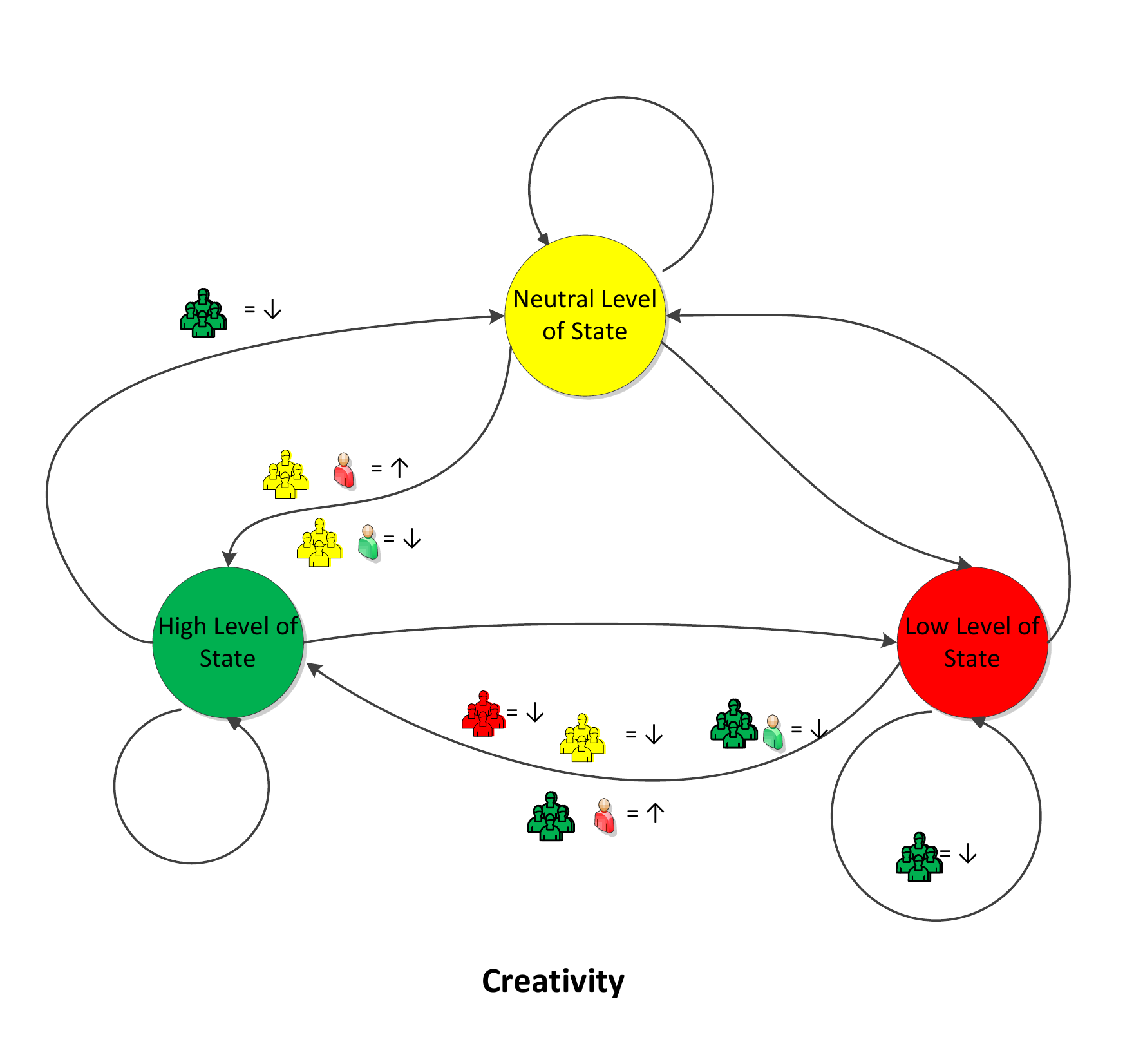}}
\caption{Level transition graph for creativity state: Nodes represent creativity level of the ego. An arrow between two circles represents the transition from one level to another. These transitions are labeled with conditions that affect the corresponding probabilities. Icons represent the creativity levels of alters and ego's trait level. Symbol $\uparrow$ (respectively $\downarrow$) indicates an increase (respectively decrease) in transition probability associated with the given combination of alters state level and ego trait level. For example, if the ego is in the low level of the creativity state, then the probability of him/her transitioning to the high level decreases with his/her interactions with alters in the high level. }
\label{creat_statePatterns}
\end{center}
\end{figure*}

\subsubsection*{Effect of Time of Day}
\paragraph{Extraversion} The spontaneous transition (N $\to$ L and H $\to$ L) during the afternoon (transition from midday to afternoon) increases with respect to midday (transition from morning to midday). This is consistent with the post hoc comparisons of ANOVA that are reported earlier in this section. That is, the spontaneous tendency is for people who behaved more extravertedly at middays to behave introvertedly in afternoons. 


\paragraph{Conscientiousness} During the afternoon, people tend to move from the low level of conscientiousness state to reach the high level (L $\to$ H). Also during the afternoon, the spontaneous transitions (N $\to$ H) increases with respect to midday. Quite generally,  people who behaved low or neutral in conscientiousness in midday tend to behave in a highly conscientious way in the afternoon. 

\paragraph{Emotional Stability} During the afternoon, the spontaneous transition (N $\to$ L) increases with respect to midday. However, the spontaneous transition (H $\to$ L) decreases with respect to midday. Therefore, during the afternoon, emotional stability level tend to decrease by only one level (N $\to$ L).

\paragraph{Creativity}  During the afternoon, people tend to move from the neutral level to the high level (N $\to$ H) with respect to midday. On the other hand, during the afternoon, the spontaneous transition (H $\to$ N) decreases  with respect to midday. 


\paragraph{High Positive Affect} During the afternoon, people move from the low level to the high level (L $\to$ H) with respect to midday. This is consistent with ANOVA results that were reported earlier for HPA. 

\paragraph{Low Negative Affect} During the afternoon, spontaneous transitions (N $\to$ L) and (H $\to$ N) increases with respect to midday. 

\subsubsection{Interaction between Time of Day and Trait}
We have anticipated that there would be an interaction between the corresponding trait and the time of the day. Therefore, we added the interaction as an additional control variable in our model. 

With respect to extraversion, the baseline probablity of ($N \to N$) stability for extroverts (according to trait level) increases from midday to the afternoon. However, the baseline probability of the same stability for introverts decreases during the same period. The probability of spontaneous ($H \to L$) tranisition increases for all egos from midday to the afternoon and more markedly so for extroverts.

With respect to agreeableness, the baseline probablity of ($N \to N$) stability for people with high score in agreeableness trait decreases from midday to the afternoon. However, the baseline probability of the same stability for people with low scores in agreeableness trait increases during the same period. 

With respect to emotional stability, the baseline probablity of ($L \to H$) transition for people with high score in emotional stability trait decreases from midday to the afternoon. However, the baseline probability of the same transition for people with low scores in emotional stability trait increases during the same period. The basline probability of ($N \to L$) transitions importantly increase from midday to afteroon for neurotic people (according to trait) and decrease for emotionally stable ones. Putting this together with the former, it seems that emotionally stable people are indeed more stable because they have a strong tendency to maintain their level they had in the morning. The spontaneous transition ($H \to L$) for emotionally stable people (according to the trait score) decreases. This confirms that  emotionally stable people (according to trait) have decreasing probability of moving from where they are from midday to the afternoon while neurotic people tend to change their level more often.

With respect to creativity, the spontaneous ($L \to H$) transition decreases in the afternoon for highly creative people (according to their trait). The opposite holds for people with low scores in the creativity trait.

With respect to high positive affect state,  the probability of spontaneous ($N \to N$) stability increases for people who have low scores in the trait in the afternoon and decreases for people who have high scores in the trait.

With respect to low negative affect state, the probability of spontaneous ($N \to N$) stability ncreases for people who have high scores in the trait in the afternoon and decreases for people who have low scores in the trait.


Second, we present our results of social influences in an increasing order according to provided details. First, we present the social influences of interaction between the social-situational factors and the dispositional factor. Second, we present the results from which we derive the social influences by reporting the direction of impact of each interaction between the two aforementioned factors: either increasing or decreasing the probability of the transition between each two levels of the state. Third, we present the coefficients of main effects and the interaction effects as a result of running our logistic regression model.

\subsection*{State-Specific Patterns}

In Fig. \ref{statePatterns}, we demonstrate the role that social-situational factors (intensity of contacts with alters in a certain level) played with respect to the dispositional factor (trait's level of the ego) towards the transitions between levels of the state in question.

\paragraph{Extraversion:} 

\begin{itemize}
\item Generally, alters acting introvertedly (according to state level) attract egos to stay or move towards their level except the egos who are already in the neutral state level and have high scores in the trait. 
\item Alters in the neutral level urge egos to act according to their traits. 
\item Surprisingly, alters in the high level of the state mostly attract egos with low trait scores to stay in the high level or move towards the alters' level whereas those alters repulse egos with high trait scores.
\end{itemize}

\paragraph{Agreeableness:} 
\begin{itemize}
   \item Generally, alters in the low level of the state encourage egos to be less agreeable except for egos who are agreeable by nature (trait) and already in the high level.
\item Alters in the neutral level urge egos with high trait scores to stay at their transient states while they mostly have an inconsistent role in case of egos with low trait scores.
\item Alters in the high level play a balancing role. They encourage egos with low trait scores to stay at their level but they repulse those egos if they are in the low level. Inversely, they push egos with high trait scores who are in their level but they attract those egos if they are in the low level of the state.


\end{itemize}


\paragraph{Conscientiousness:} 
\begin{itemize}
\item Alters in the low level of the state attract egos to stay in the low level or switch to lower levels if those egos are already not conscientious by nature. Nevertheless, the alters cannot drag conscientious egos by nature towards their level unless the egos are already in the low level.
\item Alters in the neutral level repulse egos to stay in their transient levels.
\item Alters in the high level help egos to stay in the high level or move to their level except egos who are in the neutral level and have low trait scores. 
\end{itemize}

\paragraph{Emotional Stability:} 
\begin{itemize}
   \item Alters in the low level repulse egos who are in the neutral level.
   \item Alters in the neutral level of the state repulse egos, if those egos are in the low level of the state and have high scores in the trait.
   \item Mostly, alters in the high level repulse egos with low trait scores and enocurage egos to stay at the high level if the egos already have high trait scores. 
\end{itemize}


\paragraph{Creativity:} 
\begin{itemize}
   \item Alters in the low level mostly encourage egos to stay at their level.
\item Alters in the neutral level mostly repulse egos away from their level except the egos who are in their level and have high trait scores.
\item Alters in the high level mostly encourage egos to stay or move towards their level regardless of egos' traits.
\end{itemize}

\paragraph{High Positive affect (HPA):} 
\begin{itemize}
   \item Surprisingly, alters in the low level push egos with low trait scores away from their level and attract egos with high scores in the trait to stay or move towards the low level.
\item  Alters in the neutral level push egos with low trait scores away from their level and encourage egos with high trait scores to stay at the neutral level
\item Alters in the high level push egos with low trait scores away from the high level and attract egos with high trait scores to stay at the high level of the state. 

\end{itemize}

\paragraph{Low Negative affect (LNA):} 
\begin{itemize}
   \item Mostly, alters in the low level attract egos to stay or move to their level except the egos who are in the high state level and have low trait scores
\item Alters in the neutral level mostly repulse egos away from their level.
\item Surprisingly, alters in the high level attract egos with low trait scores and repulse egos with high trait scores.
\end{itemize}

\subsection*{State Diagrams of Dynamic States}
In order to derive the social influences depicted in Fig. \ref{statePatterns}, we observed the individual effects of the interaction between social-situational factors (intensity of contacts with alters in each level) and the dispositional factor (trait level of the ego) for each transition in the state in question as illustrated in Fig. \ref{ext_statePatterns}, Fig. \ref{agr_statePatterns}, Fig. \ref{consc_statePatterns}, Fig. \ref{es_statePatterns}, Fig. \ref{creat_statePatterns}, Fig. \ref{hpa_statePatterns} and Fig. \ref{lna_statePatterns}. In the figures, we labeled each arrow between each pair  of levels of the state according to the direction of the effect of the mentioned factors (increasing ($\uparrow$) or decreasing the probability of transition ($\downarrow$)) 
We reported only the effects of factors that have statistically significant results in a given transition. Those effects could be: (1) the marginal effects of intensity of contacts with alters in a certain level of the state (e.g. intensity of contacts with alters in the high level) (2) the interaction between social-situational and dispositional factors e.g. the interaction between intensity of contacts with  alters in the high level and the trait's high level of the ego.

\subsubsection*{Extraversion}
 \paragraph{$L \to L$} The intensity of contacts with alters in the neutral level is associated with an increase in the probability that the ego will stay at the low level ($L \to L$), in case the ego has a low score in the extraversion trait.  The intensity of contacts with the same alters turns to be associated with a decrease in the probability of  ($L \to L$), if the ego has a high score in the corresponding trait. This manifests the interaction effect of the ego's trait level and the state's levels of alters with whom the ego get in touch with.  

\paragraph{$L \to H$} The intensity of contacts with alters in low and high levels of the state is associated with a decrease in the probability of this transition.

\paragraph{$N \to L$} The intensity of contacts with alters in the high level is associated with a decrease in the probability of  ($N \to L$), if the ego has a low score in the trait. The intensity of contacts with the same alters is associated with an increase in the probability of the same transition, if the ego has a high score in the trait. 

\paragraph{$N \to N$} The intensity of contacts with alters in the low level is associated with an increase in the probability of stability in the neutral level ($N \to N$) if the ego has a high score in the trait. However, the same intensity is associated with a decrease in the probability of  ($N \to N$), if the ego has a low score in the trait. 

\paragraph{$H \to L$} The intensity of contacts with alters in the high level is associated with a decrease in the probability of the transition, if the ego has a low score in the trait. The intensity of contacts with alters in the high level is associated with an increase in the probability of the transition, if the ego has a high score in the trait.

  \paragraph{$H \to H$}
The intensity of contacts with alters in the neutral level is associated with a decrease in the probability of the stability, if the ego has a low score in the trait. However, the intensity of contacts with the same alters becomes associated with an increase in the probability of the stability, if the ego has a high score in the trait. The intensity of contacts with alters in the high level is associated with an increase in the probability of the stability, if the ego has a low score in the trait. However, the intensity of contacts with alters in the high level of the state is associated with a decrease in the probability of the stability, if the ego has a high score in the trait.

\subsubsection*{Agreeableness}

  \paragraph{$L \to L$}
 The intensity of contacts with alters in the high level is associated with an increase in the probability of the stability, if the ego has a low score in the trait. However, the intensity of contacts with alters in the high level is associated with a decrease in the probability of the stability, if the ego has a high score in the trait.

  \paragraph{$L \to N$}
The intensity of contacts with alters in the high level of the state is associated with an increase in the probability of the transition, if the ego has a high score in the trait. However, the intensity of contacts with the same alters is associated with a decrease in the probability of the transition, if the ego has a low score in the trait.

  \paragraph{$L \to H$} The intensity of contacts with alters in low or high levels of the state is associated with a decrease in the probability of the transition. Also, the intensity of contacts with alters in the neutral level of the state is associated with a decrease in the probability of the transition, if the ego has a high score in the trait. However, the intensity of contacts with alters in the neutral level is associated with an increase in the probability of transition of egos with low trait scores

  \paragraph{$N \to N$} The intensity of contacts with alters in the neutral level of the state is associated with a decrease in the probability of the stability, if the ego has a low score in the trait. If the ego has a high score in the trait, then the intensity of contacts with the same alters becomes associated with an increase in the probability of stability.

  \paragraph{$N \to H$}  The intensity of contacts with alters in the neutral level of the state is associated with an increase in the probability of the transition, if the ego has a low score in the trait. However, the intensity of contacts with the same alters is associated with a decrease in the transition, if the ego has a high score in the trait.

  \paragraph{$H \to N$}
The intensity of contacts with alters in low and neutral levels of the state is associated with an increase in the probability of the transition, if the ego has a low score in the trait. However, the intensity of contacts with the same alters is associated with a decrease in the probability of the transition, if the ego has a high score in the trait. The intensity of contacts with alters in the high level of the state is associated with an decrease in the probability of the transition, if the ego has a low score in the trait. However, the intensity of contacts with the same alters is associated with an increase in the transition of egos with high trait scores.

  \paragraph{$H \to H$}
The intensity of contacts with alters in low and neutral levels of the state is associated with a decrease in the probability of the stability of egos with low scores in the trait, whereas the intensity of contacts with the alters is associated with an increase in the probability of the stability of egos with high scores in the trait. The intensity of contacts with alters in the high level of the state is associated with an increase in the probability of the stability of egos with low scores in the trait, whereas the intensity is associated with a decrease in the probability of the stability of egos with high scores in the trait.

\subsubsection*{Conscientiousness}
  \paragraph{$L \to L$}
The intensity of contacts with alters in the neutral level is associated with an increase in the probability of staying at the low level. Also, the intensity of contacts with alters in the low level is associated with an increase in the probability of stability of egos with high trait scores. The intensity of contacts with alters in the high level is associated with a decrease in the probability of stability of egos with high trait scores.

  \paragraph{$L \to N$}
The intensity of contacts with alters in the neutral level is associated with a decrease in the probability of the transition. 

  \paragraph{$L \to H$}
The intensity of contacts with alters in the neutral level is associated with a decrease in the probability of the transition of egos with high trait scores. Also, the intensity of contacts with alters in the low level is associated with a decrease in the probability of the transition of all egos.

  \paragraph{$N \to L$}
The intensity of contacts with alters in the low level is associated with a decrease in the probability of the transition of egos with high trait scores. However, the intensity of contacts with the same alters becomes associated with an increase in the probability of the transition of egos with low trait scores. The intensity of contacts with alters in the high level of the state is associated with a decrease in the probability of the transition of egos with high trait scores. However, the intensity of contacts with the same alters becomes associated with an increase in the probability of the transition of egos with low trait scores. 

  \paragraph{$N \to N$}
The intensity of contacts with alters in low and high levels of the state is associated with a decrease in the probability of the  stability of egos with low traits scores.The intensity of contacts with the same alters is associated with an increase in the probability of the stability of egos with high trait scores.

 \paragraph{$H \to L$}
The intensity of contacts with alters in the high level is associated with a  decrease in the probability of the transition. The intensity of contacts with alters in the low level is associated with an increase in the probability of the transition of egos with low trait scores.

 \paragraph{$H \to N$}
The intensity of contacts with alters in the neutral level of the state is associated with a decrease in the probability of the transition.

 \paragraph{$H \to H$}
The intensity of contacts with alters in the neutral level of the state is associated with an increase in the probability of the stability of egos with low trait scores.

\subsubsection*{Emotional Stability}

 \paragraph{$L \to H$}
The intensity of contacts with alters in the neutral level of the state is associated with a decrease in the probability of the transition of egos with high trait scores.

 \paragraph{$N \to L$}
The intensity of contacts with alters in the low level of the state is associated with a decrease in the probability of the transition.

 \paragraph{$N \to N$}
The intensity of contacts with alters in the low level of the state is associated with an increase in the probability of the stability. Also, the intensity of contacts with alters in the high level of the state is associated with an increase in the probability of the stability of egos with low trait scores.

 \paragraph{$N \to H$}
The intensity of contacts with alters in the high level of the state is associated with a decrease in the probability of the transition of egos with low trait scores.

 \paragraph{$H \to L$}
The intensity of contacts with alters in the high level of the state is associated with a decrease in the probability of the transition of egos with high trait scores whereas the intensity of contacts is associated with an increase in the probability of the transition of egos with low trait scores.

\subsubsection*{Creativity}
 \paragraph{$L \to L$} The intensity of contacts with alters in the high level of the state is associated with a decrease in the probability of the stability.

  \paragraph{$L \to H$} The intensity of contacts with alters in the low and neutral levels of the state is associated with a decrease in the probability of the transition. Also, the intensity of contacts with alters in the high level of the state is associated with a decrease in the probability of the transition of egos with high trait scores. However, the intensity of contacts with alters in the high level is associated with an increase in the probability of the transition of egos with low trait scores.

 \paragraph{$N \to H$} The intensity of contacts with alters in the neutral level is associated with an increase in the probability of the transition of egos with low trait scores whereas the intensity becomes associated with a decrease in the probability of the transition of egos with high trait scores.

 \paragraph{$H \to N$} The intensity of contacts with alters in the high level of the state is associated with a decrease in the probability of the transition.

\subsubsection*{High Positive Affect (HPA)}
 \paragraph{$L \to L$} The intensity of contacts with alters in the low level is associated with a decrease in the probability of the stability of egos with low trait scores. The intensity of contacts with the same alters becomes associated with an increase in the probability of the stability of egos with high trait scores.

 \paragraph{$L \to H$}
The intensity of contacts with alters in the low level is associated with a decrease in the probability of the transition of egos with high trait scores. The intensity of contacts with the same alters becomes associated with a increase in the probability of the transition of egos with low trait scores.

 \paragraph{$N \to L$}
The intensity of contacts with alters in the high levels is associated with an increase in the probability of the transition. The intensity of contacts with alters in the neutral level is associated with a decrease in the probability of the transition of egos with high trait scores. The intensity of contacts with the same alters becomes associated with a increase in the probability of the transition of egos with low trait scores. 

 \paragraph{$N \to N$}
The intensity of contacts with alters in the neutral level is associated with a decrease in the probability of the stability of egos with low trait scores. The intensity of contacts with the same alters becomes associated with a increase in the probability of the stability of egos with high trait scores. 
 
 \paragraph{$H \to L$} The intensity of contacts with alters in the low level is associated with a decrease in the probability of the transition of egos with low trait scores whereas the intensity becomes associated with an increase in the probability of the transition of egos with high trait scores.  The intensity of contacts with alters in the high level is associated with a decrease in the probability of the transition of egos with high trait scores whereas the intensity becomes associated with an increase in the probability of the transition of egos with low trait scores.

 \paragraph{$H \to N$} The intensity of contacts with alters in the high level is associated with a decrease in the probability of the transition. The intensity of contacts with alters in the low level is associated with a decrease in the probability of the transition of egos with low trait scores whereas the intensity becomes associated with an increase in the probability of the transition of egos with high trait scores.

\paragraph{$H \to H$} The intensity of contacts with alters in the high level is associated with an increase in the probability of the stability of egos with high trait scores,whereas the intensity becomes associated with an decrease in the probability of the stability of egos with low trait scores. The intensity of contacts with alters in the low level is associated with an increase in the probability of the stability of egos with low trait scores,whereas the intensity becomes associated with an decrease in the probability of the stability of egos with high trait scores.

\subsubsection*{Low Negative Affect (LNA)}

 \paragraph{$L \to L$} The intensity of contacts with alters in the high level is associated with an increase in the probability of the stability of egos with high trait scores whereas the intensity becomes associated with a decrease in the probability of the stability of egos with low trait scores.

 \paragraph{$L \to H$} The intensity of contacts with alters in neutral and low levels is associated with a decrease in the probability of the transition.

 \paragraph{$N \to H$} The intensity of contacts with alters in the low level is associated with a decrease in the probability of the transition.

 \paragraph{$H \to L$} The intensity of contacts with alters in the low level is associated with an increase in the probability of the transition of egos with high trait scores, whereas the intensity becomes associated with a decrease in the probability of the transition of egos with low trait scores. 

 \paragraph{$H \to N$} The intensity of contacts with alters in the neutral level is associated with an increase in the probability of the transition of egos with high trait scores, whereas the intensity becomes associated with a decrease in the probability of the transition of egos with low trait scores. 

 \paragraph{$H \to H$} The intensity of contacts with alters in the low level of the state is associated with an increase in the probability of the stability of egos with low trait scores whereas the intensity becomes associated with a decrease in the probability of the stability of egos with high trait scores.

\section*{Detailed Results}
We report our detailed results (best sub-model according to QICC) for each transition in each state in Tables  \ref{tbl:extr}, \ref{tbl:agr}, \ref{tbl:cons}, \ref{tbl:es}, \ref{tbl:creat}, \ref{tbl:hpa} and \ref{tbl:lna}. Each sub-table is labeled at the top with the corresponding transition in  a given state. The mere effects of social-situational factors (intensity with alters in each level: L, N and H) and corresponding traits of egos (T) are reported in the table, if they are statistically significant. The interaction results between the two effect are reported also ($L*T$, $N*T$ and $H*T$), if they are statistically significant. The coefficients of the control variables are reported also: the main effect of the time of the day (period) and the interaction between the time of the day and the trait (period*T).  Some reported coefficients are relatively small, therefore we used a threshold of 0.001 to consider them relevant. We focus more on the direction of the effect (increase or decrease in the probability) rather than the actual value of the effect.

\section*{Broad Speculations and Further Work}

So far, we have mainly been describing the data focusing on statistically significant effects and on their interpretation in terms of broader patterns of adaptation and complementarity. As stated in the main paper, we cannot make conclusive statements about causality without adopting additional assumptions. Nevertheless, in this section, we allow ourselves some liberty to make bolder speculations about possible implications of our findings, in case they corresponded to causal mechanisms. By definition, these observations are preliminary, and require further validation. Nonetheless, we believe they provide useful insights for further targeted studies.

\subsection*{Conscientiousness}

\paragraph{Conscientious behaviors mutually reinforce:} People acting conscientiously help each other maintain conscientious behavior, regardless of their trait. 

\paragraph{Unconscientious behaviors mutually reinforce:} As above but in the opposite direction, people acting unconscientiously help each other maintain the unconscientious behavior  regardless of their trait.

\paragraph{We cannot help you become good, but we can help you stay good:} Conscientiously behaving people fail to pull bad apples. However if the latter already engage in virtuous behaviors, conscientiously behaving people can help them keep going.

\paragraph{We cannot make you bad, unless you already chose to:} Similarly, unconscientiously behaving people fail to pull good apples. But if the latter are already going astray, the former can keep them on the wrong way.

\subsection*{Agreeableness}

\paragraph{The spiral of stubbornness:} If someone is behaving stubbornly (e.g. being difficult, argumentative, less agreeable), then interacting with people who are behaving similarly helps keep him/her in that state irrespective of his/her trait.

\paragraph{stubbornly behaving people activate others' stubbornness:} If you are stubborn by nature, and happen to behave agreeably, stay away from argumentative people, because they can stimulate your natural stubbornness.

\subsection*{Extraversion} 

\paragraph{Bringing you out of the shell:} If you are introvert by nature, hanging out with people behaving extrovertedly can help you become more outgoing.

\paragraph{Stealing the thunder effect:} If you are extrovert by nature, and acting extrovertedly in a group of extroverts, the latter might ''steal your thunder'' and suppress your extroverted behavior.

\subsection*{Creativity}
\paragraph{Creative behaviors are attractive and mutually reinforce:} If you want to behave or continue behaving in a creative way, hang out with creative people.

\paragraph{Low-creative behaviors mutually reinforce:} People behaving uncreatively help each other maintain the uncreative behavior, regardless of their trait.


\begin{thebibliography}{10}
	\expandafter\ifx\csname url\endcsname\relax
	\def\url#1{\texttt{#1}}\fi
	\expandafter\ifx\csname urlprefix\endcsname\relax\def\urlprefix{URL }\fi
	\providecommand{\bibinfo}[2]{#2}
	\providecommand{\eprint}[2][]{\url{#2}}
	
	\bibitem{asch1955opinions}
	\bibinfo{author}{Asch, S.~E.}
	\newblock \bibinfo{title}{Opinions and social pressure}.
	\newblock \emph{\bibinfo{journal}{Readings about the social animal}}
	\bibinfo{pages}{17--26} (\bibinfo{year}{1955}).
	
	\bibitem{wood2000attitude}
	\bibinfo{author}{Wood, W.}
	\newblock \bibinfo{title}{Attitude change: Persuasion and social influence}.
	\newblock \emph{\bibinfo{journal}{Annual review of psychology}}
	\textbf{\bibinfo{volume}{51}}, \bibinfo{pages}{539--570}
	(\bibinfo{year}{2000}).
	
	\bibitem{socialphysics}
	\bibinfo{author}{Pentland, A.}
	\newblock \emph{\bibinfo{title}{Social Physics}} (\bibinfo{publisher}{Penguin
		Press}, \bibinfo{year}{2014}).
	
	\bibitem{cialdini2004social}
	\bibinfo{author}{Cialdini, R.~B.} \& \bibinfo{author}{Goldstein, N.~J.}
	\newblock \bibinfo{title}{Social influence: Compliance and conformity}.
	\newblock \emph{\bibinfo{journal}{Annu. Rev. Psychol.}}
	\textbf{\bibinfo{volume}{55}}, \bibinfo{pages}{591--621}
	(\bibinfo{year}{2004}).
	
	\bibitem{Fleeson2007}
	\bibinfo{author}{Fleeson, W.}
	\newblock \bibinfo{title}{{Situation-based contingencies underlying
			trait-content manifestation in behavior}}.
	\newblock \emph{\bibinfo{journal}{Journal of Personality}}
	\textbf{\bibinfo{volume}{75}}, \bibinfo{pages}{825--861}
	(\bibinfo{year}{2007}).
	
	\bibitem{61million}
	\bibinfo{author}{Bond, R.~M.} \emph{et~al.}
	\newblock \bibinfo{title}{A 61-million-person experiment in social influence
		and political mobilization}.
	\newblock \emph{\bibinfo{journal}{Nature}} \textbf{\bibinfo{volume}{489}},
	\bibinfo{pages}{295--298} (\bibinfo{year}{2012}).
	
	\bibitem{mani2013inducing}
	\bibinfo{author}{Mani, A.}, \bibinfo{author}{Rahwan, I.} \&
	\bibinfo{author}{Pentland, A.}
	\newblock \bibinfo{title}{Inducing peer pressure to promote cooperation}.
	\newblock \emph{\bibinfo{journal}{Scientific reports}}
	\textbf{\bibinfo{volume}{3}}, \bibinfo{pages}{1--9} (\bibinfo{year}{2013}).
	
	\bibitem{schultz2007constructive}
	\bibinfo{author}{Schultz, P.~W.}, \bibinfo{author}{Nolan, J.~M.},
	\bibinfo{author}{Cialdini, R.~B.}, \bibinfo{author}{Goldstein, N.~J.} \&
	\bibinfo{author}{Griskevicius, V.}
	\newblock \bibinfo{title}{The constructive, destructive, and reconstructive
		power of social norms}.
	\newblock \emph{\bibinfo{journal}{Psychological science}}
	\textbf{\bibinfo{volume}{18}}, \bibinfo{pages}{429--434}
	(\bibinfo{year}{2007}).
	
	\bibitem{burt1987social}
	\bibinfo{author}{Burt, R.~S.}
	\newblock \bibinfo{title}{Social contagion and innovation: Cohesion versus
		structural equivalence}.
	\newblock \emph{\bibinfo{journal}{American journal of Sociology}}
	\textbf{\bibinfo{volume}{92}}, \bibinfo{pages}{1287--1335}
	(\bibinfo{year}{1987}).
	
	\bibitem{coleman1957diffusion}
	\bibinfo{author}{Coleman, J.}, \bibinfo{author}{Katz, E.} \&
	\bibinfo{author}{Menzel, H.}
	\newblock \bibinfo{title}{The diffusion of an innovation among physicians}.
	\newblock \emph{\bibinfo{journal}{Sociometry}} \bibinfo{pages}{253--270}
	(\bibinfo{year}{1957}).
	
	\bibitem{Christakis2013}
	\bibinfo{author}{Christakis, N.} \& \bibinfo{author}{Fowler, J.}
	\newblock \bibinfo{title}{Social contagion theory: examining dynamic social
		networks and human behavior}.
	\newblock \emph{\bibinfo{journal}{Statistics in Medicine}}
	\bibinfo{pages}{556--577} (\bibinfo{year}{2013}).
	
	\bibitem{Christakis2007}
	\bibinfo{author}{Christakis, N.} \& \bibinfo{author}{Fowler, J.}
	\newblock \bibinfo{title}{{The Spread of Obesity in a Large Social Network over
			32 Years}}.
	\newblock \emph{\bibinfo{journal}{New England Journal of Medicine}}
	\textbf{\bibinfo{volume}{357}}, \bibinfo{pages}{370--379}
	(\bibinfo{year}{2007}).
	
	\bibitem{coviello2014detecting}
	\bibinfo{author}{Coviello, L.} \emph{et~al.}
	\newblock \bibinfo{title}{Detecting emotional contagion in massive social
		networks}.
	\newblock \emph{\bibinfo{journal}{{PLOS ONE}}} \textbf{\bibinfo{volume}{9}},
	\bibinfo{pages}{e90315} (\bibinfo{year}{2014}).
	
	\bibitem{kramer2014experimental}
	\bibinfo{author}{Kramer, A.~D.}, \bibinfo{author}{Guillory, J.~E.} \&
	\bibinfo{author}{Hancock, J.~T.}
	\newblock \bibinfo{title}{Experimental evidence of massive-scale emotional
		contagion through social networks}.
	\newblock \emph{\bibinfo{journal}{Proceedings of the National Academy of
			Sciences}} \textbf{\bibinfo{volume}{111}}, \bibinfo{pages}{8788--8790}
	(\bibinfo{year}{2014}).
	
	\bibitem{Moturu2010}
	\bibinfo{author}{Madan, A.}, \bibinfo{author}{Moturu, S.~T.},
	\bibinfo{author}{Lazer, D.} \& \bibinfo{author}{Pentland, A.~S.}
	\newblock \bibinfo{title}{Social sensing: Obesity, unhealthy eating and
		exercise in face-to-face networks}.
	\newblock In \emph{\bibinfo{booktitle}{Wireless Health 2010}},
	\bibinfo{pages}{104--110} (\bibinfo{publisher}{ACM}, \bibinfo{address}{New
		York, NY, USA}, \bibinfo{year}{2010}).
	
	\bibitem{Fowler2010}
	\bibinfo{author}{Fowler, J.} \& \bibinfo{author}{Christakis, N.}
	\newblock \bibinfo{title}{{Cooperative Behavior Cascades in Human Social
			Networks}}.
	\newblock \emph{\bibinfo{journal}{PNAS: Proceedings of the National Academy of
			Sciences}} \textbf{\bibinfo{volume}{107}}, \bibinfo{pages}{5334--5338}
	(\bibinfo{year}{2010}).
	
	\bibitem{jordan2013contagion}
	\bibinfo{author}{Jordan, J.~J.}, \bibinfo{author}{Rand, D.~G.},
	\bibinfo{author}{Arbesman, S.}, \bibinfo{author}{Fowler, J.~H.} \&
	\bibinfo{author}{Christakis, N.~A.}
	\newblock \bibinfo{title}{Contagion of cooperation in static and fluid social
		networks}.
	\newblock \emph{\bibinfo{journal}{{PLOS ONE}}} \textbf{\bibinfo{volume}{8}},
	\bibinfo{pages}{e66199} (\bibinfo{year}{2013}).
	
	\bibitem{suri2011cooperation}
	\bibinfo{author}{Suri, S.} \& \bibinfo{author}{Watts, D.~J.}
	\newblock \bibinfo{title}{Cooperation and contagion in web-based, networked
		public goods experiments}.
	\newblock \emph{\bibinfo{journal}{{PLOS ONE}}} \textbf{\bibinfo{volume}{6}},
	\bibinfo{pages}{e16836} (\bibinfo{year}{2011}).
	
	\bibitem{macy-generosity}
	\bibinfo{author}{Tsvetkova, M.} \& \bibinfo{author}{Macy, M.~W.}
	\newblock \bibinfo{title}{The social contagion of generosity}.
	\newblock \emph{\bibinfo{journal}{{PLOS ONE}}} \textbf{\bibinfo{volume}{9}},
	\bibinfo{pages}{e87275} (\bibinfo{year}{2014}).
	
	\bibitem{Christakis2008}
	\bibinfo{author}{Christakis, N.} \& \bibinfo{author}{Fowler, J.}
	\newblock \bibinfo{title}{{The Collective Dynamics of Smoking in a Large Social
			Network}}.
	\newblock \emph{\bibinfo{journal}{New England Journal of Medicine}}
	\textbf{\bibinfo{volume}{358}}, \bibinfo{pages}{2249--2258}
	(\bibinfo{year}{2008}).
	
	\bibitem{Fowler2008}
	\bibinfo{author}{Fowler, J.} \& \bibinfo{author}{Christakis, N.}
	\newblock \bibinfo{title}{Dynamic spread of happiness in a large social
		network: Longitudinal analysis over 20 years in the framingham heart study}.
	\newblock \emph{\bibinfo{journal}{British Medical Journal}}
	\textbf{\bibinfo{volume}{337}}, \bibinfo{pages}{1--9} (\bibinfo{year}{2008}).
	
	\bibitem{Pugh2001}
	\bibinfo{author}{Pugh, S.~D.}
	\newblock \bibinfo{title}{Service with a smile: emotional contagion in the
		service encounter}.
	\newblock \emph{\bibinfo{journal}{The Academy of Management Journal}}
	\textbf{\bibinfo{volume}{44}}, \bibinfo{pages}{1018--1027}
	(\bibinfo{year}{2001}).
	
	\bibitem{Madan2010}
	\bibinfo{author}{Madan, A.}, \bibinfo{author}{Cebrian, M.},
	\bibinfo{author}{Lazer, D.} \& \bibinfo{author}{Pentland, A.}
	\newblock \bibinfo{title}{Social sensing for epidemiological behavior change}.
	\newblock In \emph{\bibinfo{booktitle}{Proceedings of the 12th ACM
			international conference on Ubiquitous computing}}, \bibinfo{pages}{291--300}
	(\bibinfo{organization}{ACM}, \bibinfo{year}{2010}).
	
	\bibitem{Howes1985}
	\bibinfo{author}{Howes, M.~J.}, \bibinfo{author}{Hokanson, J.~E.} \&
	\bibinfo{author}{Loewenstein, D.~A.}
	\newblock \bibinfo{title}{Induction of depressive affect after prolonged
		exposure to a mildly depressed individual}.
	\newblock \emph{\bibinfo{journal}{Journal of Personality and Social
			Psychology}} \textbf{\bibinfo{volume}{49}}, \bibinfo{pages}{1110--1113}
	(\bibinfo{year}{1985}).
	
	\bibitem{Hatfield1993}
	\bibinfo{author}{Hatfield, E.}, \bibinfo{author}{Cacioppo, J.} \&
	\bibinfo{author}{Rapson, R.~L.}
	\newblock \emph{\bibinfo{title}{Emotional contagion}}
	(\bibinfo{publisher}{Cambridge University Press}, \bibinfo{year}{1994}).
	
	\bibitem{Barasade2002}
	\bibinfo{author}{Barsade, S.~G.}
	\newblock \bibinfo{title}{{The Ripple Effect: Emotional Contagion and its
			Influence on Group Behavior}}.
	\newblock \emph{\bibinfo{journal}{Administrative Science Quarterly}}
	\textbf{\bibinfo{volume}{47}}, \bibinfo{pages}{644--675}
	(\bibinfo{year}{2002}).
	
	\bibitem{Hill2010}
	\bibinfo{author}{Hill, A.}, \bibinfo{author}{Rand, D.}, \bibinfo{author}{Nowak,
		M.} \& \bibinfo{author}{Christakis, N.}
	\newblock \bibinfo{title}{Emotions as infectious diseases in a large social
		network: the {SIS}a model}.
	\newblock \emph{\bibinfo{journal}{Proceedings of Royal Society B: Biological
			Sciences}} \textbf{\bibinfo{volume}{277}}, \bibinfo{pages}{3827--3835}
	(\bibinfo{year}{2010}).
	
	\bibitem{hill2010infectious}
	\bibinfo{author}{Hill, A.~L.}, \bibinfo{author}{Rand, D.~G.},
	\bibinfo{author}{Nowak, M.~A.} \& \bibinfo{author}{Christakis, N.~A.}
	\newblock \bibinfo{title}{Infectious disease modeling of social contagion in
		networks}.
	\newblock \emph{\bibinfo{journal}{{PLOS} computational biology}}
	\textbf{\bibinfo{volume}{6}}, \bibinfo{pages}{e1000968}
	(\bibinfo{year}{2010}).
	
	\bibitem{shalizi2011homophily}
	\bibinfo{author}{Shalizi, C.~R.} \& \bibinfo{author}{Thomas, A.~C.}
	\newblock \bibinfo{title}{Homophily and contagion are generically confounded in
		observational social network studies}.
	\newblock \emph{\bibinfo{journal}{Sociological Methods \& Research}}
	\textbf{\bibinfo{volume}{40}}, \bibinfo{pages}{211--239}
	(\bibinfo{year}{2011}).
	
	\bibitem{Steeg2013}
	\bibinfo{author}{Steeg, G.~V.} \& \bibinfo{author}{Galstyan, A.}
	\newblock \bibinfo{title}{{Statistical Tests for Contagion in Observational
			Social Network Studies}}.
	\newblock In \emph{\bibinfo{booktitle}{Paper presented at International
			Conference on Artificial Intelligence and Statistics}}
	(\bibinfo{year}{2013}).
	
	\bibitem{aral2009distinguishing}
	\bibinfo{author}{Aral, S.}, \bibinfo{author}{Muchnik, L.} \&
	\bibinfo{author}{Sundararajan, A.}
	\newblock \bibinfo{title}{Distinguishing influence-based contagion from
		homophily-driven diffusion in dynamic networks}.
	\newblock \emph{\bibinfo{journal}{Proceedings of the National Academy of
			Sciences}} \textbf{\bibinfo{volume}{106}}, \bibinfo{pages}{21544--21549}
	(\bibinfo{year}{2009}).
	
	\bibitem{dodds2005generalized}
	\bibinfo{author}{Dodds, P.~S.} \& \bibinfo{author}{Watts, D.~J.}
	\newblock \bibinfo{title}{A generalized model of social and biological
		contagion}.
	\newblock \emph{\bibinfo{journal}{Journal of Theoretical Biology}}
	\textbf{\bibinfo{volume}{232}}, \bibinfo{pages}{587--604}
	(\bibinfo{year}{2005}).
	
	\bibitem{friedkin2010attitude}
	\bibinfo{author}{Friedkin, N.~E.}
	\newblock \bibinfo{title}{The attitude-behavior linkage in behavioral
		cascades}.
	\newblock \emph{\bibinfo{journal}{Social Psychology Quarterly}}
	\textbf{\bibinfo{volume}{73}}, \bibinfo{pages}{196--213}
	(\bibinfo{year}{2010}).
	
	\bibitem{smilkov2014beyond}
	\bibinfo{author}{Smilkov, D.}, \bibinfo{author}{Hidalgo, C.~A.} \&
	\bibinfo{author}{Kocarev, L.}
	\newblock \bibinfo{title}{Beyond network structure: How heterogeneous
		susceptibility modulates the spread of epidemics}.
	\newblock \emph{\bibinfo{journal}{Scientific reports}}
	\textbf{\bibinfo{volume}{4}}, \bibinfo{pages}{1--7} (\bibinfo{year}{2014}).
	
	\bibitem{Zelenski2000}
	\bibinfo{author}{Zelenski, J.~M.} \& \bibinfo{author}{Larsen, R.~J.}
	\newblock \bibinfo{title}{The distribution of basic emotions in everyday life:
		A state and trait perspective from experience sampling data}.
	\newblock \emph{\bibinfo{journal}{Journal of Research in Personality}}
	\textbf{\bibinfo{volume}{34}}, \bibinfo{pages}{178--197}
	(\bibinfo{year}{2000}).
	
	\bibitem{Fleeson2004}
	\bibinfo{author}{Fleeson, W.}
	\newblock \bibinfo{title}{Moving personality beyond the person-situation debate
		the challenge and the opportunity of within-person variability}.
	\newblock \emph{\bibinfo{journal}{Current Directions in Psychological Science}}
	\textbf{\bibinfo{volume}{13}}, \bibinfo{pages}{83--87}
	(\bibinfo{year}{2004}).
	
	\bibitem{FleesonNoftle2009}
	\bibinfo{author}{Fleeson, W.} \& \bibinfo{author}{Noftle, E.}
	\newblock \bibinfo{title}{The end of the person-situation debate: an emerging
		synthesis in the answer to the consistency question}.
	\newblock \emph{\bibinfo{journal}{Social and Personality Psychology Compass}}
	\textbf{\bibinfo{volume}{2}}, \bibinfo{pages}{1667--1684}
	(\bibinfo{year}{2008}).
	
	\bibitem{costa2008revised}
	\bibinfo{author}{Costa, P.~T.} \& \bibinfo{author}{McCrae, R.~R.}
	\newblock \bibinfo{title}{The revised neo personality inventory (neo-pi-r)}.
	\newblock \emph{\bibinfo{journal}{The SAGE handbook of personality theory and
			assessment}} \textbf{\bibinfo{volume}{2}}, \bibinfo{pages}{179--198}
	(\bibinfo{year}{2008}).
	
	\bibitem{Fleeson2001}
	\bibinfo{author}{Fleeson, W.}
	\newblock \bibinfo{title}{Toward a structure- and process-integrated view of
		personality: traits as density distribution of states}.
	\newblock \emph{\bibinfo{journal}{Journal of Personality and Social
			Psychology}} \textbf{\bibinfo{volume}{80}}, \bibinfo{pages}{1011--1027}
	(\bibinfo{year}{2001}).
	
	\bibitem{friedkin2011social}
	\bibinfo{author}{Friedkin, N.~E.} \& \bibinfo{author}{Johnsen, E.~C.}
	\newblock \emph{\bibinfo{title}{Social influence network theory: a sociological
			examination of small group dynamics}}, vol.~\bibinfo{volume}{33}
	(\bibinfo{publisher}{Cambridge University Press}, \bibinfo{year}{2011}).
	
	\bibitem{Watson1988}
	\bibinfo{author}{Watson, D.}, \bibinfo{author}{Clark, L.~A.} \&
	\bibinfo{author}{Tellegen, A.}
	\newblock \bibinfo{title}{{Development and validation of brief measures of
			positive and negative affect: the PANAS scales.}}
	\newblock \emph{\bibinfo{journal}{Journal of Personality and Social
			Psychology}} \textbf{\bibinfo{volume}{54}}, \bibinfo{pages}{1063--1070}
	(\bibinfo{year}{1988}).
	\newblock \urlprefix\url{http://view.ncbi.nlm.nih.gov/pubmed/3397865}.
	
	\bibitem{chartrand1999chameleon}
	\bibinfo{author}{Chartrand, T.~L.} \& \bibinfo{author}{Bargh, J.~A.}
	\newblock \bibinfo{title}{The chameleon effect: The perception--behavior link
		and social interaction.}
	\newblock \emph{\bibinfo{journal}{Journal of personality and social
			psychology}} \textbf{\bibinfo{volume}{76}}, \bibinfo{pages}{893}
	(\bibinfo{year}{1999}).
	
	\bibitem{tiedens2003power}
	\bibinfo{author}{Tiedens, L.~Z.} \& \bibinfo{author}{Fragale, A.~R.}
	\newblock \bibinfo{title}{Power moves: complementarity in dominant and
		submissive nonverbal behavior.}
	\newblock \emph{\bibinfo{journal}{Journal of personality and social
			psychology}} \textbf{\bibinfo{volume}{84}}, \bibinfo{pages}{558}
	(\bibinfo{year}{2003}).
	
	\bibitem{Pentland2013}
	\bibinfo{author}{Pentland, A.}
	\newblock \bibinfo{title}{{The New Science of Building Great Teams}}.
	\newblock \emph{\bibinfo{journal}{Harvard Business Review}}
	\textbf{\bibinfo{volume}{90}}, \bibinfo{pages}{60--70}
	(\bibinfo{year}{2012}).
	
	\bibitem{Barrick2001}
	\bibinfo{author}{Barrick, M.~R.}, \bibinfo{author}{Mount, M.~K.} \&
	\bibinfo{author}{Judge, T.~A.}
	\newblock \bibinfo{title}{{Personality and Performance at the Beginning of the
			New Millennium: What Do We Know and Where Do We Go Next?}}
	\newblock \emph{\bibinfo{journal}{International Journal of Selection and
			Assessment}} \textbf{\bibinfo{volume}{9}}, \bibinfo{pages}{9--30}
	(\bibinfo{year}{2001}).
	
	\bibitem{Judge2002}
	\bibinfo{author}{Ilies, R.} \& \bibinfo{author}{Judge, T.~A.}
	\newblock \bibinfo{title}{{Understanding the dynamic relationships among
			personality, mood, and job satisfaction: A field experience sampling study}}.
	\newblock \emph{\bibinfo{journal}{Organizational Behavior and Human Decision
			Processes}} \textbf{\bibinfo{volume}{89}}, \bibinfo{pages}{1119--1139}
	(\bibinfo{year}{2002}).
	
	\bibitem{Bell2007}
	\bibinfo{author}{Bell, S.~T.}
	\newblock \bibinfo{title}{Deep-level composition variables as predictors of
		team performance: a meta-analysis.}
	\newblock \emph{\bibinfo{journal}{Journal of Applied Psychology}}
	\textbf{\bibinfo{volume}{92}}, \bibinfo{pages}{595--615}
	(\bibinfo{year}{2007}).
	
	\bibitem{Barsade2003}
	\bibinfo{author}{Barsade, S.}, \bibinfo{author}{Brief, A.~P.} \&
	\bibinfo{author}{Spataro, S.~E.}
	\newblock \bibinfo{title}{The affective revolution in organizational behavior:
		The emergence of a paradigm}.
	\newblock \emph{\bibinfo{journal}{Organizational Behavior: The State of the
			Science}} \bibinfo{pages}{3--52} (\bibinfo{year}{2003}).
	
	\bibitem{Olguin2009}
	\bibinfo{author}{Olgu{\'\i}n~Olgu{\'\i}n, D.} \emph{et~al.}
	\newblock \bibinfo{title}{Sensible organizations: Technology and methodology
		for automatically measuring organizational behavior}.
	\newblock \emph{\bibinfo{journal}{IEEE Transactions on Systems, Man, and
			Cybernetics, Part B (Cybernetics)}} \textbf{\bibinfo{volume}{39}},
	\bibinfo{pages}{43--55} (\bibinfo{year}{2009}).
	
	\bibitem{pentland2010honest}
	\bibinfo{author}{Pentland, A.~S.}
	\newblock \emph{\bibinfo{title}{Honest signals}} (\bibinfo{publisher}{MIT
		press}, \bibinfo{year}{2010}).
	
	\bibitem{Cattuto2010}
	\bibinfo{author}{Cattuto, C.} \emph{et~al.}
	\newblock \bibinfo{title}{Dynamics of person-to-person interactions from
		distributed rfid sensor networks}.
	\newblock \emph{\bibinfo{journal}{{PLOS ONE}}} \textbf{\bibinfo{volume}{5}}
	(\bibinfo{year}{2010}).
	
	\bibitem{Salathe2010}
	\bibinfo{author}{Salath{\'e}, M.} \emph{et~al.}
	\newblock \bibinfo{title}{A high-resolution human contact network for
		infectious disease transmission}.
	\newblock \emph{\bibinfo{journal}{Proceedings of the National Academy of
			Sciences}} \textbf{\bibinfo{volume}{107}}, \bibinfo{pages}{22020--22025}
	(\bibinfo{year}{2010}).
	
	\bibitem{Stehle2011}
	\bibinfo{author}{Stehl{\'e}, J.} \emph{et~al.}
	\newblock \bibinfo{title}{High-resolution measurements of face-to-face contact
		patterns in a primary school}.
	\newblock \emph{\bibinfo{journal}{{PLOS ONE}}} \textbf{\bibinfo{volume}{6}},
	\bibinfo{pages}{e23176} (\bibinfo{year}{2011}).
	
	\bibitem{Todd2007}
	\bibinfo{author}{Todd, P.~M.}, \bibinfo{author}{Penke, L.},
	\bibinfo{author}{Fasolo, B.} \& \bibinfo{author}{Lenton, A.~P.}
	\newblock \bibinfo{title}{Different cognitive processes underlie human mate
		choices and mate preferences}.
	\newblock \emph{\bibinfo{journal}{Proceedings of the National Academy of
			Sciences}} \textbf{\bibinfo{volume}{104}}, \bibinfo{pages}{15011--15016}
	(\bibinfo{year}{2007}).
	
	\bibitem{csikszentmihalyi2003happiness}
	\bibinfo{author}{Csikszentmihalyi, M.} \& \bibinfo{author}{Hunter, J.}
	\newblock \bibinfo{title}{Happiness in everyday life: The uses of experience
		sampling}.
	\newblock \emph{\bibinfo{journal}{Journal of Happiness Studies}}
	\textbf{\bibinfo{volume}{4}}, \bibinfo{pages}{185--199}
	(\bibinfo{year}{2003}).
	
	\bibitem{Bruno2012}
	\bibinfo{author}{Lepri, B.} \emph{et~al.}
	\newblock \bibinfo{title}{The sociometric badges corpus: A multilevel
		behavioral dataset for social behavior in complex organizations}.
	\newblock In \emph{\bibinfo{booktitle}{Privacy, Security, Risk and Trust
			(PASSAT), International Confernece on Social Computing (SocialCom)}},
	\bibinfo{pages}{623--628} (\bibinfo{organization}{IEEE},
	\bibinfo{year}{2012}).
	
	\bibitem{banerjee2013diffusion}
	\bibinfo{author}{Banerjee, A.}, \bibinfo{author}{Chandrasekhar, A.~G.},
	\bibinfo{author}{Duflo, E.} \& \bibinfo{author}{Jackson, M.~O.}
	\newblock \bibinfo{title}{The diffusion of microfinance}.
	\newblock \emph{\bibinfo{journal}{Science}} \textbf{\bibinfo{volume}{341}}
	(\bibinfo{year}{2013}).
	
	\bibitem{golder2011diurnal}
	\bibinfo{author}{Golder, S.~A.} \& \bibinfo{author}{Macy, M.~W.}
	\newblock \bibinfo{title}{Diurnal and seasonal mood vary with work, sleep, and
		daylength across diverse cultures}.
	\newblock \emph{\bibinfo{journal}{Science}} \textbf{\bibinfo{volume}{333}},
	\bibinfo{pages}{1878--1881} (\bibinfo{year}{2011}).
	
	\bibitem{killingsworth2010wandering}
	\bibinfo{author}{Killingsworth, M.~A.} \& \bibinfo{author}{Gilbert, D.~T.}
	\newblock \bibinfo{title}{A wandering mind is an unhappy mind}.
	\newblock \emph{\bibinfo{journal}{Science}} \textbf{\bibinfo{volume}{330}},
	\bibinfo{pages}{932--932} (\bibinfo{year}{2010}).
	
	\bibitem{aral2012identifying}
	\bibinfo{author}{Aral, S.} \& \bibinfo{author}{Walker, D.}
	\newblock \bibinfo{title}{Identifying influential and susceptible members of
		social networks}.
	\newblock \emph{\bibinfo{journal}{Science}} \textbf{\bibinfo{volume}{337}},
	\bibinfo{pages}{337--341} (\bibinfo{year}{2012}).
	
	\bibitem{royston2009prognosis}
	Patrick Royston, Karel~GM Moons, Douglas~G Altman, and Yvonne Vergouwe.
	\newblock Prognosis and prognostic research: developing a prognostic model.
	\newblock {\em Bmj}, 338:b604, 2009.
	
	
	\bibitem{muchnik2013social}
	\bibinfo{author}{Muchnik, L.}, \bibinfo{author}{Aral, S.} \&
	\bibinfo{author}{Taylor, S.~J.}
	\newblock \bibinfo{title}{Social influence bias: A randomized experiment}.
	\newblock \emph{\bibinfo{journal}{Science}} \textbf{\bibinfo{volume}{341}},
	\bibinfo{pages}{647--651} (\bibinfo{year}{2013}).
	
	\bibitem{Perguini2002}
	M.~Perugini and Blas L.
	\newblock The big five marker scales (bfms) and the italian ab5c taxonomy:
	Analyses from an emic-etic perspective, 2012.
	
	\bibitem{Tellegen2008}
	Auke Tellegen and Niels~G. Waller.
	\newblock Exploring personality through test construction: Development of the
	multidimensional personality questionnaire.
	\newblock In Gregory~J. Boyle, Gerald Matthews, and Donald~H. Saklofske,
	editors, {\em The SAGE Handbook of Personality Theory and Assessment: Volume
		2 --- Personality Measurement and Testing}, pages 261--293. SAGE Publications
	Ltd, 2008.
	
	
	\bibitem{Gosling03avery}
	Samuel Gosling, Peter Rentfrow, and William Swann.
	\newblock A very brief measure of the big-five personality domains.
	\newblock {\em Journal of Research in Personality}, 37(6):504--528, 2003.
	
	\bibitem{watson1988development}
	David Watson, Lee~A Clark, and Auke Tellegen.
	\newblock Development and validation of brief measures of positive and negative
	affect: the {PANAS} scales.
	\newblock {\em Journal of personality and social psychology}, 54(6):1063, 1988.
	
	\bibitem{pan2001akaike}
	Wei Pan.
	\newblock Akaike's information criterion in generalized estimating equations.
	\newblock {\em Biometrics}, 57(1):120--125, 2001.
	
\bibitem{tellegen1985structures}
Auke Tellegen.
\newblock Structures of mood and personality and their relevance to assessing
anxiety, with an emphasis on self-report.
\newblock 1985.

\bibitem{watson1984negative}
David Watson and Lee~A Clark.
\newblock Negative affectivity: the disposition to experience aversive
emotional states.
\newblock {\em Psychological bulletin}, 96(3):465, 1984.

\bibitem{yandell1997practical}
Brian S. Yandell.
	\newblock \emph{\bibinfo{title}{Practical data analysis for designed experiments}} (\bibinfo{year}{1997}).
	
\bibitem{conner2009experience}
Tamlin S. Conner and Howard Tennen and William Fleeson and Lisa Feldman Barrett
\newblock Experience sampling methods: A modern idiographic approach to personality research.
\newblock {\em Social and personality psychology compass}, 3, 2009.

\bibitem{csikszentmihalyi1987validity}
Mihaly Csikszentmihalyi and Reed Larson
\newblock Validity and reliability of the experience-sampling method.
\newblock {The journal of nervous and mental disease}, 175:9, 1987.


	
\end{thebibliography}
\end{document}